\documentclass[a4paper,iop,numberedappendix]{emulateapj}
\usepackage[colorlinks=true,linkcolor=blue,anchorcolor=blue,citecolor=blue,urlcolor=blue]{hyperref} 
\usepackage{color}
\hypersetup{pdfauthor={Keiichi Umetsu}, pdftitle={CLASH: Full Lensing Analysis of MACS1206}}
\usepackage{natbib}
  
\newcommand{\simgt}{\lower.5ex\hbox{$\; \buildrel > \over \sim \;$}}
\newcommand{\simlt}{\lower.5ex\hbox{$\; \buildrel < \over \sim \;$}}

\def\btheta{\mbox{\boldmath $\theta$}} 
\def\bnabla{\mbox{\boldmath $\nabla$}}
\def\bkappa{\mbox{\boldmath $\kappa$}}

\def\bd{\mbox{\boldmath $d$}}
\def\bs{\mbox{\boldmath $s$}}
\def\bp{\mbox{\boldmath $p$}}
\def\bs{\mbox{\boldmath $s$}}
\def\bc{\mbox{\boldmath $c$}}
\def\bkappa{\mbox{\boldmath $\kappa$}}
    
\lefthead{Umetsu et al.}
\righthead{CLASH: Full Lensing Analysis of MACS1206}  

\begin{document}

\title{CLASH: Mass Distribution in and around MACS\,J1206.2-0847 from a Full Cluster Lensing Analysis{*}}

\author{Keiichi Umetsu\altaffilmark{1}}
\author{Elinor Medezinski\altaffilmark{2}}      
\author{Mario Nonino\altaffilmark{3}}           
\author{Julian Merten\altaffilmark{4}}          
\author{Adi Zitrin\altaffilmark{5}}             
\author{Alberto Molino\altaffilmark{6}}         
\author{Claudio Grillo\altaffilmark{7}}         
\author{Mauricio Carrasco\altaffilmark{8,9}}    
\author{Megan Donahue\altaffilmark{10}}         
\author{Andisheh Mahdavi\altaffilmark{11}}      
\author{Dan Coe\altaffilmark{12}}               
\author{Marc Postman\altaffilmark{12}}          
\author{Anton Koekemoer\altaffilmark{12}}
\author{Nicole Czakon\altaffilmark{13}}         
\author{Jack Sayers\altaffilmark{13}}           
\author{Tony Mroczkowski\altaffilmark{14,4,13}} 
\author{Sunil Golwala\altaffilmark{13}}         
\author{Patrick M. Koch\altaffilmark{1}}        
\author{Kai-Yang Lin\altaffilmark{1}}           
\author{Sandor M. Molnar\altaffilmark{15}}      
\author{Piero Rosati\altaffilmark{9}}           
\author{Italo Balestra\altaffilmark{3}}         
\author{Amata Mercurio\altaffilmark{16}}        
\author{Marco Scodeggio\altaffilmark{17}}       
\author{Andrea Biviano\altaffilmark{3}}         
\author{Timo Anguita\altaffilmark{8,18}}        
\author{Leopoldo Infante\altaffilmark{8}}       
\author{Gregor Seidel\altaffilmark{18}}         
\author{Irene Sendra\altaffilmark{19}}          
\author{Stephanie Jouvel\altaffilmark{20,21}}   
\author{Ole Host\altaffilmark{20}}              
\author{Doron Lemze\altaffilmark{2}}            
\author{Tom Broadhurst\altaffilmark{19,22}}     
\author{Massimo Meneghetti\altaffilmark{23}}    
\author{Leonidas Moustakas\altaffilmark{4}}     
\author{Matthias Bartelmann\altaffilmark{5}} 
\author{Narciso Ben\'itez\altaffilmark{6}}
\author{Rychard Bouwens\altaffilmark{24}}
\author{Larry Bradley\altaffilmark{12}}  
\author{Holland Ford\altaffilmark{2}}  
\author{Yolanda Jim\'enez-Teja\altaffilmark{6}}  
\author{Daniel Kelson\altaffilmark{25}} 
\author{Ofer Lahav\altaffilmark{20}}  
\author{Peter Melchior\altaffilmark{26}}
\author{John Moustakas\altaffilmark{27}}
\author{Sara Ogaz\altaffilmark{12}}  
\author{Stella Seitz\altaffilmark{28}} 
\author{Wei Zheng\altaffilmark{2}} 
\email{keiichi@asiaa.sinica.edu.tw}
\altaffiltext{*}
 {Based in part on data collected at the Subaru Telescope,
  which is operated by the National Astronomical Society of Japan.}
\altaffiltext{1}{Institute of Astronomy and Astrophysics, Academia
Sinica, P.~O. Box 23-141, Taipei 10617, Taiwan.}  

\altaffiltext{1}{Institute of Astronomy and Astrophysics, Academia
Sinica, P.~O. Box 23-141, Taipei 10617, Taiwan.} 
\altaffiltext{2}{Johns Hopkins University} 
\altaffiltext{3}{INAF/Osservatorio Astronomico di Trieste}
\altaffiltext{4}{Jet Propulsion Laboratory, California Institute of Technology} 
\altaffiltext{5}{Universitat Heidelberg} 
\altaffiltext{6}{Instituto de Astrof\'isica de Andaluc\'ia (CSIC)} 
\altaffiltext{7}{Dark Cosmology Centre, University of Copenhagen}
\altaffiltext{8}{Pontificia Universidad Cat\'olica de Chile}
\altaffiltext{9}{ESO-European Southern Observatory} 
\altaffiltext{10}{Michigan State University} 
\altaffiltext{11}{San Francisco State University} 
\altaffiltext{12}{Space Telescope Science Institute} 
\altaffiltext{13}{California Institute of Technology} 
\altaffiltext{14}{NASA Einstein Postdoctoral Fellow}
\altaffiltext{15}{LeCosPA Center, National Taiwan University} 
\altaffiltext{16}{INAF-Osservatorio Astronomico di Capodimonte}   
\altaffiltext{17}{INAF-IASF Milano}
\altaffiltext{18}{Max-Planck-Institut f\"ur Astronomie, Heidelberg} 
\altaffiltext{19}{University of the Basque Country UPV/EHU} 
\altaffiltext{20}{University College London} 
\altaffiltext{21}{Institut de Ci\'encies de l'Espai}  
\altaffiltext{22}{Ikerbasque, Basque Foundation for Science}
\altaffiltext{23}{INAF, Osservatorio Astronomico di Bologna} 
\altaffiltext{24}{Leiden Observatory, Leiden University} 
\altaffiltext{25}{Observatories of the Carnegie Institution of Washington} 
\altaffiltext{26}{Ohio State University} 
\altaffiltext{27}{University of California at San Diego} 
\altaffiltext{28}{Universit\"ats-Sternwarte, M\"unchen} 

\begin{abstract}
We derive an accurate mass distribution of the galaxy
 cluster MACS\,J1206.2-0847 ($z=0.439$) from a combined weak-lensing
 distortion, magnification, and strong-lensing  analysis of wide-field
 Subaru $BVR_{\rm c}I_{\rm c}z'$ imaging and our recent 16-band {\em
 Hubble Space Telescope} observations taken as part of the
 Cluster  Lensing And Supernova survey with  Hubble (CLASH) program.
We find good agreement in the regions of overlap between several
weak and strong lensing mass reconstructions using a wide variety  of
 modeling methods, ensuring consistency. The Subaru data reveal the
 presence of a surrounding large scale structure with the major axis
 running approximately north-west south-east (NW-SE), aligned with the
 cluster and its brightest galaxy shapes, 
showing elongation with a
 $\sim 2:1$ axis ratio in the plane of the sky.
Our full-lensing mass profile exhibits a shallow profile slope
$d\ln\Sigma/d\ln R\sim -1$ at cluster outskirts
 ($R\simgt  1\,$Mpc\,$h^{-1}$), 
whereas the mass distribution excluding the NW-SE excess regions
steepens further out,  
well described by the Navarro-Frenk-White form.
Assuming a spherical halo, we obtain a virial mass   
$M_{\rm vir}=(1.1\pm 0.2 \pm 0.1)\times 10^{15}M_\odot\,h^{-1}$ 
and a halo concentration 
$c_{\rm vir}=6.9 \pm 1.0 \pm 1.2$ ($c_{\rm vir}\sim 5.7$ when the
 central 50\,kpc\,$h^{-1}$ is excluded), which falls in the range 
 $4\simlt\langle c \rangle \simlt 7$ of average  $c(M,z)$ predictions
 for relaxed  clusters from recent $\Lambda$ 
 cold dark matter simulations.
Our full lensing results are found to be in agreement with X-ray mass
 measurements where the data overlap, and when
 combined with {\em Chandra} gas mass measurements, yield a
cumulative gas mass fraction of $13.7^{+4.5}_{-3.0}\%$ at
$0.7\,$Mpc\,$h^{-1}(\approx 1.7\,r_{2500})$,
a typical value observed for high mass clusters. 
\end{abstract}
 

\section{Introduction} 
\label{sec:intro}

Clusters of galaxies are the largest self-gravitating systems in the
universe. 
These massive clusters 
contain rich astrophysical and cosmological information about  
the initial conditions for cosmic structure formation
and assembly of structure over cosmic time.
Statistical and detailed individual properties of clusters can therefore
provide fundamental constraints on models of cosmic structure
formation \cite[e.g.,][]{Allen+2004,Vikhlinin+2009_CCC3},
the unknown nature of dark matter 
\citep[DM, hereafter;][]{2004ApJ...606..819M,Clowe+2006_Bullet},
and possible modifications of the law of gravity 
\citep{Narikawa+Yamamoto2012},
complementing cosmic microwave background,
galaxy clustering,
and Type Ia supernova 
observations
\citep{Komatsu+2011_WMAP7,Percival+2010_BAO,Riess+1998}.

Observations of clusters have provided independent pieces of empirical
evidence for the existence of DM
\citep[e.g.,][]{Zwicky1959,2004ApJ...606..819M,Clowe+2006_Bullet,Okabe&Umetsu08,2007ApJ...668..806M}.
A prime example of this comes from combined X-ray and lensing
observations of the ``Bullet system'', which is 
understood to be 
the result of a high-speed collision of two cluster components occurring 
approximately in the plane of the sky, displaying a prominent bow shock
proceeding the cool, bullet-like gas subcluster, lying between the two
distinct clusters \citep{2004ApJ...606..819M}.  
For this system, the bulk of mass is shown to be associated with the
bimodal distribution of cluster member galaxies, supporting that DM is
effectively collisionless as galaxies on sub-Mpc scales \citep{Clowe+2006_Bullet}.
Such displacements
between the gas and mass distributions are quite common in merging systems,
and exhibit a complex variety of merging configurations
\citep{Okabe+Umetsu2008,2007ApJ...668..806M,Merten+2011}. 
   
Substantial progress has been made  through numerical simulation
in understanding the formation and structure of 
collisionless DM halos in  quasi equilibrium,
governed by nonlinear gravitational growth of
cosmic density perturbations. 
In the standard $\Lambda$ cold dark matter
($\Lambda$CDM) paradigm of  
hierarchical structure formation,  
cluster-sized DM halos 
form through successive mergers of 
smaller halos as well as through smooth accretion of matter
along surrounding filamentary structures
\citep{Colberg+2000,Shaw+2006,Gao+2012_Phoenix}.   
In this context, the hierarchical build up of clusters proceeds in a
highly anisotropic configuration where infall and merging of matter tend
to occur along preferential directions \citep{Colberg+2005}, leading to  
the emergence of the filamentary network of matter, 
as observed in large galaxy redshift surveys
\citep[e.g.,][]{Colless+2011_2dF,Tegmark+2004_SDSS,Geller+2011}.
Cluster halos are located at dense nodes where the filaments intersect,
generally triaxial reflecting the collisionless nature of DM, and
elongated in the preferential infall direction of subhalos, namely along
surrounding filaments \citep{Shaw+2006}. 

The internal structure of DM halos constitutes 
one of the most distinct predictions for the CDM paradigm.
$N$-body simulations of collisionless CDM established
a nearly self-similar form for the spherically-averaged density
profile $\langle\rho(r) \rangle$ of DM halos \citep[][hereafter, Navarro-Frenk-White
(NFW)]{1997ApJ...490..493N} over a wide range of halo masses, with 
some intrinsic variance associated with the
mass assembly histories of individual halos
\citep{Jing+Suto2000,Tasitsiomi+2004,Graham+2006,Navarro+2010,Gao+2012_Phoenix}.
The logarithmic gradient $\gamma_{\rm 3D}(r)= -d\ln{\rho}/d\ln{r}$ of
the NFW form flattens progressively toward the center, with an
inner slope flatter than a purely isothermal structure 
($\gamma_{\rm 3D}=2$) interior to the inner characteristic radius $r_s$
providing a 
distinctive, fundamental prediction for the empirical form of CDM halos.
A useful index of the degree of
concentration is $c_{\rm vir}=r_{\rm vir}/r_s$, which compares the
virial radius $r_{\rm vir}$ to $r_s$.
Halo concentration is predicted to correlate with halo mass since DM
halos that are more massive collapse later when the mean background
density of the universe is correspondingly lower
\citep{2001MNRAS.321..559B,Zhao+2003,2007MNRAS.381.1450N}.  This
prediction for the halo $c_{\rm vir}$--$M_{\rm vir}$ relation and its
evolution has been examined by several independent large scale simulations
\cite[e.g.,][]{1997ApJ...490..493N,2001MNRAS.321..559B,
2007MNRAS.381.1450N,Duffy+2008,Klypin+2011,Bhat+2011},
with sufficient detail to establish the inherent scatter of this
relation around the mean, arising from variations in the formation epoch
of individual halos of given mass
\citep{2002ApJ...568...52W,2007MNRAS.381.1450N,Zhao+2009}.

Galaxy clusters act as powerful gravitational lenses
\citep[e.g.,][]{2001PhR...340..291B,Umetsu2010_Fermi,Kneib+Natarajan2011},
providing a direct probe for testing these
well-defined predictions of halo structure 
because they are expected to have a relatively shallow mass profile with 
a pronounced radial curvature. 
A detailed examination of the $\Lambda$CDM predictions by cluster
lensing has been the focus of our preceding work
\citep{BTU+05,2007MPLA...22.2099U,BUM+08,UB2008,Umetsu+2009,Umetsu+2010_CL0024,Umetsu+2011a,Umetsu+2011b}.

Recent detailed lensing analyses have shown that
the projected cluster mass profiles
constructed from combined weak and strong lensing data
have a gradually steepening logarithmic gradient,
in agreement with the predicted
form for the family of collisionless CDM halos in virial
equilibrium 
\citep{2003A&A...403...11G,BTU+05,2007ApJ...668..643L,BUM+08,UB2008,Newman+2009_A611,Umetsu+2010_CL0024,Umetsu+2011a,Umetsu+2011b,Zitrin+2010_A1703,Zitrin+2011_A383,Oguri+2012_SGAS,Coe+2012_A2261}. 
Intriguingly, however, some of these results reveal a relatively high 
degree of mass concentration in high-mass lensing clusters
\citep[e.g.,][]{2003A&A...403...11G,2003ApJ...598..804K,BUM+08,Oguri+2009_Subaru,Zitrin+2011_A383},
lying well above the $c_{\rm vir}$--$M_{\rm vir}$ relation for
cluster-sized halos ($c_{\rm vir}\sim $4--5 for CDM halos with $M_{\rm
vir}\simgt 10^{15}M_\odot$ in the local universe)
predicted by the $\Lambda$CDM model, despite careful
attempts to correct for sizable ($\sim 50$--100$\%$)
projection and selection biases 
inherent to lensing by triaxial halos
\citep{2007ApJ...654..714H,Meneghetti+2010a,Meneghetti+2011}.  
The effects of baryons on the total mass profile are generally found to
only modify cluster concentrations at the $\sim 10\%$ level
\citep{Mead+2010_AGN,Duffy+2010}, although some studies suggest that low
mass systems ($M_{\rm vir}\simlt 5 \times 10^{14}M_\odot$) 
may be significantly affected by the effects of baryonic cooling
\citep{Fedeli2011,Oguri+2012_SGAS}. 
This apparent overconcentration of lensing clusters 
is also indicated by the generally
large Einstein radii determined from strong-lensing observations
\citep{Broadhurst+Barkana2008,Meneghetti+2010_MARENOSTRUM,Zitrin+2011_MACS}.
These lensing results could suggest either substantial additional mass
projected along the line of sight, due partly to halo
triaxiality \citep{2005ApJ...632..841O}, or an intrinsically
higher-than-predicted concentration of mass; the latter could imply that
clusters formed earlier than predicted by $N$-body simulations of the
current concordance $\Lambda$CDM cosmology.

The Cluster Lensing And Supernova survey with Hubble
\citep[CLASH,][]{Postman+2012_CLASH}\footnote{\href{http://www.stsci.edu/\%7Epostman/CLASH/}{http://www.stsci.edu/$\sim$postman/CLASH}}
has been in progress to obtain accurate cluster
mass profiles for a sizable sample of
representative clusters by combining high-quality strong- and
weak-lensing measurements,
in combination with the complementary Subaru wide-field imaging
\citep[e.g.,][]{Umetsu+2011a,Umetsu+2011b}.
CLASH is a 524-orbit multi-cycle treasury {\em Hubble Space Telescope} 
({\em HST})
program to observe 25 clusters of galaxies at $0.18<z<0.89$, each in 16
filters with the Wide Field Camera 3 \citep[WFC3;][]{Kimble+2008} and
the Advanced Camera for Surveys \citep[ACS;][]{Holland+2003_ACS},
ranging from the UV, through the optical, to the IR.
Importantly, 20 CLASH clusters were X-ray selected to be massive
and relatively relaxed.  This selection avoids the strong bias toward
high concentrations in previously well-studied clusters selected for
their strong lensing strength,
allowing us to meaningfully examine the $c$--$M$
relation over a  sufficiently wide mass and redshift range for a cluster
sample that is largely free of lensing bias \citep{Postman+2012_CLASH}. 

In this paper we present a comprehensive weak and strong lensing
analysis of the X-ray selected CLASH cluster  MACS\,J1206.2-0847
(MACS1206, hereafter; see Table \ref{tab:cluster}) at $z=0.439$ 
based on 
the Subaru wide-field $BVR_{\rm c}I_{\rm c}z'$ imaging,
combined with our recent CLASH {\em HST} imaging and 
VLT/VIMOS spectroscopic observations
presented in \citet{Zitrin+2012_M1206}, who carried out a
detailed strong-lensing analysis of the cluster. 
MACS1206 is an X-ray luminous cluster \citep{Ebeling+2009_M1206},
originally discovered  in the Massive Cluster Survey
\citep[MACS,][]{Ebeling+2001_MACS,Ebeling+2009_M1206}.
Therefore, it is an interesting target for detailed lensing analyses to
compare with well-studied, lensing-selected clusters
\citep[e.g.,][]{Umetsu+2011a,Umetsu+2011b,Oguri+2009_Subaru,Oguri+2012_SGAS}. 

The paper is organized as follows. 
In Section \ref{sec:basis} we briefly summarize the basic theory of
cluster weak gravitational lensing. 
In Section \ref{sec:subaru}, we describe details of the full
weak-lensing analysis of Subaru observations. 
In Section \ref{sec:sl}, 
we present results from several semi-independent strong-lensing analyses
to test the consistency of our strong-lens modeling.
In Section \ref{sec:wl}
we derive cluster weak-lensing profiles from Subaru data.
In Section \ref{sec:mass} 
we combine our weak-lensing measurements with inner strong-lensing based 
information from CLASH {\em HST} observations to make a full
determination of the cluster mass profiles; then, we examine the radial
dependence of the cluster mass distribution based on the full lensing
analysis. 
In Section \ref{sec:discussion}, we assess 
carefully various sources of potential systematic uncertainties
in the cluster mass and concentration measurements, and
discuss our results along with our complementary X-ray and
Sunyaev-Zel'dovich effect (SZE) observations.
Finally, a summary is given in Section \ref{sec:summary}.

Throughout this paper, we use the AB magnitude system, and
 adopt a concordance $\Lambda$CDM cosmology
with $\Omega_{m}=0.3$, $\Omega_{\Lambda}=0.7$, and
$H_0=100\,h\,$km\,s$^{-1}$\,Mpc$^{-1}$ 
with $h=0.7$.
In this cosmology, $1\arcmin$ corresponds to 238\,kpc\,$h^{-1}=341$\,kpc
 at the cluster redshift, $z=0.439$.  
We use the standard notation $M_\Delta\equiv M_{\rm 3D}(<r_\Delta)$ to
 denote the total mass enclosed within a sphere of
radius $r_\Delta$, within which the mean interior density is
$\Delta$ times the critical mass density at the cluster redshift.
We refer all our virial quantities
to an overdensity $\Delta$ of  $\Delta_{\rm vir}\approx 132$ 
based on the spherical collapse model
 \citep[Appendix A of][]{Kitayama+Suto1996}.\footnote{$\Delta_{\rm
 vir}\approx 134$ using the fitting formula given by \citet{Bryan+Norman1998}.}
All quoted errors are 68.3\%
confidence limits (CL) unless otherwise stated. 
The reference sky 
position is the center of the brightest cluster galaxy (BCG) of Zitrin 
 et al. (2012),
${\rm R.A.}=12$:06:12.15, ${\rm Decl.}=-08$:48:03.4 (J2000.0).

\section{Basic Theory of Galaxy Cluster Weak Lensing}
\label{sec:basis}

The central quantity of interest in this work
is the convergence of gravitational lensing,
$\kappa(\btheta)=\Sigma(\btheta)/\Sigma_{\rm crit}$,
which is the surface mass density projected on to the
lens plane, $\Sigma(\btheta)$,
in units of the critical surface mass density for lensing, 
\begin{eqnarray} 
\label{eq:sigmacrit}
\Sigma_{\rm crit} = \frac{c^2}{4\pi G D_l} \beta^{-1};
\ \ \ \beta(z_s) \equiv {\rm max}\left[
0,\frac{D_{ls}(z_s)}{D_s(z_s)}\right].
\end{eqnarray}
Here $D_s$, $D_l$, and $D_{ls}$ are the proper angular diameter
distances from the observer to the source, from the observer to the
lens, and from the lens to the source, respectively;
$\beta$ is the angular-diameter distance ratio associated with the
population of background sources.

The lens distortion and magnification of images are described by 
the Jacobian matrix $\cal{A}_{\alpha\beta}$ ($\alpha,\beta=1,2$) of the
lens mapping, which can be decomposed as
${\cal A}_{\alpha\beta} = (1-\kappa)\delta_{\alpha\beta}
-\Gamma_{\alpha\beta}$, 
where $\delta_{\alpha\beta}$ is Kronecker's delta, 
and $\Gamma_{\alpha\beta}$ is the
trace-free, symmetric shear matrix, 
\begin{eqnarray}
\label{eq:jacob} 
\Gamma &=&
\left( 
\begin{array}{cc} 
+{\gamma}_1   & {\gamma}_2 \\
 {\gamma}_2  & -{\gamma}_1 
\end{array} 
\right),
\end{eqnarray} 
with the components of complex gravitational shear with spin-2 nature
\citep[under coordinate rotations; see][]{2001PhR...340..291B,HOLICs2},
$\gamma=\gamma_1+i\gamma_2\equiv |\gamma|e^{2i\phi_\gamma}$. 
The $\kappa$ and $\gamma$ fields are related to each other by
\begin{equation}
\label{eq:k2g}
\triangle\kappa(\btheta) =
 \partial^\alpha\partial^\beta\Gamma_{\alpha\beta}(\btheta).
\end{equation}
The Green's function for the two-dimensional (2D) Poisson equation is
$\triangle^{-1}(\btheta,\btheta')=\ln|\btheta-\btheta'|/(2\pi)$, so that
Equation (\ref{eq:k2g}) can be readily solved \citep{1993ApJ...404..441K}.

In the strict weak-lensing limit ($\kappa,|\gamma|\ll 1$),
$\Gamma_{\alpha\beta}$ induces a small quadrupole distortion of the  
background image, which can be measured from observable ellipticities 
of background galaxy images~\citep{1995ApJ...449..460K}.
In general, the observable quantity for quadrupole weak lensing 
is not $\gamma$ but the {\it reduced} gravitational shear,
\begin{equation}
\label{eq:g}
g\equiv g_1+ig_2 =\frac{\gamma}{1-\kappa}
\end{equation}
in the subcritical regime where ${\rm det}{\cal A}>0$ 
(or $1/g^*$ in the negative parity region with ${\rm det}{\cal A}<0$).

Given an arbitrary circular loop of radius $\theta$ on the sky,
the tangential shear $\gamma_+(\theta)$ averaged around the loop
 satisfies the following identity
\citep[e.g.,][]{Kaiser1995}:
\begin{equation} 
\label{eq:loop}
\gamma_+(\theta) = \overline{\kappa}(<\theta)-\kappa(\theta),
\end{equation}
where $\kappa(\theta)$ is the azimuthal average of $\kappa$ around the
loop, and $\overline{\kappa}(<\theta)$ is the average convergence 
within the loop.
Hence, a constant
mass sheet cannot be constrained using the shear information alone,
known as the mass-sheet degeneracy
\citep[e.g.,][]{2001PhR...340..291B}.

This inherent degeneracy can be unambiguously broken
by measuring the magnification effects, which provide complementary and
independent constraints on the cluster mass distribution 
\citep{Umetsu+2011a}.
The magnification
is given by the inverse Jacobian determinant,
\begin{equation}
\label{eq:mu}
\mu(\btheta) = \frac{1}{|{\rm det}{\cal A}(\btheta)|} 
=\frac{1}{|(1-\kappa)^2-|\gamma|^2|}.
\end{equation}
The magnification $\mu(\btheta)$ can influence the observed surface
density $n_\mu(\btheta)$
of background sources, expanding the area of sky, and enhancing
the observed flux of background sources
\citep[e.g.,][]{1995ApJ...438...49B,UB2008,vanWaerbeke+2010,Rozo+Schmidt2010,Umetsu+2011a,Hildebrandt+2011,Ford+2011}. 
The former effect reduces the effective observing  
area in the source plane, decreasing the number of 
sources per solid angle; on the other hand, the latter effect
amplifies the flux of background sources, thereby increasing the number
of sources above the limiting flux. 
The net effect is known as magnification bias, and depends on the
intrinsic slope of the luminosity function of background sources as:
\begin{equation} 
\label{eq:magbias}
n_\mu(\btheta)=n_0\mu(\btheta)^{2.5s-1},
\end{equation}   
where $n_0 = dN_0(<m_{\rm cut})/d\Omega$ is the unlensed mean
number density of background sources for a 
given magnitude cutoff $m_{\rm cut}$, approximated locally as a
power-law cut with slope,
$s=d\log_{10} N_0(<m)/dm >0$.
In the strict weak-lensing limit, 
$n_\mu/n_0-1\approx (5s-2)\kappa$.
For a maximally-depleted
population of galaxies with $s=0$, $n_\mu/n_0 = \mu^{-1}\approx 1-2\kappa$
in this limit.

Alternatively, the mass-sheet degree of freedom can  be determined such that
the mean $\Sigma$ averaged over the outermost cluster region
vanishes, 
if a sufficiently wide sky coverage is available.\footnote{Or,
one may constrain the constant such that the enclosed mass
within a certain aperture is consistent with cluster mass
estimates from some other observations
\citep[e.g.,][]{Umetsu+Futamase1999}.}  

\section{Subaru Data and Analysis}
\label{sec:subaru}

In this section we present a technical description of our weak-lensing
analysis of MACS1206  based on deep Subaru multi-color images.
The data reduction and the photometry procedure 
are summarized in Section
\ref{subsec:data}.  The details of our weak-lensing shape analysis are
given in Section \ref{subsec:shape}. Our shear calibration strategy is
described in Section \ref{subsec:shearcalib}.
Details of the sample selection and lensing depth estimation are given
in Sections 
\ref{subsec:color} and \ref{subsec:depth}, respectively.

\subsection{Subaru Data and Photometry}
\label{subsec:data}

\begin{figure*}[!htb]
 \begin{center}
  \includegraphics[width=0.8\textwidth,clip]{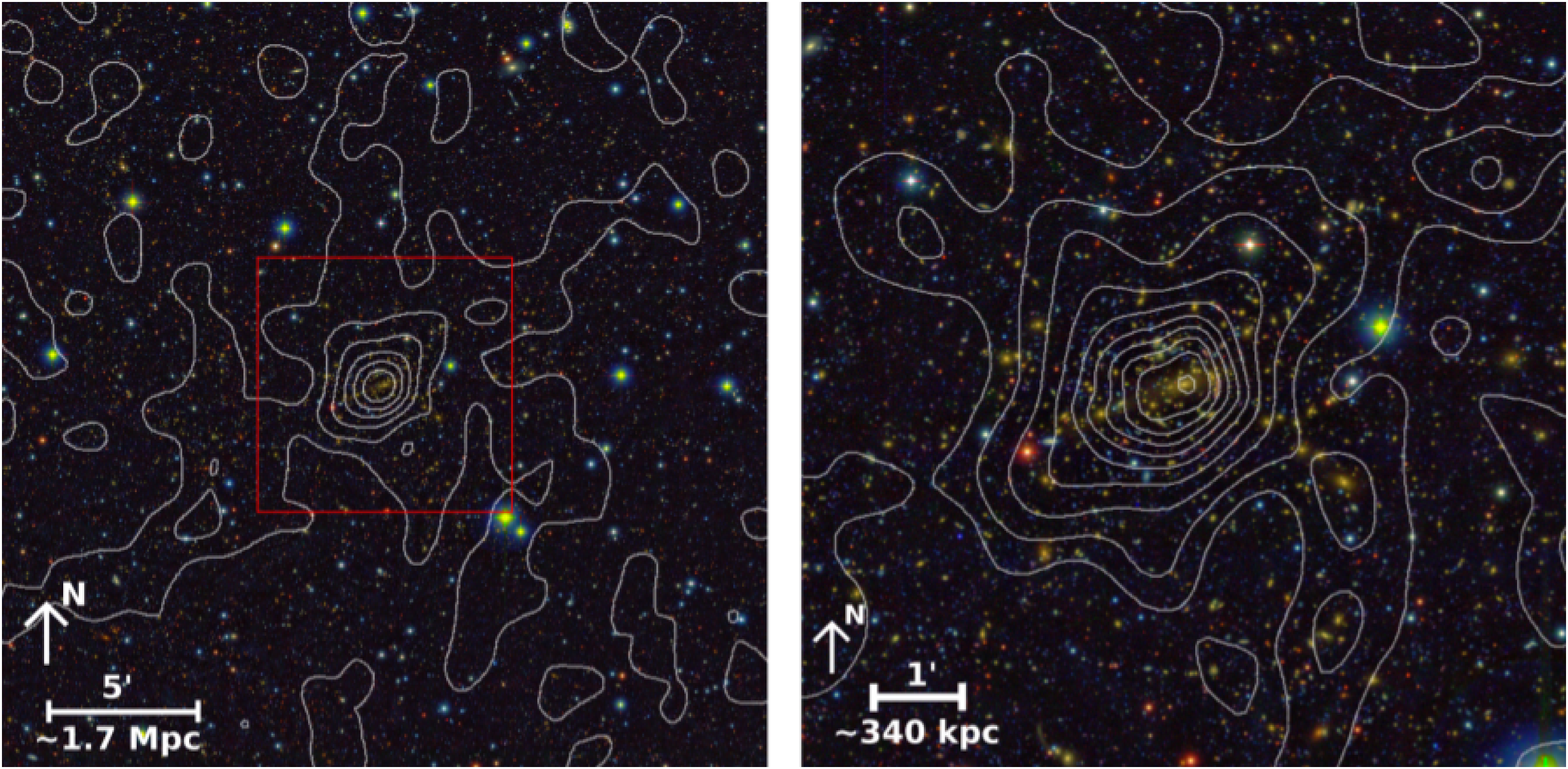}
 \end{center}
\caption{
Subaru $BVR_{\rm c}I_{\rm c}z'$ composite color images centered on the
 galaxy cluster MACS1206 ($z=0.439$), overlaid with mass contours from
 our joint strong-and-weak lensing analysis ({\sc SaWLens}) of {\it HST}
 and Subaru observations.
The image size in the left panel is $24\arcmin\times 24\arcmin$ covering
 a projected area of $5.7\times 5.7$\,Mpc\,$h^{-2}$ at the cluster
 redshift. 
In the left and right panels,
 the lowest contour levels are $\kappa=0.12$ and $0.15$,
 with increments of $\Delta\kappa=0.09$ and 0.07, respectively. 
The right panel is a zoom-in-view of the boxed region of the left
 panel, with a side length of $8\arcmin$ ($1.9\,$Mpc\,$h^{-1}$).
North is top and east is left.
\label{fig:color}
} 
\end{figure*}

We analyze deep $BVR_{\rm c}I_{\rm c}z'$ images of MACS1206 
observed with the wide-field camera Suprime-Cam 
\citep[$34^\prime\times 27^\prime$;][]{2002PASJ...54..833M}
at the prime focus of the 8.3-m Subaru
telescope.
The observations are available in 
the Subaru archive, SMOKA.\footnote{\href{http://smoka.nao.ac.jp}{http://smoka.nao.ac.jp}}
The seeing FWHM in the co-added mosaic image is
$1.01\arcsec$ in $B$ (2.4\,ks), 
$0.95\arcsec$ in $V$ (2.2\,ks),
$0.78\arcsec$ in $R_{\rm c}$ (2.9\,ks),
$0.71\arcsec$ in $I_{\rm c}$ (3.6\,ks), and
$0.58\arcsec$ in $z'$ (1.6\,ks) 
with $0.20\arcsec$ pixel$^{-1}$, covering a field of approximately
$36\arcmin \times 34\arcmin$.  
The limiting magnitudes are obtained as $B=26.5$, $V=26.5$, $R_{\rm c}=26.2$,
$I_{\rm c}=26.0$, and $z'=25.0$\,mag for a $3\sigma$ limiting detection within 
$2\arcsec$ diameter aperture.
The observation details of MACS1206 are listed in
Table \ref{tab:subaru}. 
Figure \ref{fig:color} shows Subaru $BVR_{\rm c}I_{\rm c}z'$ composite
color images of the cluster field, produced automatically using the
publicly available Trilogy software \citep{Coe+2012_A2261}.\footnote{\href{http://www.stsci.edu/\%7Edcoe/trilogy/}{http://www.stsci.edu/$\sim$dcoe/trilogy/}}

Standard reduction steps were performed using the {\tt mscred} task in
{\sc IRAF}.\footnote{{\sc IRAF} is distributed by the National Optical Astronomy 
Observatories, which is operated by the Association of Universities for
Research in Astronomy, Inc., under cooperative agreement with  the
National Science Foundation}
We closely follow the data reduction procedure outlined in
\citet{Nonino+2009} to create a co-added mosaic of Subaru Suprime-Cam
images, incorporating additional reduction steps, such as 
automated masking of bleeding of bright saturated stars.

To obtain an accurate astrometric solution for Subaru observations, we
retrieved processed MegaCam {\em griz} images from the
CFHT archive,\footnote{This research used facilities of the Canadian
Astronomy Data Centre operated by the National Research Council of
Canada with the support of the Canadian Space Agency}
and used MegaCam $r$ data (Filter Number: 9601) as a wide-field
reference image.   
A source catalog was created from the co-added MegaCam $r$ image, using
 the 2MASS catalog\footnote{This publication makes use of data products
 from the Two Micron All Sky Survey, which is a joint project of the University
 of Massachusetts and the Infrared Processing and Analysis
 Center/California Institute of Technology, funded by the National
 Aeronautics and Space Administration and the National Science
 Foundation} 
as an external reference catalog.  The extracted $r$ catalog has been
used as a reference for the SCAMP software \citep{SCAMP}
to derive an astrometric solution for the Suprime-Cam images.

The photometric zero points for the co-added Suprime-Cam images were
bootstrapped from a suitable set of reference stars identified in common
with the calibrated MegaCam data.
These zero-points were refined in two independent ways: firstly by
comparing cluster elliptical-type galaxies with the  {\it HST}/ACS
images;
subsequently by fitting SED (spectral energy distribution) templates 
with the BPZ code \citep[Bayesian photometric redshift
estimation,][]{Benitez2000,Benitez+2004}  
to Subaru photometry of $1163$ galaxies having measured spectroscopic 
redshifts from VLT/VIMOS (P. Rosati et al., in preparation).
This leads to a final photometric accuracy of $\sim 0.01$\,mag in all
five passbands (see also Section \ref{subsec:depth}). 
Five-band $BVR_{\rm c}I_{\rm c}z'$ photometry catalog was then measured
using SExtractor \citep{1996A&AS..117..393B}
in point-spread-function (PSF) matched images created by ColorPro
\citep{colorpro}, where a combination of all five bands was used as a
deep detection image. The stellar PSFs were measured from a combination
of 100 stars per band and modeled using {\sc IRAF} routines. 

For the weak-lensing shape analysis (Section \ref{subsec:shape}), we use
the $I_{\rm c}$-band data taken in 2009 January, which have the best
image quality in our data-sets (in terms of the stability and coherence
of the PSF anisotropy pattern, taken in fairly good seeing conditions).   
Two separate co-added $I_{\rm c}$-band images, each with a
total exposure time of 1.1\,ks, were produced
based on the imaging obtained at two different camera orientations
separated by 90\,degrees, in order not to degrade the shape measurement
quality. 

\subsection{Subaru Weak Lensing Shape Analysis}
\label{subsec:shape}

\begin{figure}[!htb]
 \begin{center}
\includegraphics[width=0.45\textwidth,angle=0,clip]{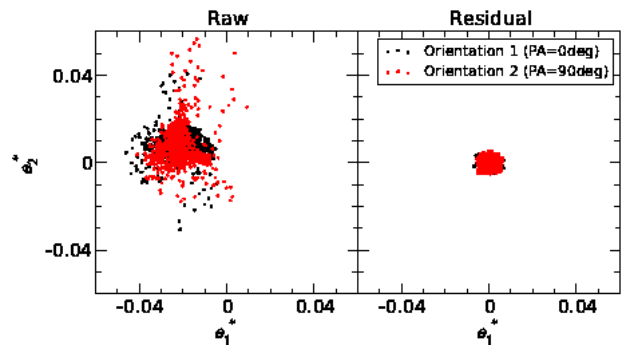}
 \end{center}
\caption{
\label{fig:PSF}
Stellar ellipticity distributions before and after the PSF anisotropy  
correction for Subaru/Suprime-Cam $I_{\rm c}$-band data
taken with camera orientations of ${\rm  PA}=0^\circ$ 
(Orientation 1; red) and 
${\rm PA}=90^\circ$ (Orientation 2; black).  
The left panel shows the raw ellipticity components 
$(e_1^*,e_2^*)$ of stellar objects, and the right panel shows
the residual ellipticity components $(\delta e_1^*, \delta e_2^*)$
after the PSF anisotropy correction.
} 
\end{figure}

For shape measurements of background galaxies, we use our weak-lensing
analysis pipeline based on the IMCAT package \citep[KSB
hereafter]{1995ApJ...449..460K}, incorporating modifications and
improvements outlined in \citet{Umetsu+2010_CL0024}.  
Our KSB+ implementation has been applied extensively to Subaru cluster
observations
\citep[e.g.,][]{BTU+05,BUM+08,2007MPLA...22.2099U,UB2008,Okabe+Umetsu2008,Umetsu+2009,Umetsu+2010_CL0024,Umetsu+2011a,Umetsu+2011b,Medezinski+2010,Medezinski+2011,Zitrin+2011_A383,Coe+2012_A2261}.  

We measure components of the complex image ellipticity, 
$e_{\alpha} = \left\{Q_{11}-Q_{22}, Q_{12} \right\}/(Q_{11}+Q_{22})$, 
from the weighted quadrupole moments of the surface brightness
$I(\btheta)$ of individual objects,
\begin{equation}
\label{eq:Qij}
Q_{\alpha\beta} = \int\!d^2\theta\,
 W({\theta})\theta_{\alpha}\theta_{\beta} 
I({\btheta})
\ \ \ (\alpha,\beta=1,2)
\end{equation} 
where $W(\theta)$ is a Gaussian window function matched to the size
($r_g$) of the object, and
the weighted object centroid is chosen as the coordinate
origin, which is iteratively refined
to accurately measure the object shapes.

Next, we correct observed ellipticities $e_\alpha$ for the PSF
anisotropy using a sample of stars in the field as references.
We select bright
($18\simlt I_{\rm c}\simlt 22$),  unsaturated 
stellar objects 
identified in a branch
of the object half-light radius ($r_h$) versus $I_{\rm c}$ diagram,
and measure the PSF anisotropy kernel of the KSB algorithm
as a function of the object size $r_g$.
Figure \ref{fig:PSF} shows
the distributions of stellar ellipticity
components ($e_{\alpha}^*$) before and after the PSF anisotropy correction.
From the rest of the object
catalog, we select as a weak-lensing galaxy sample
those objects with 
$\nu > 10$,
$r_h > \overline{r_h^*} + 1.5 \sigma(r_h^*)$,
and 
$r_g > {\rm mode}(r_g^*)$,
where $\nu$ is the KSB detection significance,
$\overline{r_h^*}$ and $\sigma(r_h^*)$
are median and rms dispersion values
of stellar sizes $r_h^*$.
The anisotropy corrected ellipticities 
$e'_\alpha$ are then corrected for the 
isotropic smearing effect as $g_{\alpha}=e'_{\alpha}/P_g$.

For each galaxy we assign  the statistical weight,
\begin{equation}
\label{eq:weight}
w_{(k)} \equiv \frac{1}{\sigma_{g(k)}^2+\alpha_g^2},
\end{equation}
where $\sigma_{g(k)}^2$ is the variance for the reduced shear estimate of
the $k$th galaxy
computed from $50$ neighbors identified in the 
$r_g$--$I_{\rm c}$ plane, and
$\alpha_g^2$ is the softening constant variance 
\citep[e.g.,][]{2003ApJ...597...98H,UB2008,Oguri+2009_Subaru,Okabe+2010_WL}.  
This weighting scheme is essential to down-weight faint and small
objects which have noisy shape measurements \citep[see Figure 4
of][]{Umetsu+2010_CL0024}.  
We choose $\alpha_g=0.4$, which is a typical value of 
the mean rms $\sigma_g$ over the background
sample \citep[see Table \ref{tab:color};][]{UB2008,Umetsu+2009,Okabe+2010_WL}.

\subsection{Shear Calibration}
\label{subsec:shearcalib}

We follow the shear calibration strategy of \citet{Umetsu+2010_CL0024}
to improve the precision in shear recovery.
This is motivated by the general tendency of KSB+ to systematically
underestimate the shear signal in the presence of measurement noise 
\citep[see][]{Umetsu+2010_CL0024,Okura+Futamase2012}.

First, we select as a sample of {\it shear calibrators}
those galaxies with $\nu>\nu_c$ and $P_g>0$.
Here we take $\nu_c=20$.
Note that the shear calibrator sample
is a subset of the target galaxy sample.
Second, 
we divide the calibrator $r_g$--$I_{\rm c}$
plane into a grid of $2\times 10$
cells each containing approximately equal numbers of calibrators, and  compute
a median value of $P_g$ at each cell.
Then, each object in the target sample is matched to the nearest
point on the ($r_g,I_{\rm c}$) calibration grid to obtain a filtered
measurement, $\langle P_g \rangle$.
Finally, we use the calibrated estimator $g_\alpha=e'_{\alpha}/\langle P_g
\rangle$ for the reduced shear. 

We have analyzed the two $I_{\rm c}$ mosaic images
separately to construct a composite galaxy shape catalog, by properly
weighting and combining the calibrated distortion measurements ($g_\alpha$)
for galaxies in the overlapping region.

We have tested our analysis pipeline using
simulated Subaru Suprime-Cam images
\citep[see Section 3.2 of][]{Oguri+2012_SGAS,2007MNRAS.376...13M}.
We find that we can recover the weak-lensing signal with good precision, typically,
$|m| \lesssim 5\%$ of the shear calibration bias,
where the range of $m$-values shows a modest dependence of calibration
accuracy on seeing conditions and PSF properties, 
and $c\sim 10^{-3}$ of the residual shear offset,
which is about one order of magnitude smaller than the typical
distortion signal in cluster outskirts ($|g|\sim 10^{-2}$). 
This level of performance is comparable to other similarly well-tested
methods
\citep{2006MNRAS.368.1323H,2007MNRAS.376...13M}.

\subsection{Sample Selection}
\label{subsec:color}

\begin{figure}[!htb]
 \begin{center}
  \includegraphics[width=0.45\textwidth,clip]{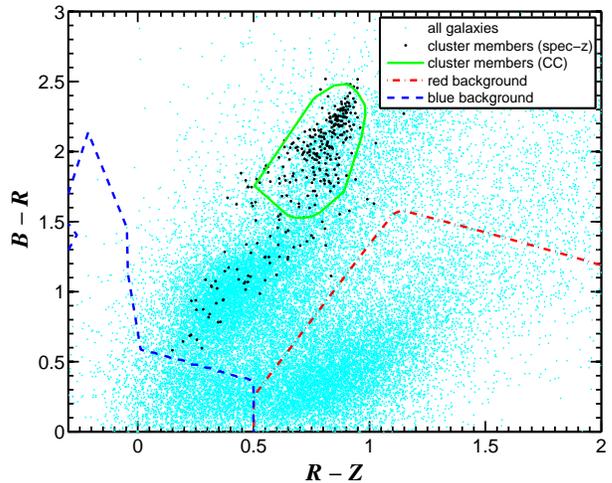}
 \end{center}
\caption{
Blue and red background galaxies are selected for 
 weak-lensing analysis (lower left blue dashed and right red dot-dashed
 regions, respectively) on the basis of Subaru $BR_{\rm c}z'$ 
 color-color-magnitude selection. 
All galaxies with $z'<24.6$\,mag (cyan) are shown in the diagram. 
 At small radius, the cluster
 overdensity is identified as the green outlined region, defining our
 green sample comprising mostly the red sequence of the cluster and a
 blue trail of later type cluster members.  The background samples are
 well isolated from the green region and satisfy other criteria as
 discussed in Section \ref{subsec:color}.  Our background selection
 successfully excludes all spectroscopically-confirmed cluster
 members (black) found within the projected cluster virial radius 
($r_{\rm vir}\approx 1.6$\,Mpc\,$h^{-1}$).
The cluster members are determined from the ongoing survey with
 VLT/VIMOS (Rosati et al., in prep.), using the algorithm of
 \citet{Mamon+2010} in the dynamical analysis which will be presented in
 a forthcoming paper (Biviano et al. in prep.).
\label{fig:CC}
} 
\end{figure}

\begin{figure}[!htb]
 \begin{center}
    \includegraphics[width=0.45\textwidth,angle=0,clip]{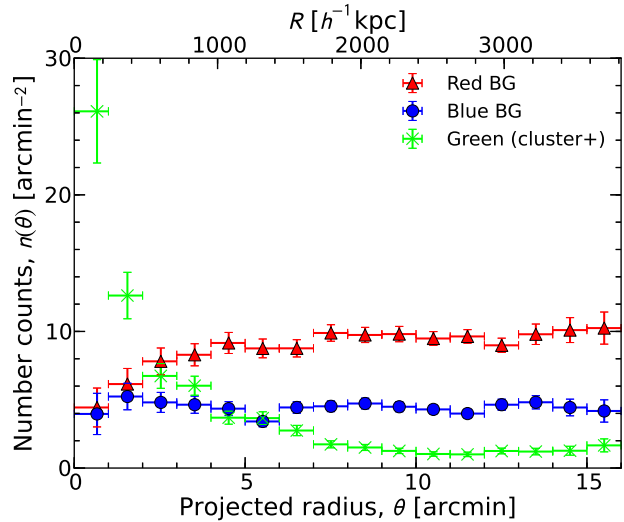}
 \end{center}
\caption{
Surface number density profiles $n(\theta)$ of Subaru 
$BR_{\rm c}z'$-selected galaxies used for the weak-lensing shape
 analysis.  The results are shown for our red (triangles), blue
 (circles), and green (crosses) samples.  
See also Figure \ref{fig:wldata}.
\label{fig:nplot}
} 
\end{figure}

\begin{figure}[!htb]
 \begin{center}
   \includegraphics[width=0.45\textwidth,angle=0,clip]{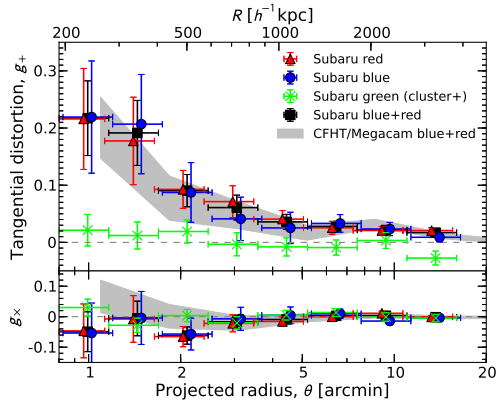}
 \end{center}
\caption{
Azimuthally-averaged radial profiles of the tangential reduced shear
 $g_+$ (upper panel) and the $45^\circ$ rotated ($\times$) component
 $g_{\times}$ (lower panel) for our Subaru red (triangles), blue
 (circles), green (crosses), and blue+red (squares) galaxy samples shown
 in Figure \ref{fig:nplot}. 
The error bars represent $68.3\%$ confidence intervals estimated by
 bootstrap resampling techniques.  The symbols for the red and blue
 samples are horizontally shifted for visual clarity. 
For a consistency check, we compare our Subaru results with CFHT/Megacam
 data based on our $grz$-selected background sample (gray area).
The $g_+$ profile for the green sample is consistent with a null signal
 at all radii, while this population is strongly clustered at small
 radius (Figures \ref{fig:nplot}), indicating that the green galaxies
 mostly consist of cluster member galaxies. 
For all of the samples, the $\times$-component is consistent with a null
 signal detection well within $2\sigma$ at all radii, indicating the
 reliability of our distortion analysis.  
\label{fig:rgb} 
}
\end{figure}  

A careful background selection is critical for a weak-lensing analysis
so that unlensed cluster members and foreground galaxies do not dilute
the true lensing signal of the background
\citep{BTU+05,Medezinski+07,UB2008,Medezinski+2010}.
This dilution effect is simply to reduce the strength of the lensing
signal when averaged over a local ensemble of galaxies \citep[by a
factor of 2--5 at $R\simlt 400\,{\rm kpc}\,h^{-1}$; see 
Figure 1 of][]{BTU+05}, particularly at small cluster radius
where the cluster is relatively dense, in proportion to the fraction of
unlensed galaxies whose orientations are randomly distributed. 

We use the background selection method of \citet{Medezinski+2010} to
define undiluted samples of background galaxies,
which relies on empirical correlations for galaxies in
color-color-magnitude space derived from the deep Subaru photometry, by 
reference to evolutionary tracks of galaxies
\citep[for details, see][]{Medezinski+2010,Umetsu+2010_CL0024} as well
as to the deep photometric-redshift survey in the COSMOS field
\citep{Ilbert+2009_COSMOS}.  

For MACS1206, we have a wide wavelength coverage 
($BVR_{\rm c}I_{\rm c}z'$) of Subaru Suprime-Cam.  
We therefore make use of the $(B-R_{\rm c})$ vs. $(R_{\rm c}-z')$
color-color (CC) diagram to carefully select two distinct background
populations which encompass the red and blue branches of galaxies.
We limit the data to $z'=24.6$\,mag in the reddest band, corresponding
approximately  to a $5\sigma$ limiting magnitude within $2\arcsec$
diameter aperture.  Beyond this limit incompleteness creeps into the
bluer bands, complicating color measurements, in particular of red
galaxies.  

To do this, we first identify in CC space an overdensity of galaxies
with small projected distance $< 3\arcmin$ ($\simlt 1$\,Mpc at
$z_l=0.439$) from the cluster center. 
Then, {\it all} galaxies within this distinctive region define the {\it
green} sample (see the green outlined region in Figure \ref{fig:CC}),
comprising mostly the red sequence of the cluster and a blue trail of
later type cluster members \citep{Medezinski+2010,Umetsu+2010_CL0024},
showing a number density profile that is steeply rising toward the center 
(Figure \ref{fig:nplot}, green crosses).  The weak-lensing signal for
this population is found to be consistent with zero at all radii (Figure
\ref{fig:rgb}, green crosses), indicating the reliability of our
procedure.  
For this population of galaxies, we find a mean photometric redshift of
$\langle z_{\rm phot}\rangle \approx 0.44$ (see Section
\ref{subsec:depth}), consistent with the cluster redshift.
Importantly, the green sample marks the region that contains a majority 
of unlensed galaxies, relative to which we select our background
samples, as summarized below. 

For the background samples, we define conservative color limits, 
where no evidence of dilution of the weak-lensing signal is visible, to
safely avoid contamination by unlensed cluster members and foreground
galaxies. 
The color boundaries for our blue and red background samples are shown
in Figure \ref{fig:CC}.
For the blue and red samples, we find a consistent, clearly rising
weak-lensing signal all the way to the center of the cluster, as shown
in Figure \ref{fig:rgb}.  

For validation purposes, we compare in CC space our color samples with a spectroscopic sample of cluster galaxies in MACS1206.
Figure \ref{fig:CC} shows  that the background selection procedure
established in our earlier work
\citep{Medezinski+2010,Umetsu+2010_CL0024,Medezinski+2011} successfully 
excludes all spectroscopically-confirmed cluster members found within
the projected cluster virial radius  
($r_{\rm vir}\approx 1.6$\,Mpc\,$h^{-1}$; see Section \ref{sec:mass}).
The cluster members are determined from the ongoing survey with
 VLT/VIMOS, part of the VLT-CLASH Large Programme 186.A-0798
(P. Rosati et al., in preparation), using the algorithm of 
\citet{Mamon+2010} in the dynamical analysis which will be presented in
a  forthcoming paper (A. Biviano et al. in preparation).
We find about 70\% of the cluster members overlap with our CC-selected green
galaxies; the rest are cluster members with bluer colors. 
We note there is a statistically inevitable fraction of interlopers even
in the dynamically-selected cluster membership as discussed in 
\citet[][their Table 1]{Wojtak+2007} and \citet[][their Figure
13]{Mamon+2010}.

As a further consistency check, we also plot in Figure \ref{fig:nplot}
the galaxy surface number density as a function of radius, $n(\theta)$,
for the blue and red samples. 
As can be seen, no clustering is observed toward the center for the
background samples, which demonstrate that there is no significant
contamination by cluster members in the samples.  
The red sample reveals a systematic decrease in their projected number
density toward the cluster center, caused by the lensing magnification
effect (Section \ref{sec:basis}).  A more quantitative magnification
analysis is given in Section \ref{subsubsec:magbias}.

To summarize, our CC-selection criteria yielded a total of 
$N=13252$, 1638, and 4570 galaxies, for the red, green, and blue
photometry samples, respectively (Table \ref{tab:color}).
For our weak-lensing distortion analysis, 
we have a subset of 8969 and 4154 galaxies in the red and blue samples
(with usable $I_{\rm c}$ shape measurements), respectively (Table
\ref{tab:wlsamples}).

\subsection{Depth Estimation}
\label{subsec:depth}

The lensing signal depends on the source redshift $z_s$ 
through the distance ratio $\beta(z_s)=D_{ls}/D_s$.
We thus need to estimate and correct for the respective depths 
$\langle\beta \rangle$ of
the different galaxy samples, when converting the
observed lensing signal into physical mass units.

For this we used BPZ (Section \ref{subsec:data}) to measure photometric
redshifts (photo-$z$s) $z_{\rm phot}$ for our  
deep Subaru $BVR_{\rm c}I_{\rm c}z'$ photometry (Section \ref{subsec:data}).
BPZ employs a Bayesian inference  where the redshift likelihood is weighted
by a prior probability, which yields the probability density $P(z,T|m)$ of a
galaxy with apparent magnitude $m$ of having certain redshift $z$ and
spectral type $T$.
In this work we used a new library (N. Benitez 2012, in preparation)
 composed of 10 SED templates originally from PEGASE
 \citep{Fioc+Rocca1997} but recalibrated using the FIREWORKS photometry
 and spectroscopic redshifts from \citet{Wuyts+2008} to optimize its
 performance.  This  library includes five templates for elliptical
 galaxies, two for spiral galaxies, and three for starburst galaxies.  
In our depth estimation we utilize BPZ's ODDS parameter, which measures 
the amount of probability enclosed within a certain interval $\Delta z$
centered on the primary peak of the redshift probability density
function (PDF), serving as a useful measure to quantify the reliability
of photo-$z$ estimates \citep{Benitez2000}.\footnote{In the present work,
we set $\Delta z = 2\times 0.03(1+z_{\rm phot})$, which is approximately
twice the width ($\sigma$) of the error distribution.}
We used our VLT/VIMOS sample of $1163$ galaxies with
spectroscopic redshifts $z_{\rm spec} (\simlt 1.5)$
to assess the performance of our photo-$z$ estimation.
From the whole sample, we find 
an rms scatter of $\sigma(\delta_z)\approx 0.027$ in the fractional error 
$\delta_z \equiv (z_{\rm phot}-z_{\rm spec})/(1+z_{\rm spec})$, 
with a small mean offset $\mu(\delta_z)=-0.0021$ and a
$5\sigma$ outlier fraction of $\approx 5.5\%$.
Using a subsample of $\sim 510$ galaxies with $0.3<z_{\rm spec}<0.5$, we
find $\sigma(\delta_z)\approx 0.031$ with $\approx 1.5\%$
of outliers.

For a consistency check, we also make use of the COSMOS catalog
\citep{Ilbert+2009_COSMOS} with robust photometry and photo-$z$
measurements for the majority of galaxies with $i'<25$\,mag.
For each sample, we apply the same CC selection to the COSMOS
photometry, and obtain the redshift distribution $N(z)$ of 
field galaxies.

For each background population,
we calculate weighted moments of the distance ratio $\beta$ as
\begin{equation}
\label{eq:moment}
\langle \beta^n \rangle =
  \frac{\int\!dz\,w(z) N(z)\beta^n(z)}{\int\!dz\,w(z)N(z)},
\end{equation}
where $w(z)$ is a weight factor; $w$ is taken to be the Bayesian ODDS
parameter for the BPZ method, and $w=1$ otherwise.
The sample mean redshift $\langle z_s\rangle$ is defined similarly to 
Equation (\ref{eq:moment}).
The first moment $\langle \beta\rangle$ represents the mean lensing
depth.\footnote{
In general, a wide spread of the redshift distribution of background
galaxies, in conjunction with the single-plane approximation,
may lead to an overestimate of the gravitational shear in
the nonlinear regime \citep{2000ApJ...532...88H}. 
Thanks to the deep Subaru photometry, we found that this bias in the
observed reduced shear is approximately $\Delta g/g \approx
(\langle\beta^2\rangle/\langle\beta\rangle^2-1)\kappa\approx 0.06\kappa$ to 
the first order of $\kappa$.  See \S~3.4 of \citet{Umetsu+2010_CL0024}
for details.}
It is useful to define the effective single-plane source redshift,
$z_{s,{\rm eff}}$, such that \citep{UB2008,Umetsu+2009,Umetsu+2010_CL0024}
\begin{equation}
\beta(z_{s,{\rm eff}})= \langle\beta\rangle.
\end{equation}

In Table \ref{tab:wlsamples} we summarize the mean depths 
$\langle\beta \rangle$ and the effective source redshifts 
$z_{s,{\rm eff}}$ for our background samples.
For each background sample,
we obtained consistent mean-depth estimates 
$\langle \beta\rangle$  (within $2\%$)
using the BPZ- and COSMOS-based methods.
In the present work, we adopt a conservative uncertainty of $5\%$ in the
mean depth for the combined blue and red sample of
background galaxies, 
$\langle\beta({\rm back})\rangle=0.54\pm 0.03$, which corresponds 
to $z_{s,{\rm eff}}=1.15\pm 0.1$.   We marginalize over this
uncertainty when fitting parameterized mass models to our weak-lensing data.

\section{Cluster Strong Lensing Analysis}
\label{sec:sl}

For a massive cluster, the strong- and weak-lensing regimes contribute
quite similar logarithmic coverage of the radial mass profile.  It is
therefore crucial to include the central strong-lensing information in a
cluster lensing analysis \citep[e.g.,][]{Umetsu+2011a,Umetsu+2011b}. 

Here we perform several complementary strong-lensing analyses using a
wide variety of modeling methods, namely  
the \citet{Zitrin+2009_CL0024} method,
{\sc Lenstool} \citep{Kneib+1996,Jullo+2007_Lenstool}, 
{\sc LensPerfect} \citep{Coe+2010,Coe+2012_A2261}, 
{\sc Pixelens} \citep{Saha+Williams2004_Pixelens,Grillo+2010_Pixelens}, 
and a joint strong-and-weak lensing reconstruction method of
\cite{Merten+2009,Pandra2011} (hereafter, {\sc SaWLens}).
All analyses here use the positions and redshifts of multiply-lensed
images identified by \citet{Zitrin+2012_M1206}.

Lens reconstruction methods are broadly classified into parametric and
non-parametric:
In the former approach, the total mass distribution of the deflector is
described in terms of a set of theoretically (and/or observationally)
motivated models, each specified by a particular functional form
characterized by a small number of free parameters. 
This involves, to some extent, the assignment of halos 
to visible galaxies assuming light approximately traces mass, while
the latter does not except for certain priors on the mass distribution.\footnote{
The latter is often based on the assumption that the lens profiles and/or
distributions can be well approximated by a pixelated mass distribution
(e.g., {\sc Pixelens} and {\sc SaWLens}).}
Among the methods used in the present work, the
\citet{Zitrin+2009_CL0024} method and {\sc Lenstool} are parametric;
{\sc LensPerfect}, {\sc Pixelens}, and {\sc SaWLens} are non-parametric.

For this work, we primarily use the detailed strong lens modeling of
\citet{Zitrin+2012_M1206} based on deep CLASH imaging and VLT/VIMOS
spectroscopy, 
as summarized in Section \ref{subsec:Z12}. 
The cluster miscentering effects are discussed in Section
\ref{subsec:offset}.
In Section \ref{subsec:calib} 
we introduce and apply a technique to self-calibrate the bin-bin
covariance matrix of the central radial mass profile derived from the
reanalysis of \citet{Zitrin+2012_M1206}.
In Section \ref{subsec:sl+} we perform several semi-independent
strong-lensing analyses on the MACS1206 {\it HST} images, utilizing various
modeling methods, in order to verify the identifications of the multiple
images and to independently assess the level of inherent systematic
uncertainties in our analyses.

\subsection{Primary Strong Lensing Model}
\label{subsec:Z12}

Here we briefly summarize our well-tested approach to strong-lens
modeling, developed by \citet{2005ApJ...621...53B} and optimized further by 
\citet{Zitrin+2009_CL0024}, which has previously uncovered large numbers
of multiply-lensed galaxies in {\it HST} images of many clusters 
\citep[e.g.,][]{2005ApJ...621...53B,Zitrin+2009_CL0024,Zitrin+2010_A1703,Zitrin+2011_MACS,Zitrin+2010_MS1358,Zitrin+2011_A383}.  
In the present work, we use a new Markov Chain Monte Carlo (MCMC)
implementation of the \citet{Zitrin+2009_CL0024} method,
where also the BCG mass is allowed to vary.\footnote{
Our very preliminary MCMC results were presented in Figure 4 of
\citet{Zitrin+2012_M1206}.}

Our {\it flexible} mass model consists of four components,
namely the BCG, cluster galaxies, a smooth DM halo, and the
overall matter ellipticity \citep[corresponding to a coherent external
shear; for details, see][]{Zitrin+2009_CL0024},
described by 7 free parameters in total.\footnote{The
\citet{Zitrin+2009_CL0024} method employs grid-based maximum-likelihood
parameter estimation in the 6-dimensional parameter space, where the
seventh parameter included in the present work is the BCG mass.}
The basic assumption adopted is that cluster galaxy light approximately
traces the DM; the latter is modeled as a smoothed version of
the former \cite[see, for details,][]{Zitrin+2009_CL0024}.
This approach to strong lensing is sufficient to accurately predict the locations
and internal structure of multiple images, since in practice the number
of multiple images uncovered readily exceeds the number of free
parameters, so that the fit is fully constrained.

\begin{figure}[htb!]
 \begin{center}
   \includegraphics[width=0.45\textwidth,angle=0,clip]{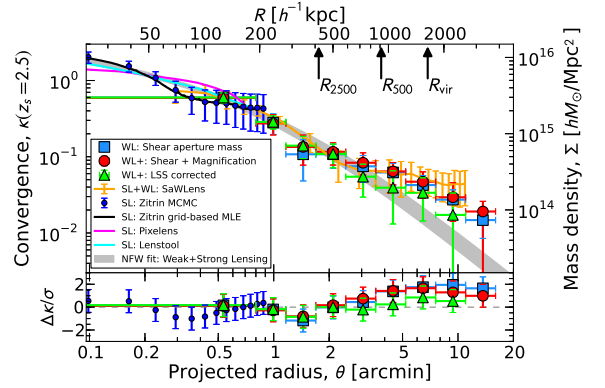}
 \end{center}
\caption{
Surface mass density profile $\kappa$ derived from our Subaru
 weak-lensing and {\it Hubble} strong-lensing measurements.  
The red circles represent our full weak-lensing constraints from joint
 shear and magnification measurements (Figure \ref{fig:wldata}),
 consistent with the purely shear-based results (squares) and the
 {\sc SaWLens} results (orange line with error bars), all showing a
 shallow radial trend with a nearly isothermal logarithmic density
 slope,  $d\ln\Sigma/d\ln R\sim -1$.  
For weak lensing, the innermost bin represents the average convergence
 $\overline{\kappa}(<\theta_{\rm min})$ interior to the 
 inner radial boundary of the weak-lensing data
 ($0.8\arcmin\le\theta\le 16\arcmin$),
$\theta_{\rm min}=0.8\arcmin$, which is about
 twice the Einstein radius for a distant background source at $z_s\sim 2$
 (see Tables \ref{tab:color} and \ref{tab:wlsamples}),
 and hence sufficiently large for our background galaxies at an
 effective source redshift of $z_{s,{\rm eff}}=1.15\pm 0.1$.
The triangles show the NE-SW mass profile excluding the large
 scale structure extending along the NW-SE direction (see Figure
 \ref{fig:MNL}), derived from a two-dimensional mass reconstruction
 using both shear and magnification data, in good agreement with the
 standard NFW form (gray area).  
The black solid line is the best-fit model of \citet{Zitrin+2012_M1206}
 based on the grid-based maximum likelihood parameter estimation.
The small blue circles with error bars represent our primary strong-lens 
 constraints on the binned mass profile derived from an MCMC
 implementation of \citet{Zitrin+2012_M1206}.  The errors are based on
 the self-calibrated covariance matrix (only every other point is shown
 here; Section \ref{subsec:calib}).   
Our mass profile results from several weak and strong lensing methods
all agree in the regions of overlap within their corresponding
 uncertainties.
For the sake of clarity, the {\sc Pixelens} and {\sc Lenstool} results
 are shown without error bars.
The bottom panel shows the respective deviations $\Delta\kappa$
 (in units of the error $\sigma$) from the best-fit NFW model.  The
 projected mass profile averaged over all azimuthal angles (squares,
 circles) shows a systematic excess at large radii with $R\simgt
 1$\,Mpc\,$h^{-1}$ ($\theta\simgt 4\arcmin$).
\label{fig:kappa}
}
\end{figure}

\citet{Zitrin+2012_M1206} identified 47 new multiple images of 12 
distant sources (including 3 candidate systems; Systems 9--11
therein), in addition to the 
known giant arc system at $z_s=1.03$ \citep{Ebeling+2009_M1206},
bringing the total known for this cluster to 50 multiply-lensed images
of 13 sources, spanning a wide redshift range of $1\simlt z_s\simlt
5.5$, spread fairly evenly over the central region, 
$3\arcsec \simlt \theta\simlt 1\arcmin$.
\citet{Zitrin+2012_M1206} used the position and redshift of 32 secure
multiple images of 9 systems to constrain the mass model.
Following \citet{Zitrin+2012_M1206}, we adopt an image
positional error of $2\arcsec$ ($\approx 1.4\arcsec$ in each
dimension), which is a typical value in the presence of uncorrelated
large scale structure (LSS) along the line of sight \citep[for details, 
see][]{Zitrin+2012_M1206,Host2012,Jullo+2010}.  
Including the BCG mass as an additional free parameter, we find here an  
acceptable fit with the minimized $\chi^2$ value ($\chi^2_{\rm min}$)
of 22.8 for 39 degrees of freedom
(dof), with an image-plane reproduction error of 1.76\arcsec. 
The new MCMC results are in good agreement with the results of
\citet{Zitrin+2012_M1206}, as shown here in Figure \ref{fig:kappa},
with only some minor differences at the innermost radii $\simlt
2\arcsec$ ($\sim 8$\,kpc\,$h^{-1}$)
dominated by the BCG \citep[see][]{Newman+2009_A611}.
The detailed central mass map reveals a fairly elliptical outer critical
curve \citep[see Figure 1 of][]{Zitrin+2012_M1206}.
For a source at $z_s=2.54$, the outer critical curve encloses 
an area with an effective Einstein radius of 
$\theta_{\rm Ein}=28\arcsec\pm 3\arcsec$; 
for the lower-redshift system with $z_s=1.03$, the effective Einstein
radius of the critical area is  
$\theta_{\rm Ein}=17\arcsec\pm 3\arcsec$ (Table \ref{tab:cluster}).

\subsection{Cluster Miscentering Effects}
\label{subsec:offset}

To obtain meaningful radial profiles, one must
carefully define the cluster center.
It is often assumed that the cluster mass centroid coincides with the
BCG position, whereas BCGs can be offset from the mass centroids of the
corresponding DM halos
\citep{Johnston+2007_SDSS1,Oguri+Takada2011,Umetsu+2011a,Umetsu+2011b}. 

Here we utilize our detailed mass model of \citet{Zitrin+2012_M1206},
which allows us to locate the peak position of the smooth DM component,
providing an independent mass centroid determination
\citep[e.g.,][]{Umetsu+2010_CL0024,Umetsu+2011a}. 
In this method, we approximate the large-scale distribution of cluster
mass by assigning a power-law mass profile to each cluster galaxy, the
sum of which is then smoothed to represent the DM distribution.
The success of this simple model in describing the
projected mass distributions of lensing clusters, as well as identifying
many sets of multiply-lensed images, assures us that
the effective DM center can be determined using multiple images
as well as the distribution of cluster member galaxies.
In this context, the DM peak location is primarily sensitive to the
degree of smoothing ($S$) and the index of the power law ($q$) of
\citet{Zitrin+2009_CL0024}. 

We find only a small offset of $\sim 1\arcsec$, or a projected offset
distance of $d_{\rm off}= 4\,$kpc\,$h^{-1}$ at $z_l=0.439$,
between the BCG and the DM peak of mass, well within the
uncertainties.   
The BCG position also coincides well with the peak of X-ray emission
within $2\arcsec$ in projection (Table \ref{tab:cluster}).
This level of cluster centering offset is fairly small as compared
to those found in other high-mass clusters, say $d_{\rm off}\approx
20$\,kpc\,$h^{-1}$ in RXJ1347-11 \citep{Umetsu+2011a}, 
often implied by other massive bright galaxies in
the vicinity of the BCG.
In the present work, we thus adopt the BCG position as the cluster
center, and limit our analysis to radii greater than 
$4\arcsec$ ($\approx 16\,$kpc\,$h^{-1}$), which is approximately the
location of the innermost strong-lensing constraint 
(see Section \ref{subsec:Z12}) and sufficiently large 
to avoid the BCG contribution. 
This inner radial limit corresponds roughly to 
$4d_{\rm off}(>2d_{\rm off})$, beyond which smoothing from the cluster
miscentering effects on the $\Sigma$ profile is sufficiently negligible
\citep{Johnston+2007_SDSS1,Umetsu+2011b,Sereno+Zitrin2012}.

\subsection{Self-Calibration of the Covariance Matrix}
\label{subsec:calib}

The MCMC approach allows for a full parameter-space extraction of the
underlying lensing signal. 
We construct from MCMC samples a central mass profile $\kappa_i$ and its
covariance matrix ${\cal C}_{ij}$ in linearly-spaced radial bins, 
spanning from $\theta=1\arcsec$ to the limit of our ACS data, 
$\theta \sim 100\arcsec$.
Note, multiple image constraints are available out to a radius of
$\approx 1\arcmin$ (Section \ref{subsec:Z12}), so that
the mass model beyond this radius
is constrained essentially by the light distribution of cluster member
galaxies, and hence the constraints there are driven by the prior.
We find that the mass profile is positively correlated from bin to bin,
especially at radii beyond $\theta_{\rm Ein}\approx 28\arcsec$
($z_s=2.5$). 
Accordingly, the ${\cal C}$ matrix is nearly singular, with very small
eigenvalues associated with large-scale modes where the constraints are weaker, 
leading to underestimated diagonal errors at
$\theta\simgt \theta_{\rm Ein}\approx 28\arcsec$ ($z_s=2.5$).  

Here, we use a regularization technique with a single degree of freedom
to calibrate the ${\cal C}$ matrix and obtain conservative errors 
for strong lensing, accounting for possible systematic errors 
introduced by the prior assumptions in the modeling.
We first perform an eigenvalue decomposition as ${\cal C}=U\Lambda U^t$,
where $\Lambda$ is a diagonal matrix of eigenvalues and $U$ is a unitary
matrix of eigenvectors. Then, we determine our regularization constant,
the minimum eigenvalue $\Lambda_{\rm min}$, by conservatively 
requiring that the
outermost $\kappa$ value, $\kappa_{\rm min}=\kappa(100\arcsec)\approx 0.22$,  
is consistent with a null detection: i.e., 
$\Lambda_{\rm min}=\kappa_{\rm min}^2=(0.22)^2$.
Replacing those less than $\Lambda_{\rm min}$
by $\Lambda_{\rm min}$ and restoring the
 ${\cal C}$ matrix with the regularized $\Lambda$ yields the desired,
 self-calibrated ${\cal C}$ matrix.
All points at $\simgt 1\arcmin$ are then excluded from our analysis.
We find that a weaker regularization with $\Lambda_{\rm min}=(0.1)^2$
only affects the halo parameters ($M_{\rm vir}$, $c_{\rm vir}$)
by less than $4\%$.

In Figure \ref{fig:kappa} we show our strong-lensing constraints on 
the central $\kappa$ profile using the self-calibrated ${\cal C}$
matrix, where the outer radial boundary is conservatively set to
$\theta=53\arcsec$ 
\citep[$\approx 2\theta_{\rm Ein}$ at $z_s=2$;
see][]{Zitrin+2012_M1206}.  
This calibration scheme produces conservative error estimates.  
Overall, the level of correction applied to the ${\cal C}$ matrix
increases with increasing radius. 
We introduce here an estimator for the total signal-to-noise ratio (S/N)
for detection, integrated over the radial range considered, 
and quantify the significance of the reconstruction,
by the
following equation \citep{UB2008}: 
\begin{equation}
\label{eq:sn}
{\rm (S/N)^2} = \displaystyle{\sum}_{i,j} \kappa_i {\cal C}^{-1}_{ij} 
\kappa_j = \bkappa^{t} {\cal C}^{-1} \bkappa.
\end{equation}
With the calibrated ${\cal C}$ matrix, we find a total S/N of 
$\approx 18$ 
for our strong-lensing $\kappa$ profile in the radial range 
$\theta\le 53\arcsec$.
We check that our results are insensitive to the choice of radial
binning scheme when the self-calibration technique is applied.

\subsection{Complementary Strong Lensing Analyses}
\label{subsec:sl+}

\begin{figure*}[htb]
 \begin{center}
   \includegraphics[width=0.8\textwidth,angle=0,clip]{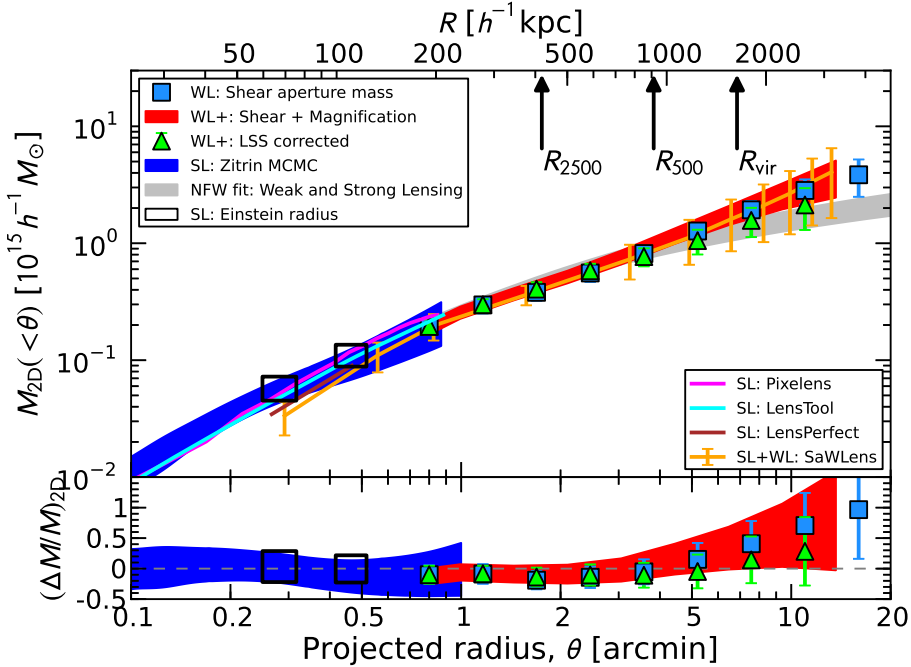}
 \end{center}
\caption{
Comparison of projected cumulative mass profiles $M_{\rm 2D}$ of
 MACS1206 obtained with different lensing methods.
The red shaded area shows our full weak-lensing constraints 
(68\% CL) derived from 
 a joint Bayesian analysis of Subaru shear and magnification
 measurements (Figure \ref{fig:wldata}), in good agreement with the
 shear aperture mass measurements (squares) obtained with a zero density
 boundary condition of $\overline{\Sigma}(16\arcmin<\theta<18\arcmin)=0$.
The triangles denote the mass profile using the NE-SW
$\Sigma$ profile of Figure \ref{fig:kappa} excluding the
NW-SE excess regions.
The two open rectangles represent model-independent Einstein-radius
constraints of 
$\theta_{\rm Ein}=17\arcsec \pm 2\arcsec$ ($z_s=1.03$) and 
$\theta_{\rm Ein}=28\arcsec \pm 3\arcsec$ ($z_s=2.54$).
The blue shaded area represents our primary strong-lens model
with $1\sigma$ uncertainty from an MCMC implementation of Zitrin et
 al. (2012), which is broadly consistent with our 
 semi-independent results  from a wide variety of four strong-lens modeling analyses
 ({\sc Pixelens}, {\sc Lenstool}, {\sc LensPerfect}, and {\sc
 SaWLens}). providing a valuable consistency check.  
Our independent strong and weak lensing profiles are in good agreement
 in the region of overlap, and together are well described by the
 standard NFW form (gray area), but increasingly exceed it at $R\simgt
 1\,$Mpc\,$h^{-1}$ out to the limit of our data.
The bottom panel shows fractional deviations $(\Delta M/M)_{\rm 2D}$
of projected mass profiles
 with respect to the best-fit NFW model (top, gray), demonstrating the
 presence of a large scale anisotropy in the mass distribution around
 the cluster.
\label{fig:m2d}
}
\end{figure*} 

We have performed complementary semi-independent strong-lensing analyses
({\sc Lenstool},
{\sc LensPerfect},
{\sc Pixelens},
{\sc SaWLens}),
using as input the sets (or subsets) of multiple images
identified by \citet{Zitrin+2012_M1206} as well as the same
spectroscopic and photometric redshift information. 
 
In our {\sc Lenstool} analysis, we parameterize the lens mass
distribution $\Sigma(\btheta)$ as a multi-component model consisting of
an elliptical NFW potential and truncated elliptical halos
\citep{Kassiola+Kovner1993} for the 86 brightest cluster members.  
All nine of the secure image systems are included as observational
constraints.
Our best solution reproduces all arc systems included and the critical
lines at $z_s=2.54$ and $1.03$ derived in \citet{Zitrin+2012_M1206},  
with an image-plane rms of $1.9\arcsec$, very similar to the value of
$\sim 1.8\arcsec$ obtained by \citet{Zitrin+2012_M1206} and typical to
parametric mass models for clusters with many multiple images 
\citep{2005ApJ...621...53B,2006MNRAS.372.1425H,2007ApJ...668..643L,Zitrin+2009_CL0024}.

In the {\sc Pixelens} analysis we model the lens mass distribution on a 
circular grid of $52\arcsec$ radius divided into 18 pixels.
We consider 200 models with decreasing projected mass profiles
(i.e., $\Sigma(R)\propto R^{-\alpha}$ with $\alpha>0$).
We use as constraints the spectroscopically-confirmed Systems 1 to 4
\citep{Zitrin+2012_M1206} of 14 multiple images, spanning the range 
3.5\arcsec to 46\arcsec in radius.  We check that adding other
multiple-image systems identified in \citet{Zitrin+2012_M1206} does not
significantly affect the {\sc Pixelens} mass reconstruction.

In the {\sc LensPerfect} analysis, we assume a prior that the projected
mass is densest near the center of the BCG and decreases outward.  
Other priors include overall smoothness and approximate
azimuthal symmetry \citep[for details, see][]{Coe+2010}.   
All secure image systems are used in this modeling, where including
the three candidate systems (\#9--11) does not change the results
significantly.  

The {\sc SaWLens} method combines central strong-lensing constraints
from multiple-image systems with weak-lensing distortion constraints in a
non-parametric manner to reconstruct the underlying lensing potential on
an adaptively refined mesh.  
For this cluster we use two levels of refinement, providing a $6\arcsec$
pixel resolution in the strong-lensing regime covered by CLASH imaging
and a $\approx 22\arcsec$ resolution in the Subaru weak-lensing field
where the background source galaxies are sparsely sampled.
The field size for the reconstruction is $25\arcmin$ on a side.
All image systems except \#10 and \#11 are included as strong-lensing
constraints.  The lens distortion measurements for the blue+red sample
are used as weak-lensing constraints.
The reconstruction errors are derived from 1000 bootstrap realizations
of the weak-lensing background catalog and 1000 samples of the redshift
uncertainties in the catalog of strong lensing features. The number of
realizations is limited by runtime constraints.

Figure \ref{fig:m2d} shows and compares the resulting 
projected integrated mass profiles  $M_{\rm 2D}(<\theta)$ 
derived from our comprehensive strong-lensing analyses, along with our
primary strong-lensing results and model-independent
Einstein-radius constraints based on
\citet{Zitrin+2012_M1206}.  All these models are broadly consistent
with the Einstein radius constraints.
The calibrated error bars of
\citet{Zitrin+2012_M1206} are roughly consistent with the spread of
the semi-independent mass profiles derived here.
This comparison shows clear consistency
among a wide variety of analysis methods with different assumptions and
systematics, which firmly supports the reliability of 
our strong-lensing analyses and calibration.

\section{Cluster Weak Lensing Analysis}
\label{sec:wl}

This section is devoted to our cluster weak-lensing analysis based on
the deep multi-color Subaru observations.
In Section \ref{subsec:MNL} we present the projected mass and galaxy
distributions in and around MACS1206. 
In Section \ref{subsec:wlprof} we derive cluster lens distortion and
magnification radial profiles from Subaru data.
In Section \ref{subsec:wlmass} we briefly summarize our Bayesian mass
inversion methods based on combined lens distortion and magnification
measurements, 
and apply to Subaru weak-lensing observations of MACS1206.

\subsection{Two-Dimensional Mass Map}
\label{subsec:MNL}

\begin{figure*}[htb]
 \begin{center}
   \includegraphics[width=0.8\textwidth,angle=0,clip]{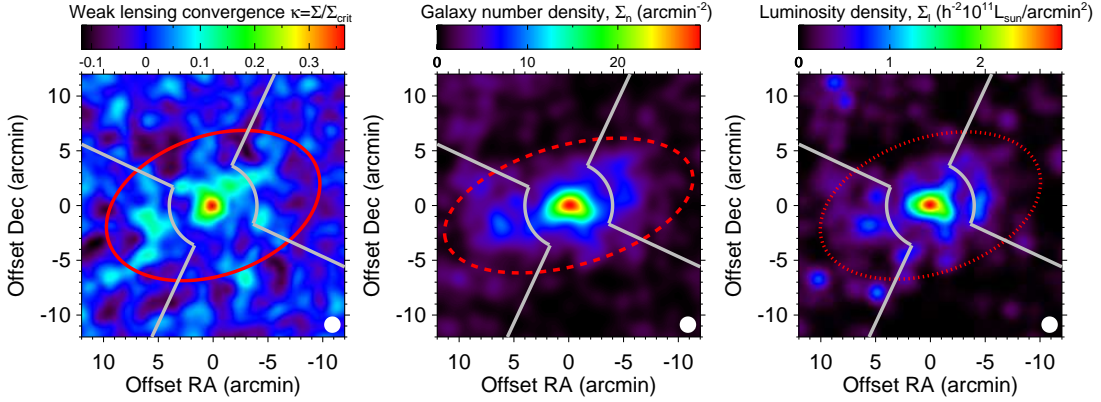}
 \end{center}
\caption{
Comparison of the surface-mass density field and the cluster galaxy
 distributions in MACS1206. 
Left: linear reconstruction of the dimensionless surface-mass density
 field, or the 
lensing convergence $\kappa(\btheta)=\Sigma(\btheta)/\Sigma_{\rm crit}$, 
reconstructed from Subaru distortion data. 
Middle: observed surface number density distribution $\Sigma_n(\btheta)$
of green galaxies,
representing cluster member galaxies. 
Right: observed $R_{\rm c}$-band surface luminosity density distribution
 $\Sigma_l(\btheta)$ of the same cluster membership. 
The solid ellipse in each panel indicates the respective mean
 ellipticity and orientation measured within a circular aperture of
 $8\arcmin$, 
which is slightly larger than the cluster virial radius ($\theta_{\rm
 vir}\approx 6.9\arcmin$).
The pair of gray solid lines in each panel defines the north-west (NW)
 and south-east (SE) excess regions.
All images are smoothed with a circular Gaussian of FWHM $1.5\arcmin$. 
The field size is $24\arcmin \times 24\arcmin$. North is to the top,
 east to the left. 
\label{fig:MNL} 
}
\end{figure*}  

Weak-lensing distortion measurements ($g$) can be used
to reconstruct the underlying projected mass density field
$\Sigma(\btheta)$ (see Equation \ref{eq:k2g}).  
Here we use the linear map-making method outlined in Section 4.4 of 
\citet{Umetsu+2009} to
derive the projected mass distribution from
the Subaru distortion data presented in Section \ref{sec:subaru}.

In the left panel of Figure \ref{fig:MNL}, 
we show the $\Sigma(\btheta)$ field in the central $24\arcmin\times
24\arcmin$ region, 
reconstructed from the blue+red sample (Section \ref{subsec:color}),
where for visualization purposes the mass map is smoothed with a
Gaussian with $1.5\arcmin$ FWHM.
A prominent mass peak is visible in the cluster center.
This first maximum in the mass map is detected at a significance level
of $9.5\sigma$, and coincides well with the optical/X-ray cluster center
within the statistical
uncertainty: 
$\Delta{\rm R.A.}=7.0\arcsec \pm 7.2\arcsec$,
$\Delta{\rm Decl.}=-1.4\arcsec \pm 7.6\arcsec$,
where $\Delta{\rm R.A.}$ and $\Delta{\rm Decl.}$ are right-ascension and
declination offsets, respectively, from the BCG center.

Also compared in Figure \ref{fig:MNL} are
member galaxy distributions in the MACS1206 field,
Gaussian smoothed to the same
resolution of $\theta_{\rm FWHM}=1.5\arcmin$.
The middle and right panels display the
number and ($K$-corrected) $R_{\rm c}$-band luminosity density fields, respectively,
of green cluster galaxies (see Table \ref{tab:color}). 

Overall, mass and light are similarly distributed in the cluster:
The cluster is fairly centrally concentrated in projection, and
associated with elongated 
LSS running north-west
south-east (NW-SE), both in the projected mass and galaxy distributions.
A more quantitative characterization of the 2D matter distribution 
around the cluster will be given in Section \ref{sec:mass}.

\subsection{Cluster Weak-Lensing Profiles}
\label{subsec:wlprof}

Now we derive azimuthally-averaged lens distortion and magnification
profiles from the Subaru data.
We calculate the weak-lensing profiles in $N$ discrete radial bins from
the cluster center (Section \ref{subsec:offset}),
spanning the range $[\theta_{\rm min},\theta_{\rm max}]$
with a constant logarithmic radial spacing
$\Delta\ln\theta=\ln(\theta_{\rm max}/\theta_{\rm min})/N$,
where the inner radial boundary 
 $\theta_{\rm min}$ is taken to be $\theta_{\rm min}=0.8\arcmin$
 ($>\theta_{\rm Ein}$).
The outer radial boundary $\theta_{\rm max}$ is
 chosen to be $\theta_{\rm max}=16\arcmin$ ($R_{\rm max}\approx
 3.8$\,Mpc\,$h^{-1}$), 
sufficiently larger than the typical virial radius $r_{\rm vir}$
of high  mass  clusters ($r_{\rm vir}\approx 1.6$\,Mpc\,$h^{-1}$ for
 MACS1206; see Section \ref{sec:mass}),
but sufficiently small with respect to the size of the
Suprime-Cam's field-of-view so  as to ensure accurate PSF anisotropy
 correction.  
The number of radial bins is set to $N=8$,
chosen such that the detection S/N (defined as in Equation \ref{eq:sn})
is of the order of unity per pixel.

\subsubsection{Lens Distortion}
\label{subsubsec:gt} 

\begin{figure}[htb]
 \begin{center}
   \includegraphics[width=0.45\textwidth,angle=0,clip]{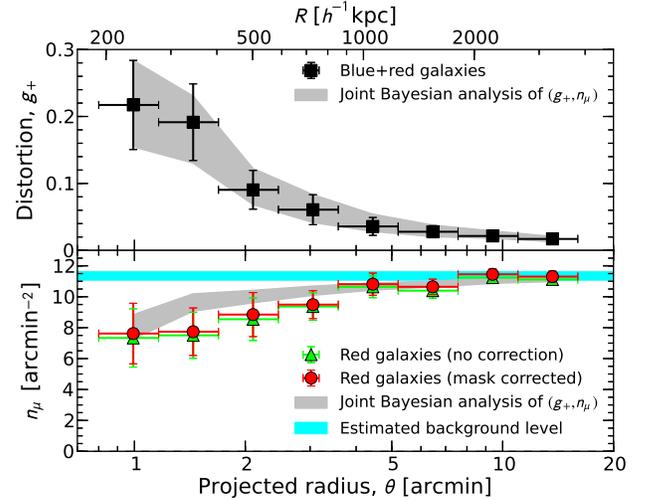}
 \end{center}
\caption{
Cluster weak-lensing radial profiles  as measured from
 background galaxies registered in deep Subaru images.
The top panel shows the tangential reduced shear profile $g_+(\theta)$
(squares) based on Subaru distortion data of
 the full background (red+blue) sample.
The bottom panel shows the count depletion profiles $n(\theta)$ due to 
magnification for a flux-limited 
 sample of red background
 galaxies. The circles and triangles show the respective results
 with and without the mask correction due to bright foreground objects 
 and cluster members. The horizontal bar represents the constraints on
 the unlensed  count normalization, $n_0$, as estimated from Subaru
 data.  Also shown in each panel is the joint Bayesian fit (68\% CL) to
 both profiles.  
\label{fig:wldata} 
}
\end{figure}  

For each galaxy, we define the tangential distortion $g_+$ and the
$45^\circ$ rotated  component, with respect to the cluster center, from
linear combinations of the distortion coefficients $(g_1,g_2)$ as 
$g_+ = -(g_1 \cos 2\phi + g_2\sin 2\phi)$ and
$g_{\times} = -(g_2 \cos 2\phi - g_1\sin 2\phi)$,
with $\phi$ being the position angle of an object with respect to the
cluster center. 
In the absence of higher-order effects, weak lensing only induces
curl-free tangential distortions, while the azimuthal averaged $\times$
component is expected to vanish. In practice, the presence of $\times$
modes can be used to check for systematic errors.

For each galaxy sample,
we calculate the weighted average of $g_+$
in a set of radial bins ($i=1,2,...,N$) as  
\begin{equation}
\label{eq:gt}
g_{+,i} \equiv g_+(\theta_i)=
\left[
\displaystyle\sum_{k\in i} w_{(k)}\, g_{+(k)}
\right]
\left[
\displaystyle\sum_{k\in i} w_{(k)}\right]^{-1},
\end{equation}
where the index $k$ runs over all objects located within the $i$th
annulus,
$\theta_i$ is the weighted center of the $i$th radial bin,
and the weight factor $w_{(k)}$ is defined by Equation \ref{eq:weight}.
We use the continuous limit of the
area-weighted center for $\theta_i$ \citep[see Appendix A of][]{UB2008}.
We perform a bootstrap error analysis to assess the uncertainty
$\sigma_{+,i}$ in the tangential distortion profile
$g_{+,i}$ \citep{Umetsu+2010_CL0024}.

In Figure \ref{fig:rgb}, we compare azimuthally-averaged radial profiles
of $g_+$ and $g_\times$ as measured from our red, blue, green, and
blue+red galaxy samples (Section \ref{subsec:color}). 
For all samples, the $\times$ component is consistent with a null
detection well within $2\sigma$ at all radii, indicating the reliability
of our distortion analysis. 
The red and blue populations show a very similar form of the radial
$g_+$ profile which declines smoothly from the cluster center.
The observed tangential distortion signal is significant with a total
detection S/N of 8.1 and 5.1 for the red and the blue sample,
respectively, both remaining positive to the limit of our data,
$\theta_{\rm max}=16\arcmin$.     
The detection significance is improved to $9.3\sigma$ using a full
composite sample of Subaru blue+red background galaxies (see the top
panel of Figure \ref{fig:wldata}). 

In Figure \ref{fig:rgb} we also compare the Subaru data with the results
obtained from CFHT/Megacam data (Section \ref{subsec:data}) using the
same analysis pipeline as described in Section \ref{sec:subaru}.  
For this we identified 15875 background galaxies 
($n_g\approx 4.5$\,galaxies\,arcmin$^{-2}$) with Megacam $grz$
photometry using our CC background selection method (Section
\ref{subsec:color}), and estimated a mean depth of 
$z_{s,{\rm eff}}\approx 1.09$,
comparable to that of the Subaru full background sample ($z_{s,{\rm
eff}}=1.15\pm 0.1$, $n_g\approx 13$\,galaxies\,arcmin$^{-2}$; 
Section \ref{subsec:depth}).  
This comparison shows excellent agreement where the data overlap,
demonstrating the robustness of our analysis.

\subsubsection{Magnification Bias}
\label{subsubsec:magbias}

For the number counts to measure magnification, we follow the
prescription of \citet{Umetsu+2011a}.
We use a sample of {\it red} galaxies (Section \ref{subsec:color}), for
which the intrinsic count slope $s$ 
at faint magnitudes is relatively flat, $s\sim 0.1$, so that a net
count depletion results \citep{BTU+05,UB2008,Umetsu+2010_CL0024,Umetsu+2011a}.
The blue background population, on the other hand, tends to have a
steeper intrinsic count slope close to the lensing invariant slope
($s=0.4$). 

The count-in-cell statistic $N(\btheta)$
is measured from a flux-limited sample of
red background galaxies
on a regular grid of equal-area cells,
each with a constant solid angle $\Delta\Omega$.
The practical difficulty here
is contamination due to the intrinsic
clustering of background galaxies, 
which locally can be larger than the lensing-induced
signal in a given cell. In order to obtain a clean measure of the
lensing signal, such intrinsic clustering needs to be down-weighted and
averaged over
\citep[e.g.,][]{1995ApJ...438...49B,UB2008}. 

To overcome this problem, we azimuthally average the red galaxy counts 
$N(\btheta)$ and obtain the average surface number density
$n_{\mu,i}\equiv  n_\mu(\theta_i)=\langle dN(\theta_i)/d\Omega\rangle$ 
as a function of radius from the
cluster center ($i=1,2,...,N$).
Here we use the approach developed in \citet{Umetsu+2011a} to 
account and correct for the masking effect due to bright cluster
galaxies, foreground objects, and 
saturated objects.
The errors $\sigma_{\mu,i}$ for $n_{\mu,i}$
include both contributions from Poisson errors in the counts
and contamination due to intrinsic clustering of red background
galaxies.
Thanks to the wide field of view Subaru/Suprime-Cam,
the normalization and slope parameters for the red sample are reliably 
estimated  as
$n_0=11.4\pm 0.3$\,galaxies$^{-2}$
and
$s=0.133\pm 0.245$
from the coverage-corrected source counts in the outer region 
($\simgt 10\arcmin$).

We show in the bottom panel of Figure \ref{fig:wldata} the resulting
magnification profile derived from our flux-limited sample of red
background galaxies ($z'<24.6$\,mag; see Table \ref{tab:color}).
A strong depletion of the red galaxy counts is seen in the central,
high-density region of the cluster and clearly detected out to $\simlt
4\arcmin$ from the cluster center.
The statistical significance of the detection of the depletion signal is
$4.4\sigma$, which is about half the S/N ratio of the tangential
distortion derived from the full background sample shown in the top
panel of Figure \ref{fig:wldata}. 
The magnification measurements with and without the masking correction
are roughly consistent with each other.

\subsection{Mass Profile Reconstruction}
\label{subsec:wlmass}

The relation between observable distortion ($g$) and underlying
convergence ($\kappa$) is nonlocal.  Hence the mass distribution
derived from distortion data alone suffers from a mass-sheet 
degeneracy (\S~\ref{sec:basis}). 

Here we construct a radial mass profile from 
complementary lens distortion and magnification
measurements, $\{g_{+,i}\}_{i=1}^N$ and $\{n_{\mu,i}\}_{i=1}^N$, 
following the Bayesian prescription given by \citet[][]{Umetsu+2011a},
effectively breaking the mass-sheet degeneracy.
A brief summary of this Bayesian method is provided in Appendix
\ref{subsec:1D}.  The model is described by a vector $\bs$ of parameters
containing the 
discrete convergence profile $\{\kappa_i\}_{i=1}^N$
in the subcritical regime ($\theta_i>\theta_{\rm Ein}$),
and the average convergence within the inner radial boundary
$\theta_{\rm min}$ of the weak-lensing data, 
$\overline{\kappa}_{\rm min}\equiv \overline{\kappa}(<\theta_{\rm
min})$, so that $\bs =\{\overline{\kappa}_{\rm min},\kappa_i\}_{i=1}^{N}$, 
being specified by $(N+1)$ parameters.

We find a consistent mass profile solution $\bs$ based on a joint Bayesian fit
to the observed distortion and magnification measurements, as shown in
Figure \ref{fig:wldata}.
The detection significance has been improved from $9.3\sigma$ to
$11.4\sigma$ by adding the magnification measurement, corresponding to
an improvement by $\sim 23\%$, compared to the lensing distortion
signal \citep{Umetsu+2011a,Coe+2012_A2261}. 

The resulting mass profile $\bs$ is shown in Figure
\ref{fig:kappa}, along with our primary strong-lensing model 
(Sections \ref{subsec:Z12}--\ref{subsec:calib}). 
Our independent strong- and weak-lensing mass profiles are in good
agreement where they overlap, and together form a well-defined
mass profile.
The outer mass profile derived from weak lensing exhibits a fairly
shallow radial trend with a nearly isothermal logarithmic density slope
in projection, $\gamma_{\rm 2D}\equiv -d\ln\Sigma/d\ln R\sim 1$.  
Note, this flat behavior is not clearly evident in the tangential
distortion profile, which is insensitive to
sheet-like mass overdensities (Section \ref{sec:basis}).  
To constrain the cluster properties from the composite halo+LSS mass
profile, this LSS contribution needs to be taken into account and
corrected for.  We will come back to this point in Sections 
\ref{subsec:elong} and \ref{subsec:lss}.   

Also shown in Figures \ref{fig:kappa} and \ref{fig:m2d}
is a purely shear-based reconstruction
using the one-dimensional (1D) method of \citet[][see also Umetsu et al. 2010]{UB2008}, 
based on the nonlinear extension of aperture mass densitometry
\citep{2000ApJ...539..540C}.  Here we have adopted a zero-density
boundary condition in the outermost radial bin, $16\arcmin\le \theta\le 18\arcmin$.
The total S/N ratio in the recovered mass profile is $\approx 9.2$, which
agrees well with $\approx 9.3$ in the $g_+$ profile (Section
\ref{subsubsec:gt}). 
Our results with different 
combinations of lensing measurements and boundary conditions, having
different systematics, are in agreement with each other.
This consistency demonstrates
that our results are robust and insensitive to 
systematic errors.

The projected cumulative mass profile $M_{\rm 2D}(<\theta)$ is given by 
integrating the density profile 
$\bs=\{\overline{\kappa}_{\rm min},\kappa_i\}_{i=1}^{N}$
\citep[see Appendices A and B of][]{Umetsu+2011a} as
\begin{equation}
M_{\rm 2D}(<\theta_i)=\pi (D_l\theta)^2 \Sigma_{\rm
 crit}\overline{\kappa}_{\rm min}
+2\pi D_l^2 \Sigma_{\rm crit}
\int_{\theta_{\rm min}}^{\theta_i}\! d\ln\theta\,\theta^2\kappa(\theta).
\end{equation}
We compare in Figure \ref{fig:m2d} the resulting
$M_{\rm 2D}$ profiles derived here from a wide variety of strong-
(Section \ref{sec:sl}) and
weak-lensing analyses, along with the
model-independent Einstein-radius constraints of
$M_{\rm 2D}(<17\arcsec)=5.8^{+1.3}_{-1.4}\times 10^{13}M_\odot\,h^{-1}$ 
at $\theta_{\rm Ein}=17\arcsec \pm 2\arcsec$ ($z_s=1.03$)
and
$M_{\rm 2D}(<28\arcsec)=1.1^{+0.2}_{-0.3}\times 10^{14}M_\odot\,h^{-1}$ 
at $\theta_{\rm Ein}=28\arcsec \pm 3\arcsec$ ($z_s=2.54$).\footnote{
\citet{Zitrin+2012_M1206} quote their full-model based estimates on the
respective integrated masses of 
$M_{\rm 2D}=6\pm 0.7\times 10^{13}M_\odot\,h^{-1}$
and
$M_{\rm 2D}=0.94\pm 0.11 \times 10^{14}M_\odot\,h^{-1}$.} 
Again, we find good agreement in the regions of overlap
among the results obtained from a variety of lensing analyses,
ensuring consistency of our lensing analysis and methods.

Unlike the non-local distortion effect, 
the magnification falls off sharply with increasing distance from
the cluster center. 
For MACS1206, we find $\kappa \simlt 1\%$ at radii $\simgt 10\arcmin$, where
the expected level of the depletion signal is 
$n_\mu/n_0-1\approx -2\kappa$ for a maximally-depleted sample with
$s=0$, indicating a depletion signal of $\simlt 2\%$ in the
outer region where we have estimated the unlensed
background counts, $n_0$.  This level of signal is smaller than
the fractional uncertainties in estimated unlensed counts $n_0$ of
$3\%$ (Section \ref{subsubsec:magbias}), thus consistent with the assumption.
Note that 
the calibration uncertainties in our observational parameters
$(n_0,s,\omega)$ have been marginalized over in our Bayesian analysis
(Appendix \ref{sec:wlmethod}).

In the presence of magnification, one probes the number counts at an
effectively fainter limiting magnitude:  
$m_{\rm cut}+2.5\log_{10}\mu(\theta)$.   
The level of magnification is on average small in the weak-lensing
regime but reaches $\mu\approx 1.6$ (at $z_{s,{\rm eff}}\approx 1.1$)
for the innermost bin in this cluster.  
Hence, we have implicitly assumed in our analysis that the power-law
behavior (Equation [\ref{eq:magbias}]) persists down to $\sim 0.5$ mag
fainter than $m_{\rm cut}$ where the count slope may be shallower.  
For a given level of count depletion, an underestimation of the
effective count slope could lead to an underestimation of $\mu$, thus
biasing the resulting mass profile.  
However, the count slope for our data flattens only slowly with depth
varying from $s\sim 0.13$ to $\sim 0.05$ from a limit of $z'=24.6$ to
$25.1$\,mag, so  that this introduces a small correction of only $\sim
10\%$ for the most magnified bins ($\mu\sim 2$).  
In fact, we have found a good consistency between the results with and
without the magnification data.

\section{Mass Profile from Joint Weak and Strong Lensing Analysis}
\label{sec:mass}

In this section, we aim to quantify and characterize the
mass distribution of MACS1206 using our comprehensive
lensing measurements derived from the deep {\it HST} and Subaru observations
described in Sections \ref{sec:sl} and \ref{sec:wl}.
Here, we compare the cluster lensing profiles with 
the theoretically and observationally motivated NFW model
\citep{1997ApJ...490..493N} to characterize the cluster mass profile.
Our use of the NFW model enables the most direct comparison with
detailed theoretical predictions for the internal structure of
DM halos based on $N$-body simulations
\citep[e.g.,][]{Duffy+2008,Klypin+2011,Prada+2011,Bhat+2011}. 
The choice of profile shape does not significantly affect the derived
halo concentrations \citep[e.g.,][]{Duffy+2008}.

To be able to constrain the inner density slope,
we consider a generalized parameterization of the
NFW model (gNFW, hereafter) of the form 
\citep{Zhao1996,Jing+Suto2000}:
\begin{equation}
\label{eq:gnfw}
\rho(r)=\frac{\rho_s}{(r/r_s)^\alpha(1+r/r_s)^{3-\alpha}},
\end{equation} 
where $\rho_s$ is the characteristic density, 
$r_s$ is the characteristic scale radius, and
$\alpha$ represents the inner slope of the density profile.
This reduces to the NFW model for $\alpha=1$.
We introduce the radius $r_{-2}$ at which the logarithmic
slope of the density is isothermal, i.e., $\gamma_{\rm 3D}=2$. For the
gNFW profile, $r_{-2}=(2-\alpha)r_s$, and thus the corresponding
concentration parameter reduces to $c_{-2}\equiv r_{\rm
vir}/r_{-2}=c_{\rm vir}/(2-\alpha)$.
We specify the gNFW model with the central cusp slope, $\alpha$, 
the halo virial mass, $M_{\rm vir}$, and the concentration,
$c_{-2}=c_{\rm vir}/(2-\alpha)$.
We employ the radial dependence of the gNFW lensing profiles given by
\citet{Keeton2001_mass}.

\subsection{Model-Independent Constraints}
\label{subsec:wl+sl}
 
First, we constrain the NFW model parameters $\bp\equiv (M_{\rm vir},c_{\rm
vir})$ by combining model-independent weak-lensing distortion,
magnification, and strong-lensing Einstein-radius measurements,  whose
systematic errors are well understood from numerical simulations 
\citep[e.g.,][]{Meneghetti+2011,Rasia+2012}.
The $\chi^2$ function for the combined Einstein-radius and weak-lensing
constraints is expressed as
\begin{equation}
\label{eq:wl+re}
\chi^2 = \chi^2_{\rm Ein} + \chi^2_{\rm WL},
\end{equation}
where the $\chi^2_{\rm Ein}$ for the Einstein-radius constraints is
defined by \citep[see][]{UB2008,Umetsu+2010_CL0024}
\begin{eqnarray}
\chi^2_{\rm Ein} = \displaystyle\sum_{i=1}^{N_{\rm Ein}}
\frac{\left[1-\hat{g}_{+,i}(\bp,z_{s,i})\right]^2}{\sigma_{+,i}^2}
\end{eqnarray}
with 
$N_{\rm Ein}$ being the number of independent Einstein-radius
constraints $\{\theta_{{\rm Ein},i}\}_{i=1}^{N_{\rm Ein}}$
from sources with different redshifts $\{z_{s,i}\}_{i=1}^{N_{\rm Ein}}$ and 
$\hat{g}_{+,i}(\bp,z_{s,i})=\hat{g}(\theta_{{\rm Ein},i}|\bp,z_{s,i})$ being the
NFW model prediction for the reduced tangential shear at
$\theta=\theta_{{\rm Ein},i}$, evaluated at the source redshift $z_{s}=z_{s,i}$.
Note, the Einstein radius marks the point of maximum distortion,
$g_+=(\overline{\kappa}-\kappa)/(1-\kappa)=1$: i.e., $\overline{\kappa}=1$
within $\theta_{\rm Ein}$. 
The $\chi^2$ function for our full weak-lensing analysis (Section
\ref{subsec:wlmass}) is described by  
\begin{equation}
\label{eq:chi2wl}
\chi^2_{\rm WL} = \displaystyle\sum_{i,j}
 \left[s_{i}-\hat{s}_{i}(\bp,z_{s,{\rm eff}})\right] 
\left({\cal C}_{\rm WL}\right)_{ij}^{-1}
 \left[s_{j}-\hat{s}_{j}(\bp,z_{s,{\rm eff}})\right],
\end{equation}
where $\bs=\{\overline{\kappa}_{\rm min},\kappa_i\}_{i=1}^{N}$ is the
mass profile reconstructed from the combined lens distortion and
magnification measurements,
$\hat{\bs}(\bp,z_{s,{\rm eff}})$ is the NFW model prediction for $\bs$,
and ${\cal C}_{\rm WL}$ is the full covariance matrix of $\bs$ defined as
\begin{equation}
{\cal C}_{\rm WL} = {\cal C} + {\cal C}^{\rm lss}
\end{equation}
with ${\cal C}$ being responsible for statistical measurement
errors (Appendix \ref{subsec:1D}) and ${\cal C}^{\rm lss}$ being the
cosmic covariance matrix responsible for the effect of 
uncorrelated LSS along the line of sight
\citep{2003MNRAS.339.1155H,Hoekstra+2011,Umetsu+2011b,Oguri+Takada2011}.\footnote{
As discussed in \citet{Oguri+2010_LoCuSS},for a ground-based
weak-lensing analysis, the shot noise is a more dominant source of the
measurement errors than the cosmic noise contamination. 
They found from a weak lensing analysis of 25 X-ray selected clusters
that the best-fit parameters are not largely biased by including the
cosmic noise covariance, but are in general consistent with each other
within statistical uncertainties. 
}
In all modeling below, the effective source redshift $z_{s,{\rm
eff}}=1.15\pm 0.1$ of our full background sample is treated as a
nuisance parameter, and its uncertainty is marginalized over.
In order to evaluate ${\cal C}^{\rm lss}$,
we assume the concordance $\Lambda$CDM cosmological model of
\citet{Komatsu+2011_WMAP7} and use the fitting formula of \citet{PD96}
to compute the nonlinear matter power spectrum.
We project the matter spectrum out to an effective source redshift of
$z_{s,{\rm eff}}=1.15$ to calculate ${\cal
C}^{\rm lss}$ for weak-lensing observations. For details, see
\citet{Umetsu+2011b}. 
For Einstein-radius measurements, we conservatively assume an rms
displacement of $2\arcsec$ due to uncorrelated LSS, as
predicted by recent theoretical work \citep[$\sim 2\arcsec$ for a
distant source at $z_s\sim 2.5$; see][]{Host2012,Jullo+2010}. This is combined in
quadrature with the measurement error in $\theta_{\rm Ein}$
(Table \ref{tab:cluster}) to estimate a total uncertainty
$\sigma_{+,i}$.\footnote{
Following \citet{UB2008}, we propagate the uncertainty in $\theta_{\rm
Ein}$ to $g_+$ assuming a singular isothermal sphere (SIS) model. At
$r\simlt r_s$, the density slope of NFW is shallower than that of SIS \citep[see Figure
1 of][]{2000ApJ...534...34W}, so that this gives a fairly conservative
estimate of $\sigma_{+}(\theta_{\rm Ein})$.}

For strong lensing,
we use double Einstein-radius constraints ($N_{\rm Ein}=2$) from 
the multiple image systems at $z_s=1.03$ and
$z_s=2.54$ (Table \ref{tab:cluster}).
For weak lensing, the cluster mass profile $\bs$  is measured  
in $N+1=9$ bins.
Hence, we have a total of 11 constraints.

The resulting constraints on the NFW model parameters are summarized in
Table \ref{tab:nfw}.

\subsubsection{Weak Lensing Constraints}
\label{subsubsec:wl+sl:wl}

First of all, when no magnification or strong-lensing information is
included, the best-fit model is obtained from a tangential reduced shear
fitting as 
$M_{\rm vir}=0.99^{+0.32}_{-0.26}\times 10^{15}M_\odot\,h^{-1}$ 
and 
$c_{\rm vir}=5.7^{+3.6}_{-2.1}$ with
$\chi^2_{\rm min}/{\rm dof}=3.3/6$.\footnote{We follow
\citet{2003MNRAS.339.1155H} to calculate the cosmic shear covariance
matrix.}

Next, when magnification bias is included to break the mass-sheet
degeneracy, we find  
$M_{\rm vir}=1.15^{+0.34}_{-0.28}\times 10^{15}M_\odot\,h^{-1}$ 
and 
$c_{\rm vir}=4.0^{+2.1}_{-1.4}$ 
($\chi^2_{\rm min}/{\rm dof}=4.5/7$), which is consistent within the large
uncertainties with the purely shear-based results, but is in favor of a larger
$M_{\rm vir}$ and a smaller $c_{\rm vir}$,
owing to the shallow outer mass profile. 
This is demonstrated in the bottom panel of Figure \ref{fig:kappa},
which shows significant deviations $\Delta\kappa$ 
from  our reference NFW model 
($M_{\rm vir}\approx 1.1\times 10^{15}M_\odot\,h^{-1}$
and
$c_{\rm vir}\approx 6.9$; see Section \ref{subsec:lss}) 
at cluster outskirts, $R\simgt 1\,$Mpc\,$h^{-1}$ ($\theta\simgt 4\arcmin$).
This large-scale excess in projected mass
is also shown in Figure \ref{fig:m2d} in terms of the integrated
projected mass profile $M_{\rm 2D}(<R)$.
Both fits here underestimate the observed Einstein radius
(see Table \ref{tab:nfw}).

\subsubsection{Combining Einstein Radius Constraints with Weak Lensing}
\label{subsubsec:wl+sl:sl}

When the inner Einstein-radius information is combined with weak
lensing, we obtain tighter parameter constraints. 
By combining all lens distortion, magnification, and Einstein-radius
constraints (Equation \ref{eq:wl+re}), we find 
$M_{\rm vir}=1.0^{+0.3}_{-0.2}\times 10^{15}M_\odot\,h^{-1}$ 
and 
$c_{\rm vir}=6.8^{+2.1}_{-1.6}$ 
($\chi^2_{\rm min}/{\rm dof}=6.9/9$), corresponding to an effective
Einstein radius of $\theta_{\rm Ein}\approx 26\arcsec$ at $z_s=2.5$.
That is, a slightly higher concentration is favored to reproduce
the observed large  Einstein radii \citep{Broadhurst+Barkana2008}.

\subsection{Mass and Galaxy Distribution Shapes in and around the Cluster}
\label{subsec:elong}

\begin{figure}[htb!]
 \begin{center}
   \includegraphics[width=0.45\textwidth,clip]{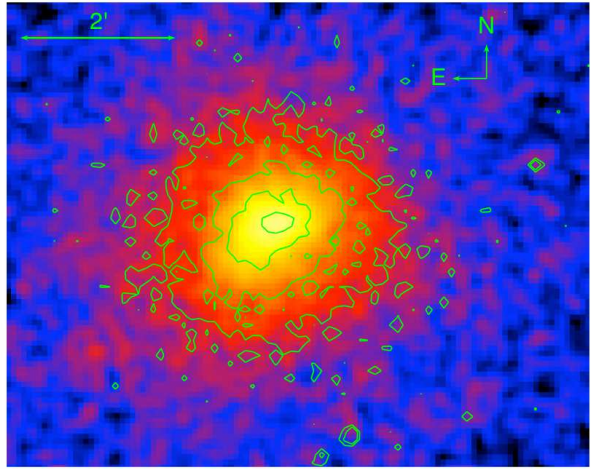}
 \end{center}
\caption{
Logarithmically-scaled XMM-{\it Newton} mosaic, exposure-corrected image
 of MACS1206 in the 0.5--2\,keV band, smoothed with a Gaussian 
of $8.0\arcsec$ FWHM. Overlaid are contours from the exposure-corrected
 {\it Chandra} 0.5--2\,keV image, smoothed with a Gaussian of
 $1.5\arcsec$ FWHM. The field size is $7.5\arcmin \times 6.0\arcmin$,
 with north to the top and east to the left. 
The scale bar shows $2\arcmin$ or about $680\,{\rm kpc} \approx 1.1 r_{2500}$.
X-ray emission is concentrated around and peaked on the BCG, but
shows some elongation within $\theta\simlt 1\arcmin$ at a position angle
around $120\deg$ east of north, aligned with the orientation of
the projected mass distribution. 
At larger distances from the cluster center,
the cluster appears fairly round in both {\it Chandra} and XMM images.
\label{fig:xraymap} 
}
\end{figure}  

The presence of surrounding large scale structure in MACS1206
has a non-negligible impact on the determination of cluster mass profile
especially at large radii (Sections \ref{subsec:wlmass} and
\ref{subsec:wl+sl}).   
It is therefore necessary to assess and correct for their effects on the
projected mass profile.  Here we use two different methods to quantify
the ellipticity and orientation of the projected mass distribution in
and around the cluster. 

First, following the prescription given by \citet{Umetsu+2009},
we introduce {\it mass-weighted} quadrupole shape moments around the
cluster center, in analogy to Equation (\ref{eq:Qij}), defined as
\begin{eqnarray}
\label{eq:qij_halo}
Q_{\alpha\beta}
= \int_{\Delta\theta\le \theta_{\rm max}}
\!d^2\theta\,  
 \Delta\theta_\alpha\Delta\theta_\beta \, \Sigma(\btheta) \ \ \ (\alpha,\beta=1,2),
\end{eqnarray}
where $\theta_{\rm max}$ is the circular aperture radius, 
and
$\Delta\theta_\alpha$ is the angular displacement vector from the
cluster center.
We construct with $\{Q_{\alpha\beta}\}$
a spin-2 ellipticity measure $e_\Sigma=|e_\Sigma|e^{2i\phi_e}$,
where the ellipticity is defined such that,
for an ellipse with major and minor axes $a$ and $b$,
it reduces to $|e_\Sigma|=1-b/a$ 
and
$\phi_e$ is the position angle of the major axis
\citep{1996A&AS..117..393B}, measured north of west here.
Similarly, the spin-2 ellipticity for the cluster galaxies
is defined using the surface number and $R_{\rm c}$-band luminosity density
fields of CC-selected cluster galaxies  (Section \ref{subsec:MNL}),
$\Sigma_n(\btheta)$ and $\Sigma_l(\btheta)$.
We calculate weighted moments using
only those pixels above the $2\sigma$ threshold with respect to the
background level
\citep[estimated with the biweight scale and location, see][]{1990AJ....100...32B}.
Practical shape measurements are done using pixelized maps
shown in Figure \ref{fig:MNL}. 

Next, we constrain the ellipticity and orientation of the projected mass
distribution by directly fitting a 2D shear map with a single elliptical
lens model.  Here, we closely follow the prescription given by
\citet{Oguri+2010_LoCuSS} to construct an
elliptical NFW (eNFW, hereafter)
model \citep[see also][]{Oguri+2012_SGAS}, 
by introducing the mass ellipticity
$|e_\Sigma|=1-b/a$ in the isodensity 
contours of the projected NFW profile $\Sigma(R)$
as $R^2\to X^2 (1-|e_\Sigma|) + Y^2 /(1-|e_\Sigma|)$ \citep[for details,
see][]{Oguri+2010_LoCuSS}.\footnote{As noted by
\citet{Oguri+2010_LoCuSS}, this elliptical model includes a triaxial
halo model \citep{Jing+Suto2000} which gives a better description of CDM
halos in $N$-body simulations than the spherical model.}  
The model shear field is computed by solving the 2D Poisson equation
\citep{Keeton2001_mass}. 
We then construct from Subaru data a lens distortion map
$(g_1(\btheta),g_2(\btheta))$ and its covariance matrix ${\cal C}_g$
(Equation \ref{eq:gCov}) on a 2D Cartesian grid with $1\arcmin$ spacing,
centered at the BCG. 
We exclude from our analysis the five innermost cells lying in the
central region, $\theta < 1\arcmin$,  
to avoid systematic errors 
(see Appendix \ref{subsec:2D}).
The halo centroid is fixed to the BCG position.
Accordingly, the eNFW model is specified by four model parameters,
$\bp=(M_{\rm vir},c_{\rm vir},|e_\Sigma|,\phi_e)$.  
The constraints on individual parameters are obtained by projecting the
2D shear likelihood function (Equation (\ref{eq:lg2}) in Appendix
\ref{subsec:2D})  to the parameter space
(or, minimizing $\chi^2$).  

In Table \ref{tab:elong}, 
we summarize our cluster ellipticity and orientation measurements.
In this analysis, we are mainly interested in the orientation of the
ellipticity, in order to correct for the effects of LSS along the axis
of elongation. 
An overall agreement is found between the shapes of mass, light, and
galaxy distributions in MACS1206, especially in terms of orientation
(Figure \ref{fig:MNL}), 
within large uncertainties (Table \ref{tab:elong}).
The mass distribution in and around the cluster is aligned well with the 
luminous galaxies in the green sample, comprising mostly of cluster
member galaxies (Section \ref{subsec:color}). 
For all cases, the position angle $\phi_e$ of the major axis is found to
be fairly constant with radius $\theta_{\rm max}$, and lies in the range 
$15\deg\simlt \phi_e\simlt 30\deg$.

In the central region, we find a projected mass ellipticity of
$|e_{\Sigma}|\sim 0.3$ and a position angle of $\phi_e\sim 14\deg$ from
the {\sc Pixelens} analysis;
we obtain consistent values for both
$|e_\Sigma|$ and $\phi_e$
from a different strong-lensing analysis
(C. Grillo et al., in preparation) using only System 7 of
\citet{Zitrin+2012_M1206}.
A similar value is found for the projected mass ellipticity of
$|e_\Sigma|=0.26\pm 0.16$ ($\phi_e\sim 19\deg$) at 
$\theta_{\rm max}=4\arcmin$ using the weak-lensing $\Sigma$ map.
From an elliptical King model fit to {\it Chandra} X-ray data (Figure
\ref{fig:xraymap}; for details, see Section \ref{subsec:multiw}), we
find an ellipticity of 
$0.30\pm 0.03$ ($a/b\approx 1.5$) and $\phi_e= 21.9\pm 1.7\deg$  
at $\theta_{\rm max}= 1.5\arcmin$.

On the other hand, we obtained higher values of ellipticity 
on large angular scales beyond the cluster virial radius, 
$\theta_{\rm vir}\equiv r_{\rm vir}/D_l\sim 7\arcmin$.
We find 
$|e_\Sigma|\sim 0.4$--$0.5$ at $\theta_{\rm max}=8\arcmin$ 
using the pixelized cluster mass, galaxy, and light distributions.
From the 2D shear fitting to a single eNFW model, the projected mass
ellipticity is constrained 
in the range $|e_\Sigma| =0.68^{+0.18}_{-0.28}$ 
($|e_\Sigma|\simgt 0.4$ or $a/b\simgt 1.7$ at $1\sigma$) 
at $\theta_{\rm max}=8\arcmin$.
This apparent increase in ellipticity with radius could be partly
explained by the additional contribution from the surrounding LSS that
is extended along the cluster major axis.
Note, the observed tendency for the shear-based method to yield higher
ellipticity estimates, compared to the mass-map based method, could be
due to the nonlocal nature of the shear field, in conjunction with our 
single-component assumption in the 2D shear fitting analysis.
Overall, this level of ellipticity is consistent within large errors
with the mean cluster ellipticity 
$\langle |e_\Sigma|\rangle=0.46\pm 0.04$  
obtained by \citet{Oguri+2010_LoCuSS} from a 2D weak-lensing
analysis of 25 X-ray luminous clusters.

In what follows, we fix the position angle of the NW-SE cluster-LSS
major axis to a reference value of $\phi_e=20\deg$, which is close to
the values derived from the {\it Chandra} X-ray data, $\Sigma_l$ and
$\Sigma$ maps.
We note that, in principle, 
the X-ray structure in a triaxial system is expected to be
tilted with respect to the total matter in projection, even in the
absence of intrinsic misalignments
\citep[see][]{Romanowsky+Kochanek1998}. 
In the present work, we define the NW and SE excess regions respectively
as NW and SE outer cone regions with $\theta>4\arcmin$ centered on the cluster center,
with opening angle $90\deg$ and position angle $\phi_e=20\deg$, as
defined by the pair of gray solid lines in each panel of Figure \ref{fig:MNL}.

\subsection{BCG-Cluster Alignment}
\label{subsec:BCG}

\begin{figure}[htb!]
 \begin{center}
   \includegraphics[width=0.45\textwidth,clip]{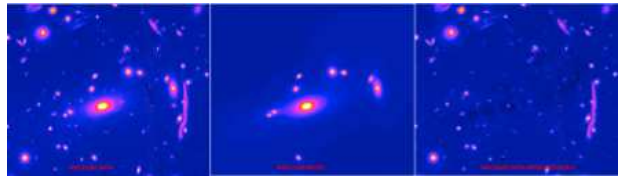}
 \end{center}
\caption{
Detailed model fits to the BCG and its nine nearby galaxies in the ACS
 F814 image ($\approx 50\arcsec\times 45\arcsec$).  The panels show the
 ACS image (left), best-fit model (middle), and image residuals (right)
 after subtraction of the model.  No systematic deviations are seen in
 the residuals between the data and the model,suggesting the BCG has not
 undergone any major merger recently.  North to the top, east to the
 left. 
\label{fig:BCG} 
}
\end{figure}  

We have also obtained CLASH constraints on the mean BCG ellipticity and
position angle derived from the ACS F814W image.  
For this we performed a detailed structural analysis on the BCG
using the {\tt snuc} task in the {\sc XVista} software package.\footnote{\href{http://astronomy.nmsu.edu/holtz/xvista/index.html}{http://astronomy.nmsu.edu/holtz/xvista/index.html}}
In Figure \ref{fig:BCG} we show the ACS F814W image,
best-fit model, and image residuals after subtraction of the model. 
No systematic deviations are seen in the residuals between the data and
the model, suggesting the BCG has not undergone any major merger
recently. 
The radial profiles of ellipticity and position angle 
were measured in several independent radial bins 
($0.2\arcsec\simlt \theta\simlt 10\arcsec$), and their respective
(sensitivity weighted) mean values were obtained as 
$\langle e_\Sigma\rangle=0.53\pm 0.03$ 
and
$\phi_e=(15.0\pm 2.3)\deg$ (Table \ref{tab:elong}).
Consistent results were found in several other {\it HST} bands
(ACS F475W to F814W and WFC3 F105W to F160W).  The mean BCG ellipticity
is found to lie in the range 0.46--0.53 with a small scatter of 0.02
across the ACS and WFC3 bands.  The BCG position angle is constrained to
be $\phi_e=(15.2\pm 0.4)\deg$, which is in excellent agreement
especially with that derived independently from the large scale
distribution $\Sigma_n$ of galaxies. 


\subsection{Effects of Surrounding Large Scale Structure}
\label{subsec:lss}

In this subsection, we look into the azimuthal dependence of the radial
projected mass distribution, $\Sigma(R,\phi)$,
to assess and correct for the effect of surrounding
LSS on the cluster mass profile measurement.
Because of the non-local nature and inherent insensitivity to sheet-like
overdensities of the shear field, it is essential to use the
combination of lens magnification and distortion to reconstruct
the projected cluster mass distribution embedded in large scale
structure.   
For this purpose, we extend the 1D Bayesian method of
\citet{Umetsu+2011a} 
into a 2D mass distribution
by combining the 2D shear pattern $g(\btheta)$ with the azimuthally-averaged
magnification measurements $n_\mu(\theta)$. 
In the 2D analysis, our model $\bs$ is a vector of parameters containing
 a set of discrete      
mass elements on a grid of $N_{\rm cell}$ independent cells,
$\bs=\{\kappa_m\}_{m=1}^{N_{\rm cell}}$.
A brief
summary of this 2D method is given in Appendix \ref{subsec:2D}.  The
details of the method will be presented in our forthcoming paper
 (K. Umetsu et al., in preparation).

By combining Subaru distortion and magnification data, 
we construct here a mass map 
over a $30\times 30$ grid with $0.8\arcmin$ spacing, covering
a $24\arcmin \times 24\arcmin$ field around the cluster 
($N_{\rm cell}=900$). 
We have $2\times 896$ distortion constraints 
$\{g_1(\btheta_m)\}_{m=1}^{N_{\rm cell}}$ and 
$\{g_2(\btheta_m)\}_{m=1}^{N_{\rm cell}}$
over the mass grid, excluding the
four innermost cells lying in the cluster central region (see Appendix
\ref{subsec:2D}), and $N=8$ radial magnification constraints
$\{n_\mu(\theta_i)\}_{i=1}^N$. 
Hence, we have a total of 1800 constraints
(900 degrees of freedom).  Additionally, we marginalize over the
calibration uncertainties in the observational parameters ($n_0,s,\omega$)
(Section \ref{subsec:depth}).  The best solution 
$\bs$ has been obtained with $\chi^2_{\rm
min}/{\rm dof}=1058/900$.  We then follow \citet{UB2008} to calculate the
radial mass distribution $\langle\Sigma(R)\rangle$
and its covariance matrix from a weighted
projection of the $\kappa$ map, 
where we conservatively limit our 2D analysis to radii smaller than
$\theta=12\arcmin$ ($R\approx 2.9\,$Mpc\,$h^{-1}$).
We check that the azimuthally-averaged radial mass profile constructed
from the $\kappa$ map reproduces our corresponding 1D results (Section
\ref{subsec:wlmass}).

We show in Figure \ref{fig:kappa} the radial mass distribution
obtained excluding the NW and SE excess regions (defined in Section
\ref{subsec:lss}; see also Figure \ref{fig:MNL}). 
This weak-lensing mass profile, corrected for the effect of surrounding LSS, 
exhibits a steeper radial trend than that averaged over all azimuthal angles.
We note, a slight remaining excess is seen at $\theta\simgt 5\arcmin$ 
($R\simgt 1.2\,$Mpc\,$h^{-1}$), which may be associated with the likely
north-south filamentary feature (Section \ref{subsec:2D}). 
By fitting the ``LSS-corrected'' mass profile with an NFW
profile, 
we find a higher concentration
$c_{\rm vir}=7.5^{+2.5}_{-1.8}$ with
$M_{\rm vir}=1.15^{+0.25}_{-0.20}\times
10^{15}M_\odot\,h^{-1}$ 
($\chi^2_{\rm min}/{\rm dof}=10.6/6$).
This model predicts an Einstein radius of 
$\theta_{\rm Ein}\approx 32\arcsec$ for $z_s=2.5$, comparable to the
observed value, $\theta_{\rm Ein}=28\arcsec\pm 3\arcsec$. 


\subsection{Full Lensing Constraints}
\label{subsec:full}

\begin{figure}[!htb]
 \begin{center}
   \includegraphics[width=0.45\textwidth,angle=0,clip]{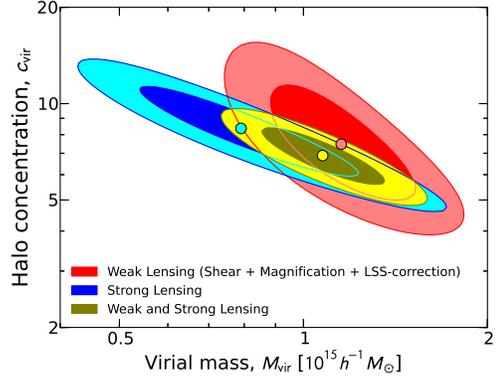}
 \end{center}
\caption{
Constraints on the NFW model parameters ($M_{\rm vir},c_{\rm vir}$), the
halo virial mass and concentration, derived from weak-lensing (red),
strong-lensing (blue), and joint weak and strong lensing (yellow)
analyses.  
 The weak-lensing results are obtained with the LSS correction
 (Section \ref{subsec:lss}).
The contours show the 68.3\% and  95.4\% confidence levels, 
estimated from $\Delta\chi^2\equiv \chi^2-\chi^2_{\rm min}=2.3$ and
 $6.17$, respectively. 
The circles indicate the respective best-fit model parameters.
For weak lensing, the source redshift uncertainty, $z_s=1.15\pm 0.1$, is 
marginalized over.
\label{fig:CM}
}
\end{figure} 


\begin{figure}[htb!]
 \begin{center}
   \includegraphics[width=0.45\textwidth,angle=0,clip]{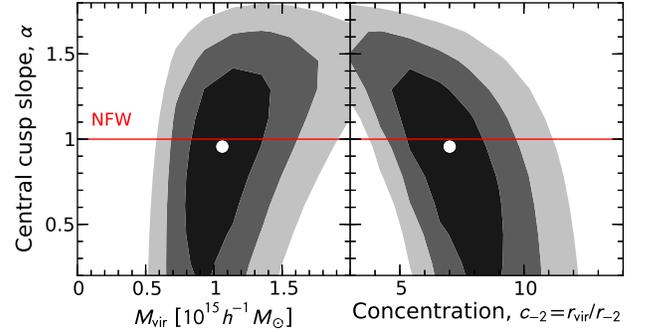}
 \end{center}
\caption{
Constraints on the gNFW model parameters, namely, the central cusp slope 
 $\alpha$, the halo virial mass $M_{\rm vir}$, and the halo
 concentration  $c_{-2}=r_{\rm vir}/r_{-2}=c_{\rm vir}/(2-\alpha)$,
when all of them are allowed to vary, derived from combined weak and
 strong lensing.
 The weak-lensing results are obtained with the LSS correction
 (Section \ref{subsec:lss}).
The left and right panels show the two-dimensional marginalized
 constraints on 
$(M_{\rm vir},\alpha)$ and ($c_{-2},\alpha$), respectively. 
In each panel of the figure, the contours show the 68.3\%, 95.4\%, and
 99.7\% confidence levels, and the circle indicates the best-fit 
model parameters.
\label{fig:gNFW}
}
\end{figure} 

As shown in Figures \ref{fig:kappa} and \ref{fig:m2d},
our weak and strong lensing data agree well in their region of overlap.
Here we further improve the statistical constraints on the halo 
parameters $\bp=(M_{\rm vir},c_{\rm vir},\alpha)$
by combining 
the joint weak-lensing distortion and magnification constraints 
$\chi^2_{\rm WL}(\bp,z_{s,{\rm eff}})$ (Section \ref{subsec:wl+sl}) with
the inner mass profile $\kappa_i$
based on the detailed strong-lensing analysis of
\citet{Zitrin+2012_M1206}. 

We write the combined $\chi^2$ function of
our full lensing constraints as
\begin{equation}
\label{eq:chi2wl+sl}
\chi^2=\chi^2_{\rm WL} + \chi^2_{\rm SL}
\end{equation}
with $\chi^2_{\rm SL}$ for strong lensing being defined as
\begin{eqnarray}
\label{eq:chi2sl}
\chi^2_{\rm SL} = \displaystyle\sum_{i,j}
 \left[\kappa_{i}-\hat{\kappa}_{i}(\bp)\right] 
\left({\cal C}_{\rm SL}\right)^{-1}_{ij}
\left[\kappa_{j}-\hat{\kappa}_{j}(\bp)\right],
\end{eqnarray}
where $\kappa_i$ is defined in 25 discrete bins over the radial range 
$[4\arcsec,53\arcsec]$ (see Section \ref{sec:sl})
and scaled to a fiducial depth $z_s=2.54$ of 
the strong-lensing observations,
matched to the spectroscopically-confirmed
five image system \citep[System 4 of][]{Zitrin+2012_M1206},
$\hat{\kappa}_i$ is the theoretical prediction for $\kappa_i$;
${\cal C}_{\rm SL}={\cal C}+{\cal C}_{\rm lss}$ is the bin-to-bin
covariance matrix for the discrete $\kappa$ profile,
with ${\cal C}$ being 
the self-calibrated covariance matrix
derived in Section \ref{subsec:calib} and
${\cal C}_{\rm lss}$ being the cosmic noise contribution.
We use a consistent single source
plane at $z_s=2.54$ to evaluate ${\cal C}^{\rm lss}$.

The resulting NFW and gNFW fits are summarized in Tables
\ref{tab:nfw} and \ref{tab:gnfw}, respectively.
For both models, we show the respective fits derived with and without
the LSS correction for the outer weak-lensing profile 
($R\simgt 1\,$Mpc\,$h^{-1}$).
We find that, when the detailed strong-lensing information is combined
with weak lensing, the LSS correction does not significantly affect the
fitting results with the adopted NFW/gNFW form.    Moreover, all these models
properly reproduce the observed location of the Einstein radius, $\theta_{\rm
Ein}\approx 28\arcsec$.

Here we summarize our primary results obtained with the LSS
correction.    
The confidence contours
on the NFW parameters ($M_{\rm vir},c_{\rm vir}$) 
are shown in Figure \ref{fig:CM}.  The constraints are strongly
degenerate when only the inner or outer mass profile is included in this
fit.    Combining complementary weak and strong lensing information
significantly narrows down the statistical uncertainties on the NFW model
parameters, placing tighter constraints on the entire mass profile
(Model \#7 of Table \ref{tab:nfw}): 
$M_{\rm vir}=1.07^{+0.20}_{-0.16}\times 10^{15}M_\odot\,h^{-1}$ 
and 
$c_{\rm vir}=6.9^{+1.0}_{-0.9}$
with $\chi^2_{\rm min}/{\rm dof}=18.0/31$, corresponding to a $Q$
value goodness-of-fit of $Q=0.970$.   
Next, when $\alpha$ is allowed to vary (Table \ref{tab:gnfw}), 
we find $M_{\rm vir}=1.06^{+0.23}_{-0.18}\times
10^{15}M_\odot\,h^{-1}$,
$c_{-2}=7.0^{+1.5}_{-1.4}$,
and
$\alpha=0.97^{+0.28}_{-0.23}$
with $\chi^2_{\rm min}/{\rm dof}=18.0/30$ and $Q=0.960$, being
consistent with the simple NFW form with $\alpha=1$.
Thus, the addition of the $\alpha$ parameter has little effect on the
fit, as shown by the quoted $\chi^2$ and $Q$ values.
The two-dimensional marginalized constraints on $(M_{\rm vir},\alpha)$
and $(c_{-2},\alpha)$ are shown in Figure \ref{fig:gNFW}.

\subsection{Impact of the Choice of Strong Lensing Models}
\label{subsec:slsys}

In this subsection we address the impact of the choice of
strong-lensing models on the determination of the halo mass and
concentration parameters in a joint weak and strong lensing analysis. 
As an alternative choice to the Zitrin et al. (2012) model,
we consider here
{\sc Pixelens} (non-parametric) and {\sc Lenstool} (parametric) 
models, in combinations with our LSS-corrected weak-lensing mass model
(Section \ref{subsec:lss}).  For each case, we define the $\chi^2$
function for strong lensing as in Equation (\ref{eq:chi2sl}), and
minimize the total $\chi^2$ function (Equation \ref{eq:chi2wl+sl}) to
estimate the NFW parameters ($M_{\rm vir},c_{\rm vir}$).

The resulting model constraints are tabulated in Table \ref{tab:sl}.
We find that both parameters
based on different strong-lensing profiles 
are consistent with each other within the statistical errors.
This also indicates consistency between these strong-lensing models and
our weak-lensing measurements, as shown in Figure \ref{fig:m2d}.
We find a tendency for {\sc Pixelens} to yield somewhat higher mass
estimates compared to other strong-lens modeling methods, as discussed
by  \citet[][their Appendix]{Grillo+2010_Pixelens}.

When the NFW (gNFW) form is assumed, the \citet{Zitrin+2012_M1206} model 
predicts a somewhat higher concentration and a lower mass than other
models as implied by its correspondingly higher central density at
$\simlt 0.2\arcmin$ (see Figure \ref{fig:m2d}).
When the inner fitting radius is increased from $4\arcsec$ to
$12\arcsec$ ($\sim 50\,$kpc\,$h^{-1}$),  
we find a fractional increase of $\sim 9\%$ in $M_{\rm vir}$ 
and a fractional decrease of $\sim 17\%$ in $c_{\rm vir}$ 
($6.9^{+1.0}_{-0.9} \to 5.7^{+1.4}_{-1.1}$).  
Including these variations as systematic uncertainties
in our mass-concentration determination, 
 the spherical NFW model for MACS1206 is constrained as
$M_{\rm vir}=(1.07^{+0.20}_{-0.16}\pm 0.10)\times 10^{15}M_\odot\,h^{-1}$
and
 $c_{\rm vir}=6.9^{+1.0}_{-0.9}\pm 1.2$, (statistical followed by
 systematic uncertainty).  
Similarly, when the central 50\,kpc\,$h^{-1}$ region is excluded from 
the fit, we have
$M_{\rm vir}=(1.17^{+0.25}_{-0.20}\pm 0.10)\times 10^{15}M_\odot\,h^{-1}$ and
$c_{\rm vir}=5.7^{+1.4}_{-1.1}\pm 1.2$.

\subsection{Alternative Mass Profile Fits}
\label{subsec:spls}

Motivated by the apparently shallow projected density profile in the
outer regions 
\citep[cf. XMMU J2235.3$-$2557 at $z=1.4$,][]{Jee+2009}, 
we consider here a softened power-law sphere (SPLS) model
\citep{Grogin+Narayan1996} as an alternative to the NFW profile, 
and perform profile fitting analyses on our full-range mass profile data  
(derived from Methods \#6 and \#7 in Table \ref{tab:nfw}; see Sections
\ref{subsec:lss} and \ref{subsec:full}).

The SPLS model has the same number of free parameters as gNFW, namely
three.  The SPLS density profile is given by  
$\rho(r)=\rho_0(1+r^2/r_c^2)^{(\eta-3)/2}$
where $\rho_0=\rho(0)$ is the central density, $r_c$ is the core radius,
and the power-law index $\eta$ is restricted to lie in the range
$0\le\eta\le 2$  \citep{Grogin+Narayan1996}.  At $r\gg r_c$,
$M(<r)\propto r^\eta$.   This reduces to a non-singular isothermal
sphere (NIS) model when $\eta=1$.
The fitting results with and without the outer LSS correction 
(Methods \#7 and \#6, respectively)
are summarized in Table \ref{tab:spls}. 

First, when $\eta$ is fixed to unity (NIS), the NIS model provides
acceptable fits, but with larger residuals ($\chi^2$) compared to the
corresponding NFW fits with the same degrees of freedom (31):
$\Delta\chi^2=\chi^2_{\rm min, NIS}-\chi^2_{\rm min, NFW}=2.3$ 
(Method \#6) and $5.9$ (Method \#7)
between the best-fit NIS and NFW models.  
Note, because of the asymptotic $M(<r)\propto r$
behavior, the assumed NIS form leads to substantially higher masses
at large radius ($r\gg r_c$) than what the NFW model predicts 
($\sim 35\%$ higher than the NFW values at $r=1.6\,$Mpc\,$h^{-1}$).

Next, when the outer slope is allowed to vary, the fit is noticeably
improved for the results with the outer LSS correction (Method \#7),
corresponding to a difference of 
$\Delta\chi^2=\chi^2_{\rm min,NIS}-\chi^2_{\rm min,SPLS}=4.4$ 
between NIS and SPLS for 1 additional degree of freedom.  For this, the 
best-fitting slope parameter is obtained as 
$\eta=0.77^{+0.13}_{-0.17}$ ($\chi^2_{\rm min}=19.5$ for 30\,dof),
corresponding to  
$2.1 \le \gamma_{\rm 3D}(r\gg r_c)\le 2.4$.
This SPLS model yields a virial mass of 
$M_{\rm vir}=(1.26\pm 0.37)\times 10^{15}M_\odot\,h^{-1}$
 ($r_{\rm vir}\approx 1.73\,$Mpc\,$h^{-1}$).

\section{Discussion}
\label{sec:discussion}

\subsection{Lensing Systematics}
\label{subsec:systematics}

Gravitational lensing probes the total mass projected on to the sky
along the line of sight, so that the lensing-based cluster mass
measurements are sensitive to projection effects 
arising from (1) additional mass overdensities (underdensities) along
the line of sight  \citep{Meneghetti+2010a,Rasia+2012}
and (2) halo triaxiality
\citep{2007ApJ...654..714H,Oguri+Blandford2009,Meneghetti+2010a,Rasia+2012}.

\subsubsection{Projection of Additional Mass Structures}

The first type of projection effects includes the cosmic noise from
distant uncorrelated LSS projected along the same line of sight
\citep{2003MNRAS.339.1155H} 
and massive structures within/around the cluster
(i.e., cluster substructures and surrounding large scale filamentary
structure).

The former can not only increase statistical uncertainties but also
produce covariance between radial bins.
Accordingly this could bias the estimates 
of cluster parameters.  Our methods take into account the estimated
contribution of cosmic covariance ${\cal C}_{\rm lss}$
in both weak and strong lensing profiles,
and allow us to
properly weight the weak and strong lensing  when performing a
combined halo fit.  In our analysis, we find that the contribution of
${\cal C}_{\rm lss}$ to the measurement errors is subdominant in both
regimes; when the weak and strong lensing constraints are combined, the
amount of degradation due to ${\cal C}_{\rm lss}$
is about $12\%$ in the total S/N ratio.
Thus the best-fit parameters are not largely affected by including
${\cal C}_{\rm lss}$, being consistent with each other 
within statistical uncertainties.

The latter represents projection effects arising from the rich,
substructured cluster
environment.  
Recently \citet{Meneghetti+2010a} and \citet{Rasia+2012} used mock
observations of simulated clusters in the 
$\Lambda$CDM cosmology to study the systematic effects in lensing and
X-ray based mass measurements, finding
that the standard tangential-shear fitting
method, assuming a single spherical NFW profile,
can underestimate the true cluster mass $M_\Delta$ 
in the presence of massive substructures, especially for low-mass
systems. 
This is understood by noting that
the azimuthally-averaged tangential shear probes the 
differential surface mass density, $\gamma_+(\theta)\propto
\overline{\Sigma}(<R)-\Sigma(R)$ (see Equation
(\ref{eq:loop})).  
\citet{Rasia+2012} found from their three most massive systems with
$M_{200}>7.5\times 10^{14}M_\odot\,h^{-1}$ that the level of bias is
$\sim -5\%$ with no noticeable radial dependence at $r=(r_{2500},
r_{1000}, r_{500})$.   
Our cluster mass estimate from the tangential-shear fitting
is $M_{\rm vir}=0.99^{+0.32}_{-0.26}\times 10^{15}M_\odot\,h^{-1}$
(Model \#1 of Table \ref{tab:nfw}), which is about $7\%$ lower than that
from our NFW model based on the full-lensing constraints
(Model \#7 of Table \ref{tab:nfw}) from our comprehensive weak-lensing
distortion, magnification, and strong-lensing analysis.
This level of underestimation seems to be consistent with the simulation
results of \citet{Rasia+2012}.

\subsubsection{Halo Triaxiality}
\label{subsubsec:triaxiality}
 
A degree of triaxiality is inevitable for collisionless gravitationally 
collapsed structures 
\citep{Jing+Suto2000,Lemze+2011}, and can affect our cluster mass
estimation  
\citep{2005ApJ...632..841O,Morandi+2011_A1689,Sereno+Umetsu2011,Sereno+Zitrin2012}.
In the context of $\Lambda$CDM, prolate halo shapes are expected to
develop by mass accretion along filaments at early stages of halo assembly; hence,
dynamically-young, cluster-sized halos tend to have a
prolate morphology  \citep{Shaw+2006,Lau+2011}. 
Accordingly, a large fraction of cluster-sized prolate halos,  
{\it in the absence of selection bias},  
is expected to be elongated in the plane of the sky
\citep{Rasia+2012}.  On average, this will lead to an
underestimation of the cluster mass in a statistical sense
when a spherical deprojection (or forward modeling
assuming a spherical halo) is applied
\citep{Rasia+2012}.  On the other hand, in the $\Lambda$CDM context,
those clusters selected by the presence of giant arcs are likely to have 
their major axes closely aligned with the line of sight
\citep{2007ApJ...654..714H,Meneghetti+2010a},  because this orientation  
boosts the projected surface mass density and hence the lensing
signal.

MACS1206 is an X-ray selected CLASH cluster \citep{Postman+2012_CLASH}, 
discovered in the MACS survey
\citep{Ebeling+2001_MACS,Ebeling+2009_M1206}. 
For MACS1206, we find a large projected mass ellipticity of
$|e_{\Sigma}|=1-b/a\simgt 0.4$ (or $a/b\simgt 1.7$ at $1\sigma$) 
at large cluster radius 
($R\simgt r_{\rm vir} \approx 1.6\,$Mpc\,$h^{-1}$)
based on the Subaru weak-lensing analysis, where its position angle 
is well aligned with the BCG, optical, X-ray, and LSS shapes in
projection space (Section \ref{subsec:lss} and Table \ref{tab:elong}).
The highly elliptical mass distribution in projection would suggest that
its major axis is not far from the sky plane, and that its true mass and 
concentration could be even higher than the projected measurements 
if the cluster size along the sight line  is shorter than its
effective size-scale ($\sqrt{ab}$) in the sky plane.    

\subsection{{\it Chandra} and XMM-{\it Newton} X-ray Observations}
\label{subsec:xray} 

\begin{figure}[htb]
 \begin{center}
   \includegraphics[width=0.45\textwidth,angle=0,clip]{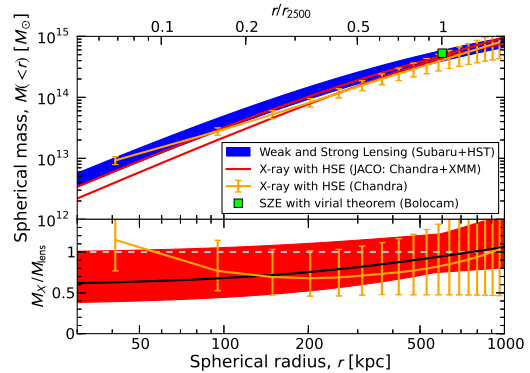}
 \end{center}
\caption{
Integrated total mass profiles $M(<r)$ as a function of spherical radius 
 $r$ derived from various observational probes (top).
The blue shaded area shows the best-fit NFW model with $1\sigma$
 uncertainty from the combined weak and strong lensing measurements
 (Figures \ref{fig:kappa} and \ref{fig:m2d}). 
The red solid lines represent the X-ray based NFW model ($1\sigma$
 confidence interval of the fit) derived using the JACO software from a 
 simultaneous fit to {\it Chandra} and XMM-{\it Newton} observations.
The enclosed masses based on {\it Chandra} data alone (solid line with
 error  bars, orange) are derived as described in the text, assuming the
 parameterized pressure profile shape from  \citet{Arnaud+2010}.  
The green square marks the Bolocam SZE mass estimate at the
 lensing-derived overdensity radius $r_{2500}$.  
The bottom panel shows the X-ray to lensing mass ratio 
 $M_{X}(<r)/M_{\rm lens}(<r)$ with $1\sigma$ uncertainty 
as a function of radius $r$.  The results are shown for both the 
{\it Chandra}-only and joint {\it Chandra}+XMM fits.  
\label{fig:xray}
}
\end{figure} 
\begin{figure}[htb!]
 \begin{center}
   \includegraphics[width=0.45\textwidth,angle=0,clip]{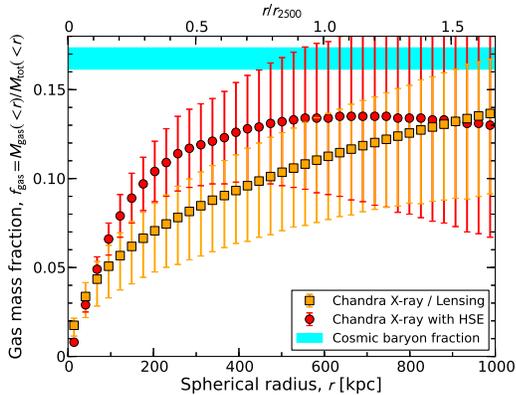}
 \end{center}
\caption{
Gas mass fraction profiles $f_{\rm gas}(<r)=M_{\rm gas}(<r)/M(<r)$ as a
 function of spherical radius $r$ derived from joint 
 Subaru weak-lensing, {\it Hubble} strong-lensing, and {\it Chandra}
 X-ray observations.   In each case the gas mass profile $M_{\rm
 gas}(<r)$ is based on the {\it Chandra} X-ray data provided in the
 ACCEPT  \citep{ACCEPT2009}.
The squares with error bars represent the results ($M_{\rm gas}/M_{\rm
 lens}$) from the combined  X-ray and lensing data without employing the
 hydrostatic equilibrium  assumption.
The circles with error bars show the {\it Chandra}-only results 
($M_{\rm  gas}/M_{X}$) based on the hydrostatic equilibrium assumption. 
The horizontal bar shows the constraints (68\% CL)
 on the cosmic baryon fraction from the WMAP seven-year data,
 $f_b=\Omega_b/\Omega_m=0.1675\pm 0.006$.
\label{fig:fgas}
}
\end{figure} 

Complementary multiwavelength observations serve as a useful guide to
the likely degree of lensing bias. 
Here we retrieved and analyzed archival {\it Chandra} and XMM-{\it
 Newton} data of MACS1206 to obtain an independent cluster
 mass estimate, as well as to constrain the physical properties of the
 X-ray gas.

We perform a simultaneous fit to {\it Chandra} and XMM data-sets
under the assumption that the intracluster gas is in hydrostatic 
equilibrium (HSE) with the overall cluster potential of the NFW form.
The tool used for this analysis is JACO \citep[Joint Analysis of Cluster 
Observations,][]{JACO}; we refer the reader to this paper for the
details of the X-ray analysis procedure, which we briefly summarize
below.

We use {\it Chandra} ObsID 3277 and XMM-{\it Newton} observation
0502430401. 
We screen periods of flaring background according to standard
procedure, resulting in usable exposure times of 23\,ks and 26\,ks,
respectively.  Appropriate coadded blank-sky fields allow us to
subtract particle background spectra for both telescopes, and the
residual (positive or negative) astrophysical background is included
and marginalized over in the global cluster gas model.  Spectra are
extracted over seven annular bins for both {\it Chandra} and XMM-{\it
Newton}. The extracted spectra extended out to a distance of
$3.7\arcmin$ (1.26\,Mpc), and contain an average of 1500 counts each.  

The model for the gas density distribution is a single $\beta$-model
multiplied by a power law of slope $\gamma$:
\begin{equation}
\rho_g(r) = \rho_0 \left(\frac{r_c}{r}\right)^\gamma \left(1 +
  \frac{r^2}{r_c^2} \right)^{-3 \beta/2}.
\end{equation}
The power-law component is required to capture the steep increase of the
density towards the center of the cluster; all parameters of the gas
distribution are fit to the data. The metallicity is allowed to vary
with radius as well, as are the parameters of the NFW mass profile. 
Model spectra are generated self-consistently in concentric spherical
shells and forward-projected onto the annular sky regions matching the
extracted annuli.  
The resulting spectra are mixed using in-orbit energy- and
position-dependent point spread functions for both {\it Chandra} and 
XMM-{\it Newton}.  
Systematic calibration uncertainties between Chandra and XMM-{\em Newton}
spectra are taken into account by adding a 4\% error
\citep[a typical correction used in][]{Mahdavi+2008_CCCP} in quadrature
to each spectral bin used for the joint fits.  This brings the joint
$\chi^2$ into the acceptable range ($\chi^2=1603$ for 1541 dof).
An MCMC procedure is used to estimate errors on the best-fit
quantities. After marginalizing over all other parameters, we measure a
total mass 
$M_{2500} = (4.45 \pm 0.28) \times 10^{14} M_\odot$, 
a gas mass 
$M_{{\rm gas},2500} = (0.54 \pm 0.02) \times 10^{14} M_\odot$, 
an NFW concentration parameter of $c_{200}=3.5 \pm 0.5$, 
an inner gas density profile slope of $0.7 \pm 0.03$, 
and a central cooling time of $ 2.1 \pm 0.1$\,Gyr.
In what follows, the examination of the X-ray results is conservatively  
limited to $r<1\,$Mpc.

In Figure \ref{fig:xray} we plot the resulting X-ray based total mass
profile, $M(<r)$, shown along with our NFW model from the full lensing
analysis.   
The results of the NFW fit are also reported in Table  \ref{tab:xray}.       
This X-ray model yields a total mass of 
$M_X=(4.6\pm 0.2)\times 10^{14}M_\odot$ at the lensing derived
overdensity radius of $r_{2500}\approx 0.60\,$Mpc. 
This is in excellent agreement with the
lensing mass at the same radius, 
$M_{\rm lens}=(4.9\pm 0.9)\times 10^{14}M_\odot$, which corresponds to
the X-ray to lensing mass ratio,  
$a_{2500}=M_X(<r_{2500})/M_{\rm lens}(<r_{2500})=0.95^{+0.23}_{-0.25}$.
The $a_{2500}$ value obtained here is in good agreement with results from
 mock observations of 20 $\Lambda$CDM clusters
 by \citet{Rasia+2012}: $a_{2500}=0.94\pm 0.02$.
At this overdensity, no significant bias was observed in detailed
 observational studies by  \citet{Zhang+2008_LoCuSS} and
 \citet{Mahdavi+2008_CCCP}, who performed a systematic comparison of
 weak-lensing and X-ray mass measurements for sizable cluster samples. 
In the bottom panel of Figure \ref{fig:xray}, we show the X-ray to
lensing mass ratio $a_\Delta$ as a function of cluster radius, in the
radial range where X-ray observations are sufficiently sensitive.   
Overall, the mass ratio is consistent with unity especially at 
$r\sim r_{2500}$.

\citet{Ebeling+2009_M1206} obtained a hydrostatic mass estimate of
$M_X=(1.7\pm 0.1)\times 10^{15}M_\odot$ at $r=2.3\,$Mpc (their
estimate for $r_{200}$) assuming an isothermal $\beta$ model with
$\beta=0.57\pm 0.02$ and their estimated temperature 
$k_{\rm B}T=11.6\pm 0.7\,$keV in the radial range $[70,1000]$\,kpc 
($M_X\propto \beta^{1/2}T$), 
which is high but consistent within the errors with 
$M_{\rm lens}(<2.3\,{\rm Mpc})=(1.4\pm 0.3)\times 10^{15}M_\odot$ 
obtained with our best NFW model based on the full lensing analysis.

Our full lensing results, when combined with X-ray gas mass measurements
 ($M_{\rm gas}$), yield a direct  estimate for the cumulative gas mass
 fraction, $f_{\rm gas}(<r)\equiv M_{\rm  gas}(<r)/M(<r)$,
free from the HSE assumption.
For this we use reduced {\it Chandra} X-ray data presented in
the Archive of {\it Chandra} Cluster Entropy  Profile Tables  
ACCEPT \citep{ACCEPT2009}.
 In Figure \ref{fig:fgas}, we plot our $f_{\rm gas}$ measurements as a
 function of cluster radius.
We find a gas mass fraction of $f_{\rm gas}(<r)=13.7^{+4.5}_{-3.0}\%$
at a radius of $r=1\,{\rm Mpc}\approx 1.7\,r_{2500} (\approx 0.8\,r_{500})$,
a typical value observed for high mass clusters
 \citep{Umetsu+2009,Zhang+2009_gas}.  
 When compared to the cosmic baryon
 fraction $f_b=\Omega_b/\Omega_m=0.1675\pm 0.006$ constrained from the WMAP
 seven-year data \citep{Jarosik+2011_WMAP7}, this indicates $f_{\rm
 gas}/f_b=0.82^{+0.27}_{-0.18}$ at this radius.
At the innermost measurement radius $r\approx 40\,$kpc where the lensing
and X-ray data overlap, we have $f_{\rm gas}(<r)=3.4^{+1.2}_{-0.8}\%$.
Thus, the hot gas represent only a minor fraction of the total lensing
mass near the cluster  center, as found for other high-mass clusters
\citep{2008MNRAS.386.1092L,Umetsu+2009}. 

Additionally, we derive a mass profile from simulated annealing fits of
the ACCEPT pressure profile \citep{ACCEPT2009}, adopting the 
\citet[][A10]{Arnaud+2010} ``universal profile''
(M. Donahue et al., in preparation). 
This {\it Chandra}-only mass profile is shown to be in good agreement
with the lensing as well as joint {\it Chandra}+XMM results (Figure
\ref{fig:xray}).   The {\it Chandra}-only gas mass fraction profile is
also shown in Figure \ref{fig:fgas}.  

We conclude, on the basis of these results and comparison with detailed
statistical studies, that the level of orientation bias in this cluster
is not significant given the large uncertainties in our lensing/X-ray
observations, as well as the possible contribution from
non-thermal pressure in the cluster core \citep[e.g.,][]{Kawaharada+2010}. 

\subsection{Bolocam SZE Observations}
\label{subsec:sze}

\begin{figure}[htb!]
 \begin{center} 
   \includegraphics[width=0.45\textwidth,angle=0,clip]{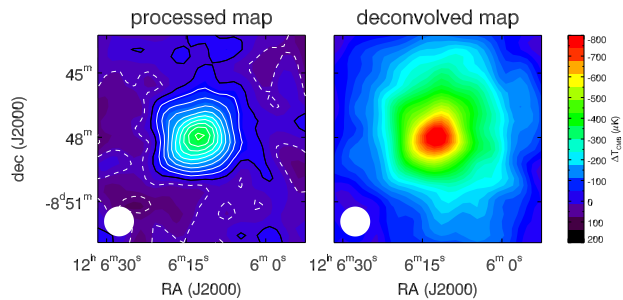}
 \end{center}
\caption{
Bolocam SZE decrement images each $10\arcmin\times 10\arcmin$, 
beam smoothed to an effective resolution of $82\arcsec$.  
Left: the processed image obtained when the data are
filtered to remove atmospheric noise. 
The solid white contours denote ${\rm S/N}=-2, -4, -6, ...$, and
the dashed contours denote ${\rm S/N}=+2, +4, ...$.
Right: image obtained when the effects of the atmospheric filtering have
 been deconvolved to obtain an unbiased image of the cluster.
\label{fig:sze}
}
\end{figure} 

We have also compared our lensing-derived results to mass estimates
determined from the Sunyaev-Zel'dovich effect (SZE) data. Using Bolocam
at the Caltech Submillimeter Observatory, we observed MACS1206 for
approximately 11\,hours in April 2011. These data were collected with
Bolocam configured at an SZE-emission-weighted band center of 140\,GHz. 
Further details of the Bolocam instrument are given in
\citet{Haig+2004}. 
We detect the cluster with an S/N value of 21.1
and a white noise rms of 24.9\,$\mu\textnormal{K}_{\small{\rm CMB}}$-arcmin. 
We reduced these data according to the procedure described in detail in
\citet{Sayers+2011}, but with the updated calibration model reported in
\citet{Sayers+2012} and some other minor modifications. 

The key steps involved in our Bolocam data reduction and cluster
modeling are summarized as follows:
We first remove sky noise from the time streams by subtracting a
template of the correlated signal over the field of view followed by a
high-pass filter \citep{Sayers+2011}.
This process results in a filtered image of the true SZE signal (see the
left panel in Figure \ref{fig:sze}), where the filtering is weakly
dependent on the cluster shape due to the correlated template
removal. 
To characterize this filtering, we process a beam-smoothed, initial
best-fit cluster profile by reverse-mapping it using our pointing
information.  These data are then processed iteratively with a new
best-fit profile using our full reduction pipeline, 
until the procedure converges.
For this analysis we use the A10 ``universal pressure profile''  which
adopts a form of the \citet{Nagai+2007} pressure profile with its slopes
fixed to the values given in A10, allowing the overall normalization and
scale radius to vary.  
   
We have derived cluster mass estimates from our SZE
data alone using the method outlined in \citet{Tony2011}.
The key innovation of this method is that, in addition to assuming
HSE, the virial theorem is used, which is no stronger an assumption than
HSE and can be derived from HSE and thermodynamics. 
This method determines the underlying total mass profile from an
SZE-determined pressure profile, with the added assumption of a constant
gas mass fraction $f_{\rm gas}$.
Cluster mass estimates derived with this method have been shown to be
consistent with X-ray derived results using data from the
Sunyaev-Zel'dovich Array \citep[SZA,][]{Tony2011}, and SZA followup of
blind SZE detections using the Atacama Cosmology Telescope
\citep[ACT,][]{Reese+2012}. 
The SZE-only mass estimates for MACS1206 are given in Table
\ref{tab:sze}, which presents $M_\Delta$ and $r_\Delta$ values at
overdensities $\Delta=2500$ and 500 derived from our Bolocam data alone,
under the assumptions made. 
This table also contains Bolocam-derived mass estimates at the
lensing-derived values of $r_{2500}$ and $r_{500}$.   
We note that the values in Table \ref{tab:sze} include an estimate of
our systematic errors on the SZE-derived masses, which we describe in
detail below. 

The dominant source of uncertainty in our mass estimates, as discussed
in \citet{Tony2011}, stems from the uncertainty in the assumed value of
a radially-constant $f_{\rm gas}(r)$. Masses derived under this
assumption scale as $\propto f_{\rm gas}^{-1/2}$. We adopt the value 
$f_{\rm gas} = 0.13$, and marginalize over uncertainties for a range 
$f_{\rm gas} = [0.1, 0.17]$, consistent with our X-ray determined 
gas fraction measurements at radii near $r_{2500}$ 
(see Figure \ref{fig:fgas}). 
An additional source of systematic uncertainty is the absolute
calibration of the Bolocam maps, which is about 5\% and results in a
$\simlt 5$\% uncertainty in our derived masses. 
Finally, we include a $\pm1.5\%$ systematic at $r_{2500}$, and
$\pm 5\%$ systematic at $r_{500}$, due to our particular choice of
parameterization for the pressure profile. These values are roughly
consistent with those shown in \citet{Tony2011} for different
parameterization of the exponents in the pressure profiles. 

By comparison to the lensing mass estimates, we find an SZE to lensing
mass ratio of $a_{2500}=1.08\pm 0.29\pm 0.22$ (statistical followed by
systematic at $68\%$ confidence)
at the lensing-derived overdensity radius $r_{2500}$ of 0.60\,Mpc 
(Table \ref{tab:sze}).  
Hence, our lensing mass estimate is in agreement with both the X-ray and
SZE mass estimates at $r_{2500}$.  At a lower overdensity of 
$\Delta =500$, we find an SZE to lensing mass ratio of 
$a_{500}=1.55\pm 0.30\pm 0.26$ at
the lensing derived radius $r_{500}$ of 1.3\,Mpc, roughly consistent
with unity within large errors.

\subsection{Dynamical and Physical Conditions of the Cluster}
\label{subsec:multiw}

MACS1206 is an X-ray luminous cluster at a redshift of $z=0.439$, or a
cosmic time of $t\sim 9\,$Gyr.
The cluster appears relatively relaxed in projection in both optical and
X-ray images, with a pronounced X-ray peak at the BCG position 
\citep{Ebeling+2009_M1206,Postman+2012_CLASH}.   
This cluster was classified to be relaxed by \citet{Gilmour+2009}
on the basis of a visual examination of its X-ray morphology.
Our detailed morphology analysis shows no sign of significant recent
merging activity around the BCG, which is also supported by our
strong-lensing analysis, finding no significant offset between the DM
center of mass, BCG, and X-ray peak (Section \ref{subsec:offset}).
A good agreement between the lensing and X-ray mass estimates 
(Section \ref{subsec:xray}) indicates that the hot gas is not far
from a state of hydrostatic equilibrium in the cluster potential well.

However, some evidence of merger activity along the line of sight was
suggested by the high velocity dispersion of 
$\sigma_v\approx 1580$\,km\,s$^{-1}$ based on 38 redshift measurements
\citep{Ebeling+2009_M1206}.
Recently, a much larger spectroscopic sample of cluster members
has been obtained for this cluster using VLT/VIMOS
(P. Rosati et al., in preparation).
Defining membership is crucial for a dynamical analysis
since interlopers by projection effects can largely bias the derived
projected velocity dispersion, especially at large radii where the
number density of cluster members is low \citep{Wojtak+2007}.  
Using a secure sample of $>400$ cluster members identified
in the projected phase space 
\citep[e.g.,][]{2006A&A...452...75B,Lemze+2009}
the velocity dispersion profile is found to decrease outward fairly
rapidly from $\sim 1500\,$km\,s$^{-1}$ in the central region to 
$\sim 800$\,km\,s$^{-1}$ at a projected distance of $R\sim
2\,$Mpc.  Accordingly, the dynamical mass estimate is in agreement with
the lensing estimate (A. Biviano et al., in preparation). 
This may argue against a strong deviation from dynamical relaxation.

The present {\it Chandra} analysis yields a gas temperature of 
$10.8\pm 0.7$\,keV averaged in the radial range [$70,700$]\,kpc.
Assuming that the galaxies and the gas are confined in the same
gravitational potential well, this is consistent with a
line-of-sight velocity dispersion of $\sim 1300$\,km\,s$^{-1}$, which is
again in agreement with the observed value.
This may also suggest that the cluster is not far
from equilibrium.
The cluster appears fairly round in both {\it Chandra} and XMM images
at large distances from the cluster center, 
as demonstrated in 
Figure \ref{fig:xraymap}.
X-ray emission is concentrated around and peaked on the BCG, but
shows some elongation within $\theta\simlt 1\arcmin$ at a position angle
around $120\deg$ east of north, aligned with the orientation of
the projected mass distribution. 
The surface brightness profile is fairly smooth; but there might be
some tiny hints of discontinuities (see the ACCEPT catalog).\footnote{\href{http://www.pa.msu.edu/astro/MC2/accept/clusters/3277.html}{http://www.pa.msu.edu/astro/MC2/accept/clusters/3277.html}} 
However, a much deeper observation is required to confirm them.

Finally, morphological analysis of Bolocam data has been performed in
an identical way to the procedure used in \citet{Sayers+2011}.
We find an ellipticity of $(10\pm 7)\%$ with a position angle of
$55 \pm 27\deg$ north of west from elliptical A10 model
fits to our Bolocam SZE data.  The fits include all data within a
$14\times 14$ arcmin square, corresponding to a fairly large aperture of
$\theta_{\rm max}\approx 9\arcmin > \theta_{\rm vir}\sim 7\arcmin$.
Of the approximately 50 clusters observed with Bolocam and fit with an
elliptical A10 model, MACS1206 is one of the more circularly symmetric
model fits.  

\subsection{Comparison with $\Lambda$CDM Predictions}
\label{subsec:lcdm}

\begin{figure}[htb!]
 \begin{center}
  \includegraphics[width=0.45\textwidth,clip,angle=0]{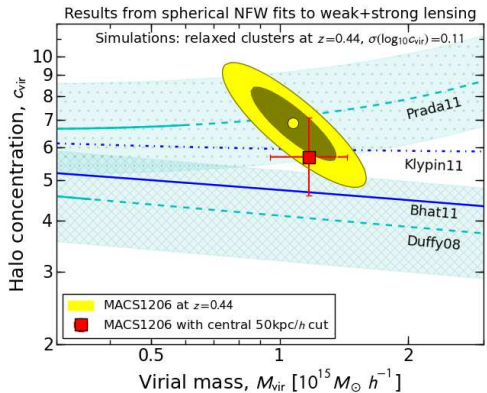}
 \end{center}
\caption{
Constraints on the halo mass and concentration parameters 
$(M_{\rm vir},c_{\rm vir})$ for the X-ray selected CLASH cluster
 MACS1206 ($z=0.439$) derived from spherical NFW fits to
 combined weak and strong lensing observations, compared to $\Lambda$CDM
 predictions for relaxed populations of simulated cluster-sized halos at
 $z=0.44$ \citep[except $z=0.5$ for][]{Klypin+2011}.
Our results are shown with and without the central 50\,kpc\,$h^{-1}$ cut
 (Section \ref{subsec:slsys}) applied to the \citet{Zitrin+2012_M1206}
 based strong-lensing model.
The $N$-body predictions by \citet{Duffy+2008} and
 \citet{Prada+2011} are shown in light blue, including $1\sigma$ lognormal
 scatter (0.11 in $\log_{10}c_{\rm vir}$; $\sim 29\%$) indicated by the 
 respective hatched areas.  Portions of these lines are dashed to
 indicate extrapolations to higher masses.  Average results from two
 additional simulations \citep{Klypin+2011,Bhat+2011} are shown in blue
 for clarity.  \citet{Duffy+2008} and \citet{Bhat+2011} derived results
 for dynamically relaxed cluster subsamples, yielding concentrations
 $\sim 10\%$ higher than for the full samples.  This 10\% factor has
 been applied to the results from the other simulations.
\label{fig:lcdm}
}
\end{figure}

In Figure \ref{fig:lcdm}, we summarize our full lensing constraints on 
the mass and concentration parameters of MACS1206, along with recent
$\Lambda$CDM predictions for {\it relaxed} cluster-sized halos based on
$N$-body  simulations \citep{Duffy+2008,Klypin+2011,Bhat+2011,Prada+2011}.
Our range of allowed concentration values 
($4.6\le c_{\rm vir}\le 7.9$ at $1\sigma$; 
see Section \ref{subsec:slsys}) 
span the high end and average expectations 
($4\simlt \langle c_{\rm vir}\rangle\simlt 7$) from
$\Lambda$CDM simulations   
\citep{Duffy+2008,Zhao+2009,Klypin+2011,Bhat+2011,Prada+2011}.
Average concentrations for relaxed clusters are found to be
$\sim 10\%$ higher and have lower scatter than those for the full
population of halos \citep{Duffy+2008,Bhat+2011}.
A relatively high concentration of MACS1206
may also be indicated by the large Einstein
radius $\theta_{\rm Ein}\approx 28\arcsec (17\arcsec)$ 
at $z_s=2.5$ $(1.0)$ 
\citep{Ebeling+2009_M1206,Zitrin+2012_M1206}.

Care must be taken when comparing these predictions for
spherically-averaged halo structure  with our
lensing results, which are obtained from an NFW fit to the
projected lensing measurements assuming a spherical halo.  
In the previous subsection 
(Section \ref{subsec:systematics}), we have shown that our lensing
results are in good agreement with the X-ray derived mass profiles
(see Figures \ref{fig:xray} and \ref{fig:fgas}) in the region of overlap 
 ($\simlt 1$\,Mpc), as well as with the Bolocam SZE mass estimates
(Section \ref{subsec:sze}), suggesting that the level
 of orientation bias (see Section \ref{subsubsec:triaxiality}) is not
 significant in this cluster.

Additionally, the effects of baryonic physics can impact the 
inner halo profile 
\citep[at $r\simlt 0.0\,5r_{\rm vir}$,][]{Duffy+2010},
and thus modify the gravity-only  $c$--$M$ relation, especially for less
massive halos 
\citep[$M_{\rm vir} \simlt 4\times 10^{14}M_\odot$\,$h^{-1}$;
see][]{Bhat+2011}.
Using cosmological hydrodynamical simulations including the back
reaction of baryons on DM,
\citet{Duffy+2010} found a $< 20\%$ increase in 
the halo concentration for cluster-sized halos
($M_{\rm vir}<6\times 10^{14}M_\odot$\,$h^{-1}$ at $z=0$).  
When excluding the central $50$\,kpc\,$h^{-1}$ 
($\approx 0.03\,r_{\rm vir}$) region from our primary
strong-lensing mass model \citep{Zitrin+2012_M1206}, we find a $\approx
17\%$ decrease in the best-fit concentration parameter derived from our
full-lensing analysis (Section \ref{subsec:slsys}), as demonstrated in
Figure \ref{fig:lcdm}.
We note, the CLASH clusters are massive 
\citep[$5\times 10^{14}<M_{\rm vir}/M_\odot<3\times 10^{15}$,
see][]{Postman+2012_CLASH}, and hence expected to be less affected by
baryonic effects.  

For this cluster, the lensing-derived {\it total} mass distribution is
consistent with the NFW form 
($\alpha=\gamma_{\rm 3D}(r\to 0)=0.96^{+0.31}_{-0.49}$), 
as found for several relaxed clusters: 
A611 \citep{Newman+2009_A611};
A383 \citep{Zitrin+2011_A383}; 
A1703 \citep[$\alpha\approx 0.9$][]{Richard+2009_A1703,Oguri+2009_Subaru}; 
a stacked full-lensing analysis of A1689, A1703, A370, and Cl0024+17
\citep[$\alpha=0.89^{+0.27}_{-0.39}$,][]{Umetsu+2011b}.   
Multiwavelength observations can be used to
measure gas and stellar density profiles for subtraction from
lensing-derived total mass profiles to yield DM-only mass profiles
\citep{2008MNRAS.386.1092L,Newman+2009_A611}, 
allowing for a more direct comparison with CDM predictions from
gravity-only simulations.
We defer this analysis to a forthcoming paper.

\section{Summary and Conclusions}
\label{sec:summary}

In this paper, we have presented a comprehensive lensing analysis,
combining independent measurements of weak-lensing distortion,
magnification, and strong lensing of the massive X-ray selected cluster
MACS1206 at $z=0.439$. 
This is based on wide-field Subaru $BVR_{\rm c}I_{\rm c}z'$ imaging,
combined with detailed strong-lensing information obtained from 
deep CLASH {\em HST} 16-band imaging and VLT/VIMOS spectroscopy
\citep{Zitrin+2012_M1206}.

The deep Subaru multi-band photometry is used to separate background,
foreground and cluster galaxy populations using the selection
techniques established in our earlier
work \citep{Medezinski+2010,Umetsu+2010_CL0024}, allowing us to obtain a
reliable weak lensing signal free from significant contamination of
unlensed cluster and foreground galaxies. 
By combining complementary distortion and magnification measurements,
we constructed a model-free mass distribution out to well beyond
the virial radius ($r_{\rm vir}\approx 1.6\,$Mpc\,$h^{-1}$).
In addition to breaking the mass-sheet degeneracy inherent in shape
distortion measurements, the magnification
measurements also increase the overall significance by $\sim 23\%$
(Section \ref{subsec:wlmass}).

We have also obtained an improved inner mass distribution from a
reanalysis of the \citet{Zitrin+2012_M1206} data using our new MCMC
implementation of the \citet{Zitrin+2009_CL0024} method.
We introduced a technique to self-calibrate the bin-to-bin covariance
matrix of the inner mass profile (Section \ref{subsec:calib}),
accounting for possible systematic errors inherent in the analysis.  
This is a crucial step for a joint analysis  to combine constraints in
different regimes of signal strength.
The inner radial boundary for the mass profile is chosen to be
sufficiently large to avoid smoothing from cluster miscentering effects
\citep{Johnston+2007_SDSS1}. 
The derived inner mass profile is shown to be consistent with our 
semi-independent results from a wide variety of four strong-lensing analyses
 ({\sc Lenstool}, {\sc Pixelens}, {\sc LensPerfect}, and {\sc SaWLens};
 see Section \ref{subsec:sl+}),
and overlap well with the Subaru-based outer mass profile, ensuring
consistency in both the weak and strong regime.  

The Subaru data reveal the presence of an elongated large scale
structure around the cluster, in both the distribution of galaxies and
from the mass distribution, with the major axis running NW-SE, aligned  
well with the cluster and BCG shapes, showing elongation with a
$\sim 2:1$ axis ratio in the plane of the sky 
(Section \ref{subsec:elong}).
The azimuthally-averaged projected mass profile from our full-lensing
analysis exhibits a shallow profile slope $d\ln\Sigma/d\ln R\sim -1$ at
cluster outskirts ($R\simgt 1\,$Mpc\,$h^{-1}$), 
whereas the mass distribution excluding the NW-SE excess regions
steepens further out, well described by the standard NFW form 
(Section \ref{subsec:lss}).
Assuming a spherical halo, we have obtained a virial mass
$M_{\rm vir}/10^{15}M_\odot\,h^{-1}=1.07^{+0.20}_{-0.16} ({\rm stat.})\pm 0.10 ({\rm syst.})$
and a halo concentration
$c_{\rm vir}=6.9^{+1.0}_{-0.9}({\rm stat.})\pm 1.2({\rm syst.})$,
which is somewhat high but falls in the range 
$4\simlt \langle c\rangle\simlt 7$ of average $c(M,z)$
predictions for relaxed clusters from recent $\Lambda$CDM simulations. 
When the innermost 50\,kpc\,$h^{-1}$ is excluded from the fit, we find a
slightly lower concentration $c_{\rm vir}=5.7^{+1.4}_{-1.1}({\rm
stat.})\pm 1.2({\rm syst.})$, a decrease of approximately 17\%  (Section
\ref{subsec:slsys}).

We have shown that our full-lensing mass profile is in agreement with
{\em Chandra}+XMM X-ray data in the region of overlap (Figure
\ref{fig:xray}). 
The hydrostatic X-ray to lensing mass ratio,
$a_\Delta=M_{X}(<r_\Delta)/M_{\rm lens}(<r_\Delta)$, 
 is consistent with unity
especially at $r\sim r_{2500}$ with $a_{2500}=0.95^{+0.23}_{-0.25}$.
Our full lensing results, when combined with {\em Chandra} gas mass
measurements, yield a gas mass fraction estimate free from the HSE
assumption.  We find a cumulative gas mass fraction of 
$f_{\rm gas}(<r)=13.7^{+4.5}_{-3.0}\%$ at 
$r\approx 1.7\,r_{2500}$, 
a typical value observed for high mass clusters. 
Overall good agreement is also obtained with SZE-only cluster mass
estimates based on Bolocam observations (Section \ref{subsec:sze}).

The CLASH survey is producing substantial improvements in both the
quality and quantity of direct empirical constraints on cluster-sized
DM halos \citep{Postman+2012_CLASH,Zitrin+2011_A383,Coe+2012_A2261,Zheng+2012},
for an X-ray selected sample of relaxed clusters, selected
free of lensing selection bias, as well as for a lensing-selected sample  
of high-magnification clusters.
The CLASH imaging, in combination with Subaru weak-lensing data,
allows us to make precise measurements of the mass distributions of
individual clusters over the full range of cluster radius, and to help
understand the possible evolutionary and tidal effects of connecting
filaments and local clusters on the mass distribution of the central
cluster, for a detailed comparison with the standard $\Lambda$CDM
cosmology and a wider examination of alternative scenarios.
With the full sample of CLASH clusters, we will
be able to establish the representative mass profile of massive clusters
in gravitational equilibrium,
and robustly test models of structure formation.


\acknowledgments
We thank our referee for a careful reading of the manuscript 
and and for providing useful comments. 
We acknowledge useful discussions with Nobuhiro Okabe, Masamune Oguri,
and Mauro Sereno.  We are grateful for comments by Cheng-Jiun Ma. 
We thank Nick Kaiser for making the IMCAT package publicly available.  
We thank G. Mark Voit for having contributed to the ACCEPT-based
X-ray mass measurements in advance of publication. 
We are grateful for the hospitality of the  Spitzer Science Center at Caltech,
where part of this work was done.

The CLASH Multi-Cycle Treasury Program is based on observations made
with the NASA/ESA {\em Hubble Space Telescope}. 
The Space Telescope Science Institute is operated by
the Association of Universities for Research in Astronomy, Inc.~under
NASA contract NAS 5-26555. 
ACS was developed under NASA contract NAS 5-32864.
This research is supported in part by NASA grant HST-GO-12065.01-A,
National Science Council of Taiwan grant NSC100-2112-M-001-008-MY3,
and PRIN INAF 2010.
KU acknowledges support from the Academia Sinica Career Development
Award.
Part of this work is based on data collected at the Very Large
Telescope at the ESO Paranal Observatory, under Programme ID 186.A-0798.
PR, CG, IB, and SS acknowledge partial support by the DFG cluster of 
excellence Origin and Structure of the Universe.
The Bolocam observations were partially supported by the Gordon and Betty
Moore Foundation. JS was supported by NSF/AST0838261 and NASA/NNX11AB07G;
NC was partially supported by a NASA Graduate Student Research Fellowship.
AZ is supported by the ``Internationale Spitzenforschung II/2'' of the
Baden-W\"urttemberg Stiftung.
CG acknowledges support from  the Dark Cosmology Centre which is funded
by the Danish National Research Foundation. 
IS holds a PhD FPI fellowship contract from the Spanish Ministry of
Economy and Competitiveness and is also supported by the mentioned
ministry through research project FIS2010-15492.
Support for TM was provided by NASA through the Einstein Fellowship
Program, grant PF0-110077. 

 
\bibliography{Keiichi,Lensref,CMB}



\begin{deluxetable}{lc}
\tablecolumns{2}
\tablecaption{
 \label{tab:cluster}
Properties of the galaxy cluster MACS1206
} 
\tablewidth{0pt}  
\tablehead{ 
 \multicolumn{1}{c}{Parameter} &
 \multicolumn{1}{c}{Value} 
} 
\startdata
ID .............................................. &  MACS J1206.2-0847 \\
Optical center position (J2000.0) & \\
\ \ \ \ R.A. ...................................... & 12:06:12.15\\
\ \ \ \ Decl. ..................................... & -08:48:03.4\\
X-ray peak position (J2000.0) & \\
\ \ \ \ R.A. ...................................... & 12:06:12.28\\
\ \ \ \ Decl. ..................................... & -08:48:02.4\\
Redshift .................................... & $0.4385$\\
X-ray temperature  (keV) ..........& $10.9\pm 0.6$\\
Einstein radius ($\arcsec$) ....................& $28\pm 3$ ($17\pm 2$) 
at $z_s=2.54$ ($1.03$)
\enddata
\tablecomments{
The cluster MACS\,J1206.2-0847 ($z = 0.4385$) was discovered in the
 Massive Cluster Survey (MACS) as described by Reference [1].
The optical cluster center is defined as the center of the BCG from
 Reference [2].
Units of right ascension are hours, minutes, and
seconds, and units of declination are degrees, arcminutes, and
arcseconds. 
The X-ray properties are taken from Reference
 [3]. See also Reference [1].
The BCG is located within $\approx 2\arcsec$ (a projected separation of
 $\approx 9$\,kpc\,$h^{-1}$) of the X-ray emission peak.
The Einstein radii are constrained by detailed strong
 lens modeling by  Reference  [2].  
}
\tablerefs{ 
 [1] \cite{Ebeling+2009_M1206};
 [2] \cite{Zitrin+2012_M1206};
 [3] \cite{Postman+2012_CLASH};
 }
\end{deluxetable}



\begin{deluxetable}{cccc}
\tablecolumns{4}
\tablecaption{
 \label{tab:subaru}
Subaru Suprime-Cam data.
} 
\tablewidth{0pt} 
\tablehead{ 
 \multicolumn{1}{c}{Filter} &
 \multicolumn{1}{c}{Exposure time\tablenotemark{a}} &
 \multicolumn{1}{c}{Seeing\tablenotemark{b}} &
 \multicolumn{1}{c}{$m_{\rm lim}$\tablenotemark{c}}  
\\
 \colhead{} &
 \multicolumn{1}{c}{(ks)} &
 \multicolumn{1}{c}{(arcsec)} &
 \multicolumn{1}{c}{(AB mag)} 
} 
\startdata  
 $B$  & 2.4 & 1.01 & 26.5\\
 $V$  & 2.2 & 0.95 & 26.5\\
 $R_{\rm c}$  & 2.9 & 0.78 & 26.2 \\ 
 $I_{\rm c}$  & 3.6 & 0.71 & 26.0 \\
 $z'$  & 1.6 & 0.58 & 25.0 \\
\enddata
\tablenotetext{a}{Total exposure time.} 
\tablenotetext{b}{Seeing FWHM in the full stack of images.}
\tablenotetext{c}{{}Limiting magnitude for a $3\sigma$ detection within a
 $2\arcsec$ aperture.}
\end{deluxetable}


\begin{deluxetable}{ccccc}
\tabletypesize{\footnotesize}
\tablecolumns{5} 
\tablecaption{
 \label{tab:color}
Galaxy color selection.
}  
\tablewidth{0pt}  
\tablehead{ 
 \multicolumn{1}{c}{Sample} &
 \multicolumn{1}{c}{Magnitude limits\tablenotemark{a}} &
 \multicolumn{1}{c}{$N$} &
 \multicolumn{1}{c}{$n_g$\tablenotemark{b}} & 
 \multicolumn{1}{c}{$\langle z_s\rangle$\tablenotemark{c}} 
\\
 \colhead{} & 
 \multicolumn{1}{c}{(AB mag)} &
 \colhead{} &
 \multicolumn{1}{c}{(arcmin$^{-2}$)} &
 \colhead{}
} 
\startdata  
 Red      & $21.5<z'<24.6$ &  13252 & 9.9 & 1.16 \\
 Green    & $z'<24.6$      &  1638  & 3.4 & 0.44 \\
 Blue     & $22.0<z'<24.6$ &  4570  & 4.3 & 1.95 \\
\enddata 
\tablenotetext{a}{Magnitude limits for the galaxy sample.}
\tablenotetext{b}{Mean surface number density of source background galaxies.} 
\tablenotetext{c}{Mean photometric redshift of the sample obtained with the BPZ code.}
\end{deluxetable}


\begin{deluxetable}{ccccccccc}
\tabletypesize{\footnotesize}
\tablecolumns{10} 
\tablecaption{
 \label{tab:wlsamples}
Galaxy samples for weak-lensing shape measurements.
}  
\tablewidth{0pt} 
\tablehead{ 
 \multicolumn{1}{c}{Sample} &
 \multicolumn{1}{c}{$N$} &
 \multicolumn{1}{c}{$n_g$\tablenotemark{a}} & 
 \multicolumn{1}{c}{$\sigma_g$\tablenotemark{b}} & 
 \multicolumn{2}{c}{$z_{s,{\rm eff}}$\tablenotemark{c}} &
 \multicolumn{2}{c}{$\langle D_{ls}/D_s\rangle$\tablenotemark{d}} &
\\
 \colhead{} & 
 \colhead{} &
 \multicolumn{1}{c}{(arcmin$^{-2}$)} &
 \colhead{} &
 \multicolumn{1}{c}{\scriptsize M1206} &
 \multicolumn{1}{c}{\scriptsize COSMOS} &
 \multicolumn{1}{c}{\scriptsize M1206} &
 \multicolumn{1}{c}{\scriptsize COSMOS} 
} 
\startdata  
 Red      & 8969 & 9.2&  0.42 & 
1.05 & 1.05 & 
0.51 & 0.51\\
 Blue    &  4154 & 4.3 & 0.48 & 
1.55  & 1.58 & 
0.62 & 0.63\\
 Blue+red & 13123 & 13.4 & 0.44 & 
1.15 & 1.12 & 
0.54 & 0.53\\
\enddata 
\tablecomments{
}
\tablenotetext{a}{Mean surface number density of source background galaxies.} 
\tablenotetext{b}{Mean rms error for the shear estimate per galaxy, 
 $\sigma_g\equiv (\overline{\sigma_g^2})^{1/2}$.}
\tablenotetext{c}{Effective source redshift corresponding to the mean
 depth $\langle\beta\rangle$ of the sample.}
\tablenotetext{d}{Distance ratio 
 averaged over the redshift distribution of the sample,
 $\langle\beta\rangle$.}
\end{deluxetable}

 
\begin{deluxetable}{ccccccl} 
\tabletypesize{\scriptsize}
\tablecolumns{7}
\tablecaption{\label{tab:nfw} Best-fit NFW model parameters for MACS1206.} 
\tablewidth{0pt} 
\tablehead{ 
 \multicolumn{1}{c}{$M_{\rm vir}$\tablenotemark{a}} &
 \multicolumn{1}{c}{$c_{\rm vir}$} &
 \multicolumn{1}{c}{$\chi^2/{\rm dof}$} &
 \multicolumn{1}{c}{$\theta_{\rm Ein}$\tablenotemark{b}} &
 \multicolumn{2}{c}{$N$\tablenotemark{c}} &
 \multicolumn{1}{c}{Method}
\\ 
 \multicolumn{1}{c}{$(10^{15}M_\odot\,h^{-1})$} &
 \multicolumn{1}{c}{} &
 \multicolumn{1}{c}{} &
 \multicolumn{1}{c}{($\arcsec$)}  &
 \multicolumn{1}{c}{WL} &
 \multicolumn{1}{c}{SL} &
 \multicolumn{1}{c}{} 
}
\startdata 
 $0.99^{+0.32}_{-0.26}$ &
 $5.7^{+3.6}_{-2.1}$ &
 $3.3/6$ &
 $21$ &
 8 & 0 &
 1) WL tangential distortion (\S \ref{subsubsec:wl+sl:wl}))
\\
 $1.15^{+0.34}_{-0.28}$  & 
 $4.0^{+2.1}_{-1.4}$ &
 $4.5/7$ &
 $14$ &
9 & 0 &
 2) WL tangential distortion $+$ magnification  (\S \ref{subsubsec:wl+sl:wl}))
\\
 $1.15^{+0.25}_{-0.20}$  & 
 $7.5^{+2.5}_{-1.8}$ &
 $10.6/6$ &
 $32$ &
 8 & 0 &
 3) WL(\#2) $+$ LSS correction\tablenotemark{d} (\S \ref{subsec:lss}))
\\
\hline
 $0.88^{+0.25}_{-0.21}$ &
 $8.0^{+2.3}_{-1.7}$ &
 $3.9/8$ &
 $28$ &
 8 & 2 &
 4) WL(\#1) $+$ Einstein radius\tablenotemark{e} (\S \ref{subsubsec:wl+sl:sl}))
\\
 $0.97^{+0.28}_{-0.23}$ &
 $6.8^{+2.2}_{-1.6}$ &
 $6.9/9$ &
 $26$ &
 9 & 2 &
 5) WL(\#2) $+$ Einstein radius  (\S \ref{subsubsec:wl+sl:sl}))
\\
\hline
 $1.14^{+0.22}_{-0.18}$  & 
 $6.6^{+1.0}_{-0.9}$ &
 $24.2/31$ & 
 $28$ &
 8 & 25 &
 6) WL(\#2) $+$ SL\tablenotemark{f}  (\S \ref{subsec:full}))
\\
 $1.07^{+0.20}_{-0.16}$  & 
 $6.9^{+1.0}_{-0.9}$ &
 $18.0/31$ & 
 $28$ &
 8 & 25 &
 7) WL(\#3) $+$ SL $=$ our primary NFW result  (\S \ref{subsec:full}))
\\
\enddata
\tablecomments{
All our methods take into account the cosmic covariance from
 distant, uncorrelated large scale structure (LSS) projected along the line of
 sight.  For weak lensing, the source redshift uncertainty ($z_{s,{\rm
 eff}}=1.15\pm 0.1$) of our background sample has been marginalized over.
}
\tablenotetext{a}{The virial overdensity is 
$\Delta_{\rm vir}\approx 132$ times the critical
 density of the universe at $z=0.439$ in the adopted
 cosmology  ($\Omega_m=0.3, \Omega_\Lambda=0.7$).}
\tablenotetext{b}{Effective Einstein radius for a source at $z_s=2.5$
 predicted by the model. The observed value is $28\arcsec \pm
 3\arcsec$.}
\tablenotetext{c}{Respective numbers of weak and strong lensing constraints.}
\tablenotetext{d}{Excluding the elongated LSS around the cluster 
extending along the NW-SE direction
 (see Figure \ref{fig:MNL}).}
\tablenotetext{e}{Combining with double Einstein-radius constraints of $\theta_{\rm
 Ein}=17\arcsec\pm 2\arcsec$ at $z_s=1.03$ and $\theta_{\rm
 Ein}=28\arcsec\pm 3\arcsec$ at $z_s=2.54$.  Additionally, an rms
 displacement of 2\arcsec is assumed for each system due to uncorrelated LSS projected
 along the line of sight, and is combined in quadrature with the
 respective measurement error to estimate a total uncertainty.}
\tablenotetext{f}{Combining with the inner strong-lensing based mass profile
 derived from an MCMC implementation of \citet{Zitrin+2012_M1206}
  (Sections \ref{subsec:Z12}--\ref{subsec:calib}).
 The outer fitting radius is limited to less than
 $12\arcmin$ for direct comparison with Method \#7 based on the
 LSS-corrected weak-lensing profile (Section \ref{subsec:lss}).}
\end{deluxetable}

 
\begin{deluxetable}{ccccc} 
\tabletypesize{\scriptsize}
\tablecolumns{4}
\tablecaption{\label{tab:elong} Ellipticity and position angle measurements.} 
\tablewidth{0pt} 
\tablehead{ 
 \multicolumn{1}{c}{Method} &
 \multicolumn{1}{c}{$\theta_{\rm max}$\tablenotemark{a}} &
 \multicolumn{1}{c}{Ellipticity\tablenotemark{b}} &
 \multicolumn{1}{c}{PA\tablenotemark{c}}\\
 \multicolumn{1}{c}{} &
 \multicolumn{1}{c}{($\arcmin$)} &
 \multicolumn{1}{c}{} &
 \multicolumn{1}{c}{($\deg$)} &
}
\startdata  
 BCG            & $10\arcsec$& $0.53^{+0.03}_{-0.03}$ & $15.0^{+2.3}_{-2.3}$ \\ 
{\it Chandra} X-ray & $1.5\arcmin$& $0.30^{+0.03}_{-0.03}$ & $21.9^{+1.7}_{-1.7}$ \\
Galaxy \# density & $8\arcmin$ & $0.53^{+0.04}_{-0.04}$ & $15.7^{+1.3}_{-5.9}$ \\
Galaxy light & $8\arcmin$ & $0.41^{+0.06}_{-0.06}$ & $19.0^{+5.9}_{-5.4}$ \\ 
WL mass map   & $8\arcmin$ & $0.37^{+0.13}_{-0.13}$ & $19.4^{+8.5}_{-17.7}$\\
WL 2D shear fit & $8\arcmin$ & $0.68^{+0.18}_{-0.28}$ & $28.6^{+5.8}_{-7.9}$ \\
\tablenotetext{a}{Circular aperture radius.} 
\tablenotetext{b}{Ellipticity modulus defined such that, for an ellipse
 with major and minor axes $a$ and $b$, it reduces to $1-b/a$.}
\tablenotetext{c}{Position angle of the major axis measured north of west.}
\end{deluxetable} 

 
\begin{deluxetable}{cccccc} 
\tabletypesize{\scriptsize}
\tablecolumns{6}
\tablecaption{\label{tab:gnfw} Best-fit generalized-NFW model parameters for MACS1206.} 
\tablewidth{0pt} 
\tablehead{ 
 \multicolumn{1}{c}{Method\tablenotemark{a}} &
 \multicolumn{1}{c}{$M_{\rm vir}$} &
 \multicolumn{1}{c}{$c_{-2}$\tablenotemark{b}} &
 \multicolumn{1}{c}{$\alpha$\tablenotemark{c}} &
 \multicolumn{1}{c}{$\chi^2/{\rm dof}$} &
 \multicolumn{1}{c}{$\theta_{\rm Ein}$\tablenotemark{d}} 
\\ 
 \multicolumn{1}{c}{} &
 \multicolumn{1}{c}{$(10^{15}M_\odot\,h^{-1})$} &
 \multicolumn{1}{c}{} &
 \multicolumn{1}{c}{} &
 \multicolumn{1}{c}{} &
 \multicolumn{1}{c}{($\arcsec$)}  
}
\startdata 
 \#6 &
 $1.17^{+0.29}_{-0.22}$ & 
 $6.3^{+1.5}_{-1.5}$    &
 $1.09^{+0.28}_{-0.42}$ &
 $24.1/30$ & 
 $28$ 
\\
 \#7 &
 $1.06^{+0.23}_{-0.18}$ & 
 $7.0^{+1.5}_{-1.4}$    &
 $0.96^{+0.31}_{-0.49}$ &
 $18.0/30$ &
 $28$ 
\\
\enddata
\tablecomments{
See for details Section \ref{subsec:full}.
}
\tablenotetext{a}{Fitting method in Table \ref{tab:nfw}.}
\tablenotetext{b}{Effective concentration parameter for gNFW,
 $c_{-2}\equiv r_{\rm vir}/r_{-2}=c_{\rm vir}/(2-\alpha)$.}
\tablenotetext{c}{Central cusp slope of gNFW.}\
\tablenotetext{d}{Effective Einstein radius for a source at $z_s=2.5$
 predicted by the model. The observed value is $28\arcsec \pm
 3\arcsec$.}
\end{deluxetable} 

 
\begin{deluxetable}{ccccl} 
\tabletypesize{\scriptsize}
\tablecolumns{5}
\tablecaption{\label{tab:sl} Impact of the choice of strong lensing
 models in the full lensing analysis.}
\tablewidth{0pt} 
\tablehead{ 
 \multicolumn{1}{c}{$M_{\rm vir}$} &
 \multicolumn{1}{c}{$c_{\rm vir}$} &
 \multicolumn{1}{c}{$\chi^2/{\rm dof}$} &
 \multicolumn{1}{c}{$\theta_{\rm Ein}$\tablenotemark{a}} &
 \multicolumn{1}{c}{Method\tablenotemark{b}}
\\ 
 \multicolumn{1}{c}{$(10^{15}M_\odot\,h^{-1})$} &
 \multicolumn{1}{c}{} &
 \multicolumn{1}{c}{} &
 \multicolumn{1}{c}{($\arcsec$)} &
 \multicolumn{1}{c}{} 
}
\startdata 
 $1.07^{+0.20}_{-0.16}$  & 
 $6.9^{+1.0}_{-0.9}$ &
 $18.0/31$ &
 $28$ &
 WL(\#3) + Zitrin+12\tablenotemark{c}
\\
 $1.17^{+0.25}_{-0.20}$  & 
 $5.7^{+1.4}_{-1.1}$ &
 $16.0/26$ &
 $25$ &
 WL(\#3) + Zitrin+12 + 50\,kpc$/h$ cut\tablenotemark{d} 
\\
 $1.37^{+0.26}_{-0.22}$  & 
 $5.8^{+0.9}_{-0.8}$ &
 $15.4/20$ & 
 $29$ &
 WL(\#3) + {\sc Pixelens}
\\
 $1.26^{+0.20}_{-0.17}$  & 
 $6.0^{+0.9}_{-0.8}$ &
 $11.6/31$ &
 $28$ &
 WL(\#3) + {\sc Lenstool} 
\\
\enddata
\tablecomments{See for details Section \ref{subsec:slsys}.}
\tablenotetext{a}{Effective Einstein radius for a source at $z_s=2.5$
 predicted by the model.}
\tablenotetext{b}{Combination of strong and weak lensing mass models
 used for the fitting.  For all cases, Method \#3 of Table
 \ref{tab:nfw} is used for weak lensing.}
\tablenotetext{c}{This corresponds to our best model (Model \#7) 
 of Table \ref{tab:nfw}.}
\tablenotetext{d}{Now applying a central $12\arcsec$ ($\sim
 50\,$kpc\,$h^{-1}$) cut to the \citet{Zitrin+2012_M1206} based strong
 lensing model.}  
\end{deluxetable} 

 
\begin{deluxetable}{c|ccccc|ccccc} 
\tabletypesize{\scriptsize}
\tablecolumns{11}
\tablecaption{\label{tab:spls} Best-fit SPLS model parameters for MACS1206.} 
\tablewidth{0pt} 
\tablehead{ 
 \multicolumn{1}{c|}{Method\tablenotemark{a}} &
 \multicolumn{1}{c}{$\kappa_0$} &
 \multicolumn{1}{c}{$r_c$} &
 \multicolumn{1}{c}{$\eta$} &
 \multicolumn{1}{c}{$\chi^2/{\rm dof}$} &
 \multicolumn{1}{c|}{$M_{\rm vir}$} &
 \multicolumn{1}{c}{$\kappa_0$} &
 \multicolumn{1}{c}{$r_c$} &
 \multicolumn{1}{c}{$\eta$} &
 \multicolumn{1}{c}{$\chi^2/{\rm dof}$} &
 \multicolumn{1}{c}{$M_{\rm vir}$} 
\\ 
 \multicolumn{1}{c|}{} &
 \multicolumn{1}{c}{} &
 \multicolumn{1}{c}{(kpc\,$h^{-1}$)} & 
 \multicolumn{1}{c}{} &
 \multicolumn{1}{c}{} &
 \multicolumn{1}{c|}{($10^{15}M_\odot\,h^{-1}$)} &
 \multicolumn{1}{c}{} &
 \multicolumn{1}{c}{(kpc\,$h^{-1}$)} & 
 \multicolumn{1}{c}{} &
 \multicolumn{1}{c}{} &
 \multicolumn{1}{c}{($10^{15}M_\odot\,h^{-1}$)} 
}
\startdata  
 \#6&
 $3.57^{+0.70}_{-0.54}$ &
 $23.1^{+5.1}_{-4.4}$ &
 1 & 
 $26.5/31$ &
 $1.78\pm 0.56$ &
 $3.17^{+0.60}_{-0.48}$ &
 $33.0^{+10.9}_{-8.4}$ & 
 $0.84^{+0.11}_{-0.14}$ &
 $24.0/30$ &
 $1.41\pm 0.34$
\\
\#7 &
 $3.62^{+0.71}_{-0.56}$ &
 $22.4^{+5.0}_{-4.4}$ &
 1 &
 $23.9/31$ &
 $1.74\pm 0.55$ &
 $3.07^{+0.57}_{-0.47}$ &
 $36.8^{+12.8}_{-9.4}$ &
 $0.77^{+0.13}_{-0.17}$ &
 $19.5/30$ &
 $1.26\pm 0.37$\\
\enddata
\tablecomments{
The convergence profile of the softened power-law sphere
 (SPLS) model, $\rho(r)=\rho_0(1+r^2/r_c^2)^{(\eta-3)/2}$,
is given by
 $\kappa(\theta)=\kappa_0\left(1+\theta^2/\theta_c^2\right)^{(\eta-2)/2}$ where
 $\theta_c=r_c/D_l$ and
 $\kappa_0=B\left(\frac{1}{2},1-\frac{\eta}{2}\right)\rho_0 r_c/\Sigma_{\rm crit}$
with $B$ being the standard Euler beta function.
Here $\Sigma_{\rm crit}$
 is evaluated for a source at a reference redshift of $z_s=2.54$.
For details, see Section \ref{subsec:spls}.
}
\tablenotetext{a}{Fitting method in Table \ref{tab:nfw}.}
\end{deluxetable} 

 
\begin{deluxetable}{ccccl} 
\tabletypesize{\scriptsize}
\tablecolumns{5}
\tablecaption{\label{tab:xray} Comparison with X-ray cluster mass estimates.} 
\tablewidth{0pt} 
\tablehead{ 
 \multicolumn{1}{c}{Data} &
 \multicolumn{1}{c}{$M_{2500}$} &
 \multicolumn{1}{c}{$c_{2500}$} &
 \multicolumn{1}{c}{$r_{2500}$} &
 \multicolumn{1}{c}{$\theta_{\rm Ein}$\tablenotemark{a}} 
\\ 
 \multicolumn{1}{c}{} &
 \multicolumn{1}{c}{$(10^{14}M_\odot)$} &
 \multicolumn{1}{c}{} &
 \multicolumn{1}{c}{(Mpc)} &
 \multicolumn{1}{c}{($\arcsec$)} 
}
\startdata  
 Chandra  &
 $4\pm 1$  & 
 $1.8\pm 1.5$ &
 $0.6\pm 0.1$ &
 23\\
 Chandra+XMM &
 $4.5\pm 0.3$ & 
 $0.9\pm 0.3$ & 
 $0.58\pm 0.02$ &
 20\\
 WL+SL\tablenotemark{b} &
 $4.9\pm 1.3$  & 
 $1.8\pm 0.3$ &
 $0.60\pm 0.06$ &
 28\\
\enddata
\tablecomments{
See for details Section \ref{subsec:xray}.
All quantities here are given in physical units assuming the concordance
 $\Lambda$CDM cosmology ($h=0.7, \Omega_m=0.3, \Omega_\Lambda=0.7$).
}
\tablenotetext{a}{Effective Einstein radius for a source at $z_s=2.5$
 predicted by the model.}
\tablenotetext{b}{Model \#7 of Table \ref{tab:nfw} based on the full weak
 and strong lensing constraints.}
\end{deluxetable} 

 
\begin{deluxetable}{ccccc} 
\tabletypesize{\scriptsize}
\tablecolumns{5}
\tablecaption{\label{tab:sze} Bolocam SZE-derived cluster mass estimates.} 
\tablewidth{0pt} 
\tablehead{ 
 \multicolumn{1}{c}{Overdensity\tablenotemark{a}} &
 \multicolumn{2}{c}{Bolocam-derived $r_\Delta$\tablenotemark{b}} &
 \multicolumn{2}{c}{Lensing-derived $r_\Delta$\tablenotemark{c}}\\
 \multicolumn{1}{c}{$\Delta$} &
 \multicolumn{1}{c}{$r_\Delta$} &
 \multicolumn{1}{c}{$M(<r_\Delta)$} &
 \multicolumn{1}{c}{$r_\Delta$} &
 \multicolumn{1}{c}{$M(<r_\Delta)$}\\
 \multicolumn{1}{c}{} &
 \multicolumn{1}{c}{(Mpc)} &
 \multicolumn{1}{c}{($10^{14}M_\odot$)} &
 \multicolumn{1}{c}{(Mpc)} &
 \multicolumn{1}{c}{$(10^{14}M_\odot)$}
}
\startdata  
$2500$ & $0.63^{+0.01 +0.06}_{-0.02 - 0.05}$ & $5.8^{+0.4 + 1.7}_{-0.4 -1.4}$ &
         $0.60$                 & $5.3^{+0.2 +0.8}_{-0.2 -0.7}$ \\
$500$ & $1.67^{+0.09 + 0.12}_{-0.08 -0.12}$  & $21.2^{+3.7 +5.1}_{-3.0 -4.3}$& 
        $1.31$                  & $15.7^{+1.2 +2.3}_{-1.1 -2.1}$ \\
\enddata
\tablecomments{
For each value the first error estimate represents our
measurement uncertainty and the second error estimate represents our
uncertainty due to systematics in our fitting method, flux calibration,
and choice of parameterization.
See for details Section \ref{subsec:sze}.
All quantities here are given in physical units assuming the concordance
 $\Lambda$CDM cosmology ($h=0.7, \Omega_m=0.3, \Omega_\Lambda=0.7$).
}
\tablenotetext{a}{Mean interior overdensity with  respect to the
 critical density of the universe at the cluster redshift $z=0.439$.} 
\tablenotetext{b}{Bolocam cluster mass estimates at the Bolocam-SZE
 derived values of overdensity radius $r_\Delta$.}
\tablenotetext{c}{Bolocam cluster mass estimates at the lensing-derived
 values of overdensity radius $r_\Delta$.}
\end{deluxetable}


\appendix

\section{Combining Lens Distortion and Magnification}
\label{sec:wlmethod}

\subsection{One-Dimensional Method}
\label{subsec:1D}
  
We first summarize the Bayesian method
of \citet{Umetsu+2011a}
for a direct reconstruction of the cluster mass profile 
from combined radial distortion and magnification profiles. 

In the Bayesian framework, we sample from the posterior PDF
of the underlying signal  $\bs$ given the data
$\bd$, $P(\bs|\bd)$. 
Expectation values of any statistic of the signal $\bs$ shall converge
to the expectation values of the a posteriori marginalized PDF,
$P(\bs|\bd)$.  
For a mass profile analysis, $\bs$ is a vector containing 
the discrete convergence profile,  
$\kappa_i\equiv \kappa(\theta_i)$ with $i=1,2,..,N$ in the
subcritical regime ($\theta_i>\theta_{\rm Ein}$), and the average
convergence within 
the inner radial boundary $\theta_{\rm min}$ of the weak lensing data, 
$\overline{\kappa}_{\rm min}\equiv \overline{\kappa}(<\theta_{\rm min})$, 
so that $\bs =\{\overline{\kappa}_{\rm min},\kappa_i\}_{i=1}^{N}$, being   
specified by $(N+1)$ parameters.

Bayes' theorem states that
\begin{equation}
P(\bs|\bd) \propto P(\bs) P(\bd|\bs),
\end{equation}
where ${\cal L}(\bs)\equiv P(\bd|\bs)$ is the 
likelihood of the data
given the model ($\bs$), and $P(\bs)$ is the prior probability
distribution for the model parameters.
The ${\cal L}(\bs)$ function for
combined weak lensing observations is given as a product of the
two separate likelihoods, ${\cal L}={\cal L}_{g_+}{\cal L}_\mu$,
where ${\cal L}_{g_+}$ and ${\cal L}_\mu$ are the likelihood functions for
tangential distortion and magnification bias, respectively.
The log-likelihood functions 
for the weak-lensing observations $\{g_{+,i}\}_{i=1}^{N}$ and
$\{n_{\mu,i}\}_{i=1}^{N}$ are given respectively (ignoring constant
terms) as
\begin{eqnarray}
\label{eq:lg}
l_{g_+}(\bs) &\equiv& -\ln{{\cal L}_g}=
\frac{1}{2}
\displaystyle\sum_{i=1}^{N}
\frac{[g_{+,i}-\hat{g}_{+,i}(\bs)]^2}{\sigma_{+,i}^2},\\
\label{eq:lmu}
l_\mu(\bs) &\equiv& -\ln{{\cal L}_\mu}=
\frac{1}{2}
\displaystyle\sum_{i=1}^{N}
\frac{[n_{\mu,i}-\hat{n}_{\mu,i}(\bs)]^2}{\sigma_{\mu,i}^2},
\end{eqnarray}
where 
$\{\hat{g}_{+,i}\}_{i=1}^{N}$ and
$\{\hat{n}_{\mu,i}\}_{i=1}^{N}$ 
are the theoretical predictions
for the corresponding observations.  The total likelihood 
$l_{\rm 1D}(\bs)\equiv -\ln{\cal L}$
of the combined observations is obtained as
\begin{equation}
l_{\rm 1D} = l_{g_+} +l_\mu.
\end{equation}
Here we consider a simple flat prior with a lower bound of $\bs=0$.
Additionally, we account for the uncertainty in 
the calibration parameters, $\bc=(n_0,s,\omega)$, namely
the normalization and slope parameters ($n_0,s$) of
the background counts and the relative lensing depth
$\omega\equiv \langle\beta({\rm red})\rangle/\langle\beta({\rm back})\rangle$ 
between the background
samples used for the magnification and distortion measurements.

We use the Markov chain Monte Carlo (MCMC) technique
with Metropolis-Hastings sampling to constrain our mass model 
$\bs$.
The covariance matrix ${\cal C}$ of $\bs$ is obtained from MCMC
samples.

\subsection{Two-Dimensional Method}
\label{subsec:2D}

Here we extend the 1D method of \citet{Umetsu+2011a} to 
a 2D mass distribution $\kappa(\btheta)$, by combining 2D distortion
data with the azimuthally-averaged magnification information.
For this analysis, the signal $\bs$ is a vector of parameters containing
discrete mass elements on a 2D Cartesian grid of independent cells:
$\bs=\{\kappa_m\}_{m=1}^{N_{\rm cell}}$.
The $\gamma(\btheta)$ field can be written as a linear combination of the
parameters $\bs$ (Equation (\ref{eq:k2g})).  
Then, the distortion $g(\btheta)$ and magnification $\mu(\btheta)$
fields can be uniquely specified in the subcritical regime (Section
\ref{sec:basis}).

In analogy to Equation
(\ref{eq:gt}), we calculate the weighted average $g_{\alpha,m}\equiv
g_{\alpha}(\btheta_m)$ ($\alpha=1,2$) of individual distortion
estimates,
and its covariance matrix,
\begin{equation}
\label{eq:gCov}
{\rm Cov}[g_{\alpha,m}, g_{\beta,n}]\equiv
\left(C_g\right)_{\alpha\beta,mn}
=
\frac{1}{2}
\sigma^2_{g}(\btheta_m)
\delta_{mn}
\delta_{\alpha\beta},
\end{equation}
where $\sigma^2_g(\btheta_m)$ is 
the standard error of the weighted mean distortion, $g(\btheta_m)$.  
Accordingly, the 2D shear log-likelihood function 
$l_g(\bs)\equiv -\ln{{\cal L}_g}$
is written as
\begin{equation}
\label{eq:lg2}
l_g(\bs) = \frac{1}{2}
 \sum_{m,n=1}^{N_{\rm cell}}
 \sum_{\alpha,\beta=1}^{2}
[g_{\alpha,m}-\hat{g}_{\alpha,m}(\bs)]
\left({\cal W}_g\right)_{\alpha\beta,mn}[g_{\beta,n}-\hat{g}_{\beta,n}(\bs)],
\end{equation}
where  $\hat{g}_{\alpha,m}(\bs)$ is the theoretical prediction for 
$g_{\alpha,m}$, and $({\cal W}_g)_{\alpha\beta,mn}$
is the shear weight matrix,
\begin{equation} 
\left({\cal W}_g\right)_{\alpha\beta,mn} =
M_m M_n \left({\cal C}_g^{-1}\right)_{\alpha\beta,mn},
\end{equation} 
with
$M_m$ being a mask weight, defined such that $M_m=0$ if the
$m$th cell is masked out and $M_m=1$ otherwise.
In practice, we exclude from our analysis innermost cells
which lie in the cluster central region, where the surface mass density
can be close to or greater than the
critical value (i.e., $\kappa\simgt 1$).
Furthermore, this is crucial to  minimize contamination
by unlensed cluster member galaxies
(see Section \ref{subsec:color}).

Now we combine 2D distortion data with magnification
information to obtain the total log-likelihood $l_{\rm
2D}(\bs)$ as
\begin{equation}
\label{eq:l2D}
l_{\rm 2D} = l_{g} + l_\mu,
\end{equation}
where $l_\mu$, given by Equation (\ref{eq:lmu}),
imposes a set of azimuthally-integrated constraints on the underlying
$\kappa$ field.  Since the degree of magnification
is locally related to $\kappa$, this
will essentially provide the (otherwise unconstrained) normalization of
$\kappa(\btheta)$ over a set of concentric rings where count
measurements $n_{\mu,i}$ are available.
Note, no assumption is made of azimuthal symmetry or isotropy of the
cluster mass distribution. 

This 2D inversion problem involves estimation of a large number of
parameters $\bs$; typically, $N_{\rm cell}\simgt 1000$ when 
distortion data are binned into subarcminute pixels.
We use in our implementation 
the conjugate-gradient method
\citep{1992nrfa.book.....P} to find the best solution.
We include Gaussian priors on the calibration nuisance parameters
$\bc=(s,n_0,\omega)$, 
given by means of quadratic penalty terms with mean values and variances
directly estimated from data.  
The log posterior PDF, $F = -\ln{P(\bs|\bd)}$, is
expressed as a linear sum of $l_{\rm 2D}(\bs)$ 
and the prior terms on $\bc$.
The best-fit parameters are determined with a maximum-likelihood
estimation, by minimizing the function $F$ with respect to 
$\bp\equiv (\bs,\bc)$, a vector containing the mass and calibration
parameters. 
Here we employ an analytic expression for the gradient function $\bnabla
F(\bp)$ obtained in the nonlinear subcritical regime.
To quantify the errors on the mass reconstruction, we
evaluate the Fisher matrix at the maximum likelihood estimate
$\bp=\hat{\bp}$ as 
\begin{equation}
{\cal F}_{mn} = 
\left\langle \frac{\partial^2 F(\bp)}
{\partial p_m \partial p_n}
\right\rangle\Big|_{\bp=\hat{\bp}}
\end{equation}
where the angular brackets represent an ensemble average, and the
indices $(m,n)$ run over all model parameters.
We estimate the covariance matrix ${\cal C}$ of $\bs$ as
\begin{equation}
{\cal C}_{mn} = \left({\cal F}^{-1}\right)_{mn}.
\end{equation}

\end{document}